\documentclass{aa}
\usepackage{natbib}
\usepackage{graphicx}
\usepackage[varg]{txfonts}
\usepackage{color}
\usepackage{longtable,lscape}

\begin{document} 

\title{Interplay between pulsation, mass loss, and third dredge-up: More
  about Miras with and without technetium\thanks{Tables~\ref{taba1} and \ref{taba2} are only available in electronic form at the CDS via anonymous ftp to cdsarc.u-strasbg.fr (130.79.128.5) or via http://cdsweb.u-strasbg.fr/cgi-bin/qcat?J/A+A/.}}

\author{
S. Uttenthaler\inst{1}, I. McDonald\inst{2}, K. Bernhard\inst{3,4}, S. Cristallo\inst{5,6},
D. Gobrecht\inst{7}
}

\institute{Kuffner Observatory, Johann-Staudstra\ss e 10, 1160 Vienna; 
\email{stefan.uttenthaler@gmail.com}
\and
Jodrell Bank Centre for Astrophysics, Alan Turing Building, Manchester, M13 9PL, UK;\\
\email{iain.mcdonald-2@jb.man.ac.uk}
\and
Bundesdeutsche Arbeitsgemeinschaft f\"ur Ver\"anderliche Sterne e.V. (BAV), Berlin, Germany;\\
\email{klaus.bernhard@liwest.at}
\and
American Association of Variable Star Observers (AAVSO), Cambridge, MA, USA
\and
INAF -- Osservatorio Astronomico, 64100 Italy
\and
INFN -- Sezione di Perugia, Italy
\and
Instituut voor Sterrenkunde, Celestijnenlaan 200D, bus 2401, 3001 Leuven, Belgium;  \email{david.gobrecht@kuleuven.be}
}

\date{Received July 09, 2018; accepted December 08, 2018}

\abstract
    {We follow-up on a previous finding that AGB Mira variables containing the third
    dredge-up indicator technetium (Tc) in their atmosphere form a different
    sequence of $K-[22]$ colour as a function of pulsation period than Miras without Tc.
    A near- to mid-infrared colour such as $K-[22]$ is a good probe for the {\it \textup{dust}}
    mass-loss rate of the stars. Contrary to what might be expected, Tc-poor Miras
    show {\it \textup{redder}} $K-[22]$ colours (i.e.\ higher dust mass-loss rates) than
    Tc-rich Miras at a given period.}
    {Here, the previous sample is extended and the analysis is expanded towards other
    colours and dust spectra. The most important aim is to investigate if the same two
    sequences can be revealed in the {\it \textup{gas}} mass-loss rate.}
    {We analysed new optical spectra and expanded the sample by including more stars from the
    literature. Near- and mid-IR photometry and {\it ISO} dust spectra of our stars were
    investigated where available. Literature data of gas mass-loss rates of Miras and
    semi-regular variables were collected and analysed.}
    {Our results show that Tc-poor Miras are redder than Tc-rich Miras in a broad range
    of the mid-IR, suggesting that the previous finding based on the $K-[22]$ colour is
    not due to a specific dust feature in the 22\,$\mu{\rm m}$ band. We establish a
    linear relation between $K-[22]$ and the gas mass-loss rate. We also find that the
    13\,$\mu{\rm m}$ feature disappears above $K-[22]\simeq2.17$\,mag, corresponding to
    $\dot{M}_{\rm g}\sim2.6\times10^{-7}\,M_{\sun}\,yr^{-1}$. No similar sequences of Tc-poor
    and Tc-rich Miras in the gas mass-loss rate vs.\ period diagram are found,
    most probably owing to limitations in the available data.}
    {Different hypotheses to explain the observation of two sequences in the
    \mbox{$P$ vs. $K-[22]$} diagram are discussed and tested, but so far, none of them
    convincingly explains the observations. Nevertheless, we might have found an hitherto
    unknown but potentially important process influencing mass loss on the TP-AGB.}
    \keywords{stars: AGB and post-AGB -- stars: late-type -- stars: evolution -- stars: mass-loss -- stars: oscillations}
\authorrunning{S.\ Uttenthaler et al.}
\titlerunning{More about Miras with and without Tc}
\maketitle

\section{Introduction}

For low- to intermediate-mass stars $(1-8M_{\sun})$, the final stage of energy production
through nuclear burning is the asymptotic giant branch (AGB). In this phase, stars become
very luminous, and their atmospheres are very cool. Three important processes
characterise stars on the AGB: i) pulsations of the outer atmosphere; ii) internal
nucleosynthesis and mixing; and iii) mass loss.

The pulsations have their origin deep within the convective envelope of the star
and cause it to appear strongly variable in the visual range of the spectrum. 
The long-period variables (LPVs), consisting of Miras, semi-regular (SRVs), and irregular (Lb)
variables, pulsate on timescales of 80 to more than 1000 days and have visual amplitudes larger
than 0.1\,mag. It has been shown that LPVs follow period-luminosity sequences depending on
their pulsation mode \citep{Wood99,Whi08,Rie10,Wood15}. Mira variables almost certainly
pulsate in the fundamental mode, whereas SRVs tend to pulsate in the first or higher overtones
\citep{Tra17}. It should be noted, however, that semi-regular variables of type {\it a}
(SRa) have comparatively regular light curves and probably pulsate in the fundamental mode as well:
their amplitude is only too low for them to be classified as Miras \citep[$\Delta V<2\fm5$,][]{Sam17}.

The elemental abundances at the visible photosphere of evolved AGB stars are altered by
the mixing of nuclear-burning products to the outer layers of the star following thermal
pulses (TPs). This process is called third dredge-up (3DUP) and typically occurs several
times on the AGB. One important nuclear product is carbon, which is usually less abundant
than oxygen in the stellar atmosphere but can be enriched so much by 3DUP that it becomes
more abundant (by number of atoms) than oxygen. This has important consequences on the
spectral appearance of the stars because the more abundant species will determine which
molecules, oxygen- or carbon-bearing, will form \citep{Cri07,LE10}. Intermediate-mass
stars $(\gtrsim4M_{\sun})$ may avoid becoming carbon-rich by converting carbon to
nitrogen in the CN-cycle at the hot base of the convective envelope. This process is
known as hot bottom burning \citep[HBB; e.g.][]{SB92}. Another result of HBB is that
these stars can become strongly enriched in lithium \citep{GH13}.

In addition to carbon, heavy elements are synthesised deep within the star by the slow
neutron-capture process (s-process). The element technetium (Tc) is an important indicator
of s-process and 3DUP because it has only radioactively unstable isotopes. The
longest-lived isotope that is produced in the s-process, $^{99}$Tc, has a half-life of
$\sim2\times10^{5}$ years, much shorter than the stellar lifetime. Absorption lines of
Tc can thus be used to distinguish AGB stars that have produced Tc in situ and undergone
a 3DUP event from those that have not \citep[e.g.][]{Utt07}. Hence, Tc-poor is synonymous
for stars before the 3DUP stage (pre-3DUP), whereas Tc-rich is synonymous for stars that
have undergone 3DUP events (post-3DUP).

The process that eventually determines the fate of the star is mass loss from its surface,
rather than nuclear burning in its interior. The stellar pulsations are thought to play an
important role in the mass-loss process because pulsation-induced shock waves lift gas to
cooler, but still dense layers of the atmosphere where dust grains can form and condense.
The dust grains then absorb or scatter photons from the stellar surface, so gaining
momentum away from the star. They then collide with atoms and molecules in the gas phase,
thus transferring momentum to them and triggering the outflow \citep{Hoef08}. While a
relation between pulsation and mass loss is widely accepted, the details of the mass-loss
process are still a matter of debate in particular for oxygen-rich stars
\citep{Wil00,Hoef08,Eri14,Hoef16}.
In general, the poor knowledge of the mass-loss process is one of the major uncertainties
in the modelling of AGB star evolution \citep[e.g.][]{SGC06}. For details and the current
understanding of AGB mass loss, we refer to the recent review by \citet{HO18}.

Following from this description, the three processes (pulsation, 3DUP, and mass loss)
influence each other. Indeed, some correlation between pulsation period and mass-loss
rate has been found. For example, \citet{MZ16} find that a threshold period of
$\sim60$\,d exists above which the dust mass-loss rate, as indicated by a near- to mid-infrared
colour, increases significantly, while \citet{Glass09} find a similar discontinuity
using different mid-IR bands. \citet{MZ16} also demonstrated that the mass-loss rate
is less well correlated to the luminosity of nearby cool giant stars, indicating that
radiation pressure on dust has little influence on the mass-loss rate at this
evolutionary stage. This is important because most stellar evolution calculations
assume a Reimers-like mass-loss law \citep{Rei75}, which is directly related to the
stellar luminosity.

A second increase in mass-loss rate as a function of pulsation period is found between
300 and 400\,d of period \citep{VW93,Mat05}. Probably, this transition coincides with the
timescale for dust to effectively form and grow, so that the radiation pressure on dust
grains can drive the mass loss. However, a more linear correlation between mass-loss
rate and period of Miras is not always obvious, see\ \citet{Jura93}, for instance. Such a correlation
may become apparent only when Miras with very long pulsation period ($P\gtrsim600$\,d)
are included \citep{Groen09} or when the selection is restricted to C-rich Miras
\citep{Whi06}. Most recently, \citet{Gol16} have found only a weak dependence of mass-loss
rate on pulsation period for intermediate-mass (OH/IR) stars in the late super-wind
phase.

The relation between chemistry (3DUP) and mass loss has been investigated by a few works.
For example, \citet{GJ98} found that S-type Miras follow the same period -- gas mass-loss rate
relation as C-type Miras. However, gas mass-loss rates derived from radio CO rotational
lines have large uncertainties \citep{Ram08}, and effects of 3DUP become visible in a
stellar spectrum before it becomes an S-type star \citep[e.g.][]{Utt11}.

In a previous work \citep[][hereafter Paper~I]{Utt13}, it was shown that Mira variables
form two separate sequences in the pulsation period versus $K-[22]$ colour diagram
(henceforth called the \mbox{$P$ vs.\ $K-[22]$} diagram) if a distinction is made for
the presence of Tc. Here, $[22]$ is the 22\,$\mu{\rm m}$ band of the Wide-field Infrared
Survey Explorer ({\it WISE}) space observatory \citep{Wri10}. The $K-[22]$ colour is
an indicator of the dust mass-loss rate of the stars. Such a clear correlation between
period and mass-loss rate has previously not been found in these stars. Interestingly, at
a given period, the pre-3DUP Miras were found to have a {\it \textup{higher}} dust mass-loss rate
than the post-3DUP Miras. This result is somewhat counter-intuitive because one would
expect the post-3DUP Miras to be more evolved than the pre-3DUP Miras, thus that they
would have higher mass-loss rates. So far, there is no satisfactory explanation for
these observations.

This paper is a follow-up work on Paper~I. It is motivated along several
lines. First of all, since the publication of Paper~I, the sample was expanded by
analysing new observations. Moreover, more Tc-rich SRVs have been included in the present
work because it was noted in Paper~I that these stars form a sequence of increasing
$K-[22]$ colour at short periods ($P\lesssim200$\,d). SRVs probably represent an
evolutionary stage before a star switches to the fundamental pulsation mode of the
Mira phase \citep{LH99,Tra17}, making a comparison interesting. Second, the
observations from Paper~I were analysed and discussed in more detail, for example\ by
presenting diagrams that involve other combinations of near- to mid-IR colours, as
well as Infra-red Space Observatory ({\it ISO}) dust spectra of stars with and
without Tc.

The most important motivation comes from the fact that the $K-[22]$ colour used
in the previous work is only an indicator of the {\it \textup{dust}} mass-loss rate. Basically,
$K-[22]$ probes the dust column density, which can be translated into a dust mass-loss
rate if the expansion velocity is known. This can be measured from rotational
lines of CO in the gas phase, for instance, which are observable at radio wavelengths. More importantly, dust
is only a minor (although important) fraction of the total mass lost from a cool giant
star; most of the mass is lost in the form of {\it \textup{gas}}. It may be assumed that the
gas mass-loss rate ($\dot{M}_{\rm g}$) equals the \textup{{\it total}} mass-loss rate. The
latter assumption is justified by the fact that typical dust-to-gas mass ratios in
AGB stars, depending on the individual star, are of the order 10$^{-4}$-10$^{-2}$. A
question that naturally arises is whether the two sequences identified in the
\mbox{$P$ vs.\ $K-[22]$} diagram also show up if the gas mass-loss rate is inspected
instead of the dust column density. Thus, we here aim to expand our analysis to the
gas mass-loss rate and its relation to the pulsation period on the one hand, and the
occurrence of 3DUP on the other. This could help to identify the underlying processes
that cause the two separate sequences in the \mbox{$P$ vs.\ $K-[22]$} diagram.
%

The paper is structured in the following way: In Sect.~\ref{data} the data used
in this work are described. These data are analysed in Sect.~\ref{analysis}, where
results are also presented. Hypotheses for explaining the observations are discussed
in Sect.~\ref{dicuss}. In Sect.~\ref{bulgesec} our results are applied to a
sample of Galactic bulge Miras to search for the occurrence of 3DUP in that
stellar population. Finally, the results and conclusions are summarised in
Sect.~\ref{conclusio}.

\section{Data}\label{data}

The data set compiled for Paper~I also forms the basis of the present paper.
The sample was extended in two ways: First, 25 Tc-rich SRVs were added from the
works of \citet{Lit87}, \citet{Van98}, \citet{LH99}, \citet{VJ99}, and
\citet{Van00}. Second, we searched the spectra from the ESO Large Programme
X-shooter Spectral Library \citep[XSL;][]{Chen14} for AGB stars that could be
analysed for their Tc content.

\subsection{Stars from the X-shooter library}\label{xsldat}

All available Phase-3 science-ready spectra of M-, S-, and C-type stars observed in the
XSL were downloaded from the ESO archive. These spectra were obtained with the X-shooter
spectrograph \citep{Ver11} mounted to UT2 of ESO's Very Large Telescope on Cerro Paranal,
Chile. Dwarf and supergiant stars were excluded from the study (as expected, none of them
was found to exhibit the Tc lines in the subsequent analysis). Of the remaining stars,
many had a signal-to-noise ratio (S/N) in the wavelength range of interest
(4200 -- 4300\,\AA) that was too low to decide on the presence or absence of the Tc lines.

The spectra were first shifted to the rest-frame by measuring the radial velocity in the
wavelength range 5365 -- 5435\,\AA, using a cross-correlation technique. This interval
was chosen because the very red sample stars emit much more flux here than in the vicinity
of the Tc lines. After correcting for the radial velocity shift, we searched for the
\ion{Tc}{i} lines at 4238.191, 4262.270, and 4297.058\,\AA\  \citep{Bozman} by visually
inspecting and comparing the spectra and by applying the flux-ratio method introduced
by \citet{Utt07}. In this method, the mean flux in a narrow region around a nearby
quasi-continuum point is divided by the mean flux in a few  wavelength points around the
laboratory wavelengths of the Tc lines. Tc-rich stars are identified by having enhanced
ratios in all three lines, clearly separating them from Tc-poor ones. In principle,
there are subordinate lines of Tc at even shorter wavelengths that have been identified
previously \citep{MG56}, but these were not analysed here because they would yield
little additional information due to the low flux of the stars and the comparably low
spectral resolution.

\begin{figure}
  \centering
  \includegraphics[width=\linewidth,bb=82 370 551 698, clip]{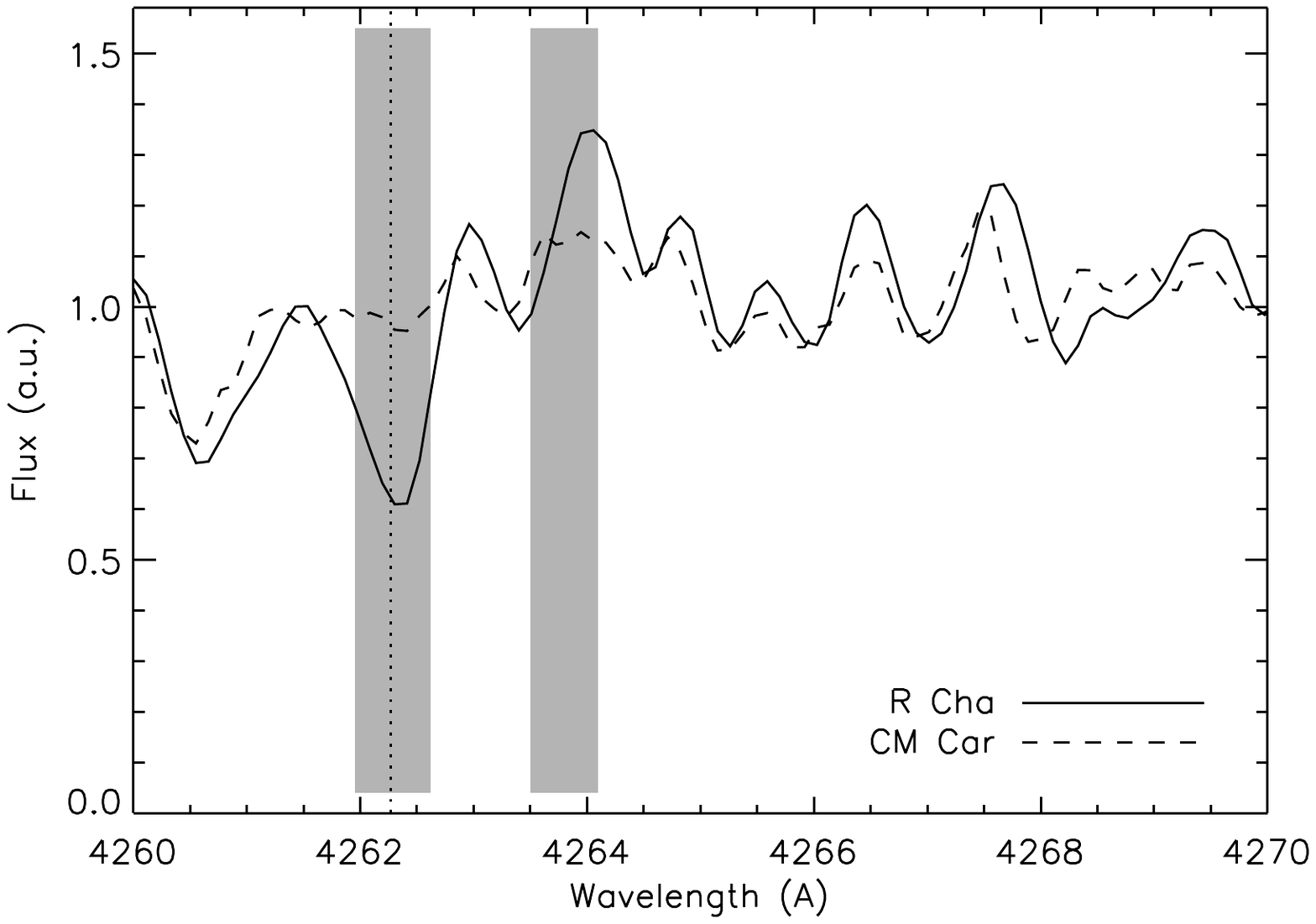}
  \caption{Demonstration of the detection of Tc lines in the XSL spectra. This
    example shows a portion of the spectra of R~Cha (solid line) and CM~Car
    (dashed line) around the Tc 4262.270\,\AA\ transition. The vertical dotted
    line indicates the laboratory wavelength of this transition \citep{Bozman}.
    The grey underlaid areas depict the wavelength ranges that were used to
    determine the mean fluxes in the line and continuum region, respectively,
    for the flux ratio method.}\label{demoTcdetect}
\end{figure}

The XSL spectra have a resolving power of $R=\lambda/\Delta\lambda=9\,800$ at the
wavelength of the Tc lines, which is lower than what has previously been used to
determine the presence of Tc lines in red giant stars. Although already somewhat
at the limit of viability, this resolution is still sufficient to distinguish
stars with Tc from those without. This is demonstrated in Fig.~\ref{demoTcdetect},
where a portion of the spectrum of the Miras R~Cha and CM~Car is shown around the
4262.270\,\AA\ Tc line. The vertical dotted line indicates the laboratory
wavelength of this transition. Clearly, R~Cha ($P=337.8$\,d, $K-[22]=1.811$) has
an absorption at the wavelength of the Tc line, whereas CM~Car ($P=339.0$\,d,
$K-[22]=2.708$) lacks this line. Except for this, the spectra are very similar.
The small flux excess in the quasi-continuum point of R~Cha seen here may
influence the flux ratio determined for this particular line, but the
classification into Tc-poor and Tc-rich is always done based on all three lines
and on visual inspection of the spectra. This pair of Miras with comparable
pulsation periods illustrates the strongly differing $K-[22]$ colours of Tc-rich
and Tc-poor stars (cf.\ Paper~I).

Eventually, we were able to decide on the presence of Tc in 18 giant stars with high enough
S/N. These stars were added to the original sample of Paper~I and are identified by
reference 15 in Col.~3 of Table~\ref{taba1}. Two of the stars in the XSL sample are
located in the Large Magellanic Cloud (LMC), namely SHV 0520261-693826 (abbreviated
SHV~0520 in Table~\ref{taba1}) and HV~12667. Thus, these are the first oxygen-rich stars
outside the Milky Way Galaxy whose Tc content and hence 3DUP activity has been determined.
While SHV~0520 does contain Tc in its atmosphere, HV~12667 does not. HV~12667 shows a
6708\,\AA\ Li line of significant strength, thus is likely an intermediate-mass
($M\gtrsim4\,M_{\sun}$) AGB star that undergoes HBB.

Furthermore, we also found that the C-type stars X~Cnc, HD~198140, and IRAS~10151-6008 do 
{\it \textup{not}} show lines of Tc. This confirms the result by \citet{Lit87} for X~Cnc. Hence,
these stars are {\it \textup{extrinsic}} carbon stars that owe their carbon enrichment to mass
transfer from a now extinct binary companion. Since we aim at investigating genuine AGB
stars and their mass-loss and pulsation properties, we excluded these stars from further
analysis. The carbon star Y~Hya was also observed by XSL, but the S/N of the spectrum is too low to decide on the presence of Tc.

Finally, the S-type star SU~Mon is contained in XSL and its spectrum has a very high S/N.
However, we were unable to determine the presence of Tc because its spectral lines
appear to be extremely broadened for an unknown reason. IR photometry of SU~Mon suggests
that it is an intrinsic S-star \citep{Groen93}, but since we are unable to confirm this
with spectral observations, this star was also excluded from the sample.

%
%

The data of the new sample stars such as pulsation period, near-IR magnitudes,
and spectral types have been collected in the same fashion as was done for
Paper~I. We refer to Sect.~2 of that paper for details. For most of the new
stars, {\it J}- and {\it K}-band magnitudes were available only from the Two
Micron All Sky Survey (2MASS) catalogue \citep{2MASS}, whereas for many of the
sample stars of Paper~I, measurements at two or more epochs were available. We
also included observations in the 65 and 90\,$\mu{\rm m}$ bands of the
{\it Akari} satellite \citep{Mur07} to increase the wavelength coverage by the
data and to improve the data quality at the long-wavelength end compared to
Infrared Astronomical Satellite \citep[{\it IRAS},][]{Neu84} data.
Throughout the paper, magnitudes are stated up to the third digit after
the decimal point, as in the catalogues, even if the accuracy due to the
stellar variability is certainly poorer than that.

The collected data are presented in Table~\ref{taba1} in the appendix. This
table supersedes the one from Paper~I published at the CDS. The extended sample contains
18 new stars from XSL and 25 Tc-rich SRVs to give a new total of 240 stars, or
a sample size increase of $\sim22$\% compared to Paper~I.

\subsection{Mass-loss rates from radio CO observations}\label{co_data}

In order to satisfy the main motivation of this work, that is, to determine whether the
two separate sequences of Miras can also be revealed if the gas mass-loss rate is
used instead of a dust mass-loss rate indicator, we also collected such measurements
of AGB stars from the literature. Because we are interested in measurements that are
independent from IR photometry, we only collected gas mass-loss rates determined
from radio CO rotational lines, while those obtained from dust radiative transfer
models were neglected. In total, data of 135 Mira stars with known period were
collected in Table~\ref{taba2}. This is a comprehensive but by no means complete
sample.
For each star, preference was given to the most recent mass-loss rate determination
in the hope that more recent observations would be more sensitive and more
recent works would apply more sophisticated methods for determining the rates. The
listed rate might not be the most accurate for each individual star, but this
procedure was deemed advantageous for the sample as a whole.
Nevertheless, for stars where more than one measurement was found in the literature,
usually these agreed reasonably well.

Data for SR and Lb variables were adopted only from the list of \citet{Olo02}, which contains 65 stars.
This was preferred because this is a homogeneous analysis of a sizeable sample, whereas a
literature compilation would be more inhomogeneous. These data are not reproduced in
Table~\ref{taba2}, which contains only data of Mira stars. Instead, we refer to Table~3 of
\citet{Olo02}. Pulsation periods of the stars have been updated, in particular, with data
from the All Sky Automated Survey \citep[{\it ASAS},][]{Poj97}. Since there appears to be
no real difference between SRV and Lb \citep{LO09}, we refer to this simply as the
SRV sample.

It should be mentioned that despite all the care taken, CO mass-loss rate
determinations suffer from a number of uncertainties. These trace back to uncertainties
in distance (hence luminosity), inhomogeneities in the outflow (clumpiness,
asymmetries), the dependence of the CO:H$_2$ ratio on the chemical type and metallicity,
the derivation of CO mass-loss rate from the observations, and variable dissociation
by interstellar UV radiation. \citet{Ram08} estimated that these determinations are
uncertain to within a factor of three.

\section{Analysis and results}\label{analysis}


\subsection{Miras in the $P$ vs.\ $K-[22]$ diagram}\label{Miras}

\begin{figure}
  \centering
  \includegraphics[width=\linewidth,bb=12 2 490 340, clip]{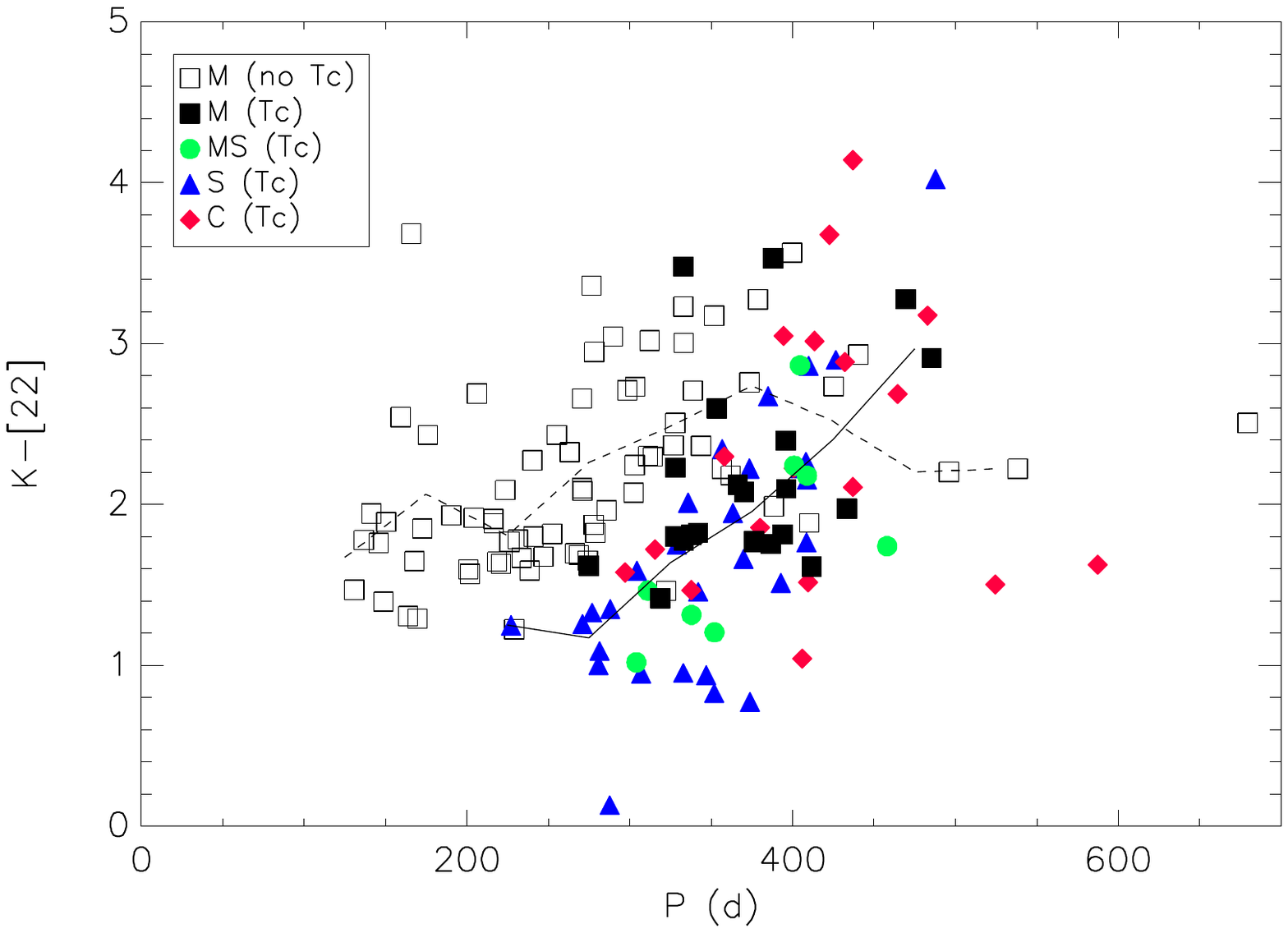}
  \caption{\mbox{$P$ vs.\ $K-[22]$} diagram of the sample Miras. The relation of
  spectral type and Tc content to colours and symbols is explained in the legend
  in the upper left corner of the plotting area. The dashed and solid lines show
  the run of the mean $K-[22]$ colour for Tc-poor and Tc-rich stars,
  respectively, in period bins of 50\,d.}\label{K-22}
\end{figure}

Figure~\ref{K-22} shows the updated version of the \mbox{$P$ vs.\ $K-[22]$} diagram
presented in Paper~I. Most of the newly added stars confirm the separation of Tc-poor
and Tc-rich Miras reported earlier. Miras with periods $<200$\,d will have
masses $<1M_{\odot}$ \citep{Feast09} and are thus expected to never undergo 3DUP.
In the period interval $\sim200 - 400$\,d, the Tc-poor Miras are clearly redder
than the Tc-rich ones, likely indicating a higher (dust) mass-loss rate at a given
period. This is supported by the mean $K-[22]$ colour shown in Fig.~\ref{K-22}.

Tc-rich Miras appear in substantial numbers only above a period of
$\approx275$\,d. This may be used to infer the age of these stars and hence the
initial mass required for a star to experience 3DUP during its AGB evolution.

It is useful to discuss a few outliers in Fig.~\ref{K-22} to understand their
physical origin. As noted in Paper~I, the stars R~Cen, R~Nor, and possibly W~Hya,
are intermediate-mass star candidates undergoing HBB. The Tc-poor object at
$P\approx680$\,d is the LMC star HV~12667 and is newly added to the sample from XSL
observations. As mentioned in Sec.~\ref{xsldat}, this star has a strong Li line,
indicating that it is an intermediate-mass star undergoing HBB similar to those
identified by \citet{Smi95} in the Magellanic Clouds. These few stars are the
reason why the mean $K-[22]$ colour of the Tc-poor stars is lower than that of
Tc-rich ones at $P\gtrsim450$\,d. As has been shown by \citet{GH13}, 
intermediate-mass stars on the early AGB have high Li abundances that are 
accompanied by weak or no s-process element enhancements because HBB is strongly
activated during the first TPs, but the $^{22}$Ne neutron source needs many more
TPs to be efficiently activated. Moreover, more 3DUP episodes are needed for the
s-elements to become detectable at the stellar surface. In addition, the Tc
abundance in intermediate-mass stars may be slightly reduced because of the
reduced effective half-life time of $^{99}$Tc at temperatures in excess of $10^8$\,K
\citep{Schatz83,Mathews86}.

Tc-rich Mira variables that are confirmed to be members in a binary system, such
as $o$~Cet and R~Aqr \citep{Whi87}, appear to be outliers in the sense that
they have redder $K-[22]$ colours than other Tc-rich Miras at comparable periods.
These stars have been excluded from computing the run of the mean colour (dashed
and solid lines). W~Aql is a binary AGB star as well, but it does not deviate so
significantly from the sequence of the other stars, although it is the reddest
S-type Mira and the second-reddest star in the whole sample. On the other hand,
the Tc-poor outlier R~Cet, which has been mentioned in Paper~I,  might also be a
binary star. It is the reddest star with a period shorter than 200\,d. The
unusually high amount of dust could be caused by a binary companion. The two
longest-period carbon stars LX~Cyg and BH~Cru are blue outliers, most probably
because  they just recently turned into carbon stars \citep{LE85,Utt16}.

\begin{figure}
  \centering
  \includegraphics[width=\linewidth,bb=18 36 484 340,clip]{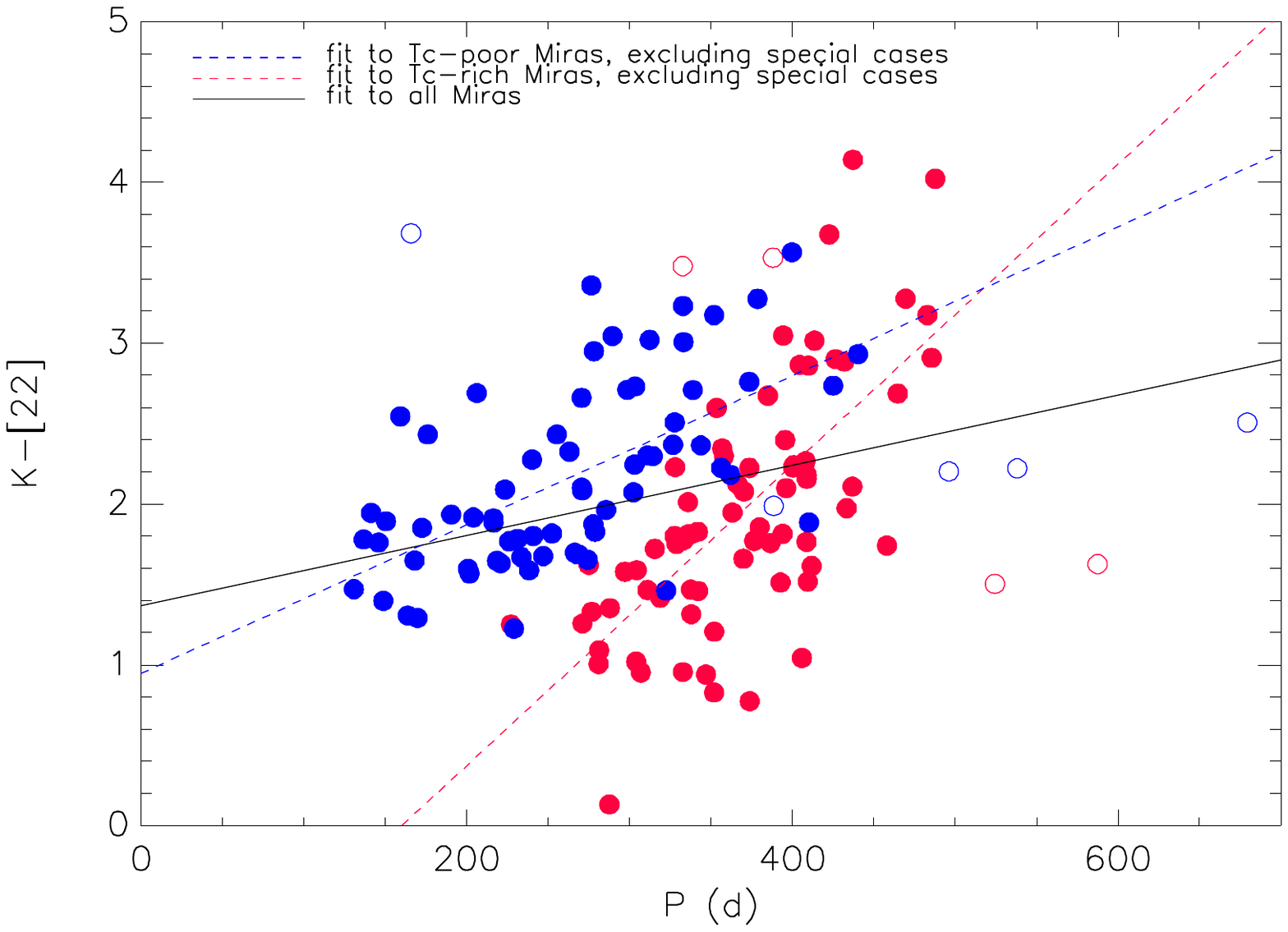}
  \includegraphics[width=\linewidth,bb=18  4 484 335,clip]{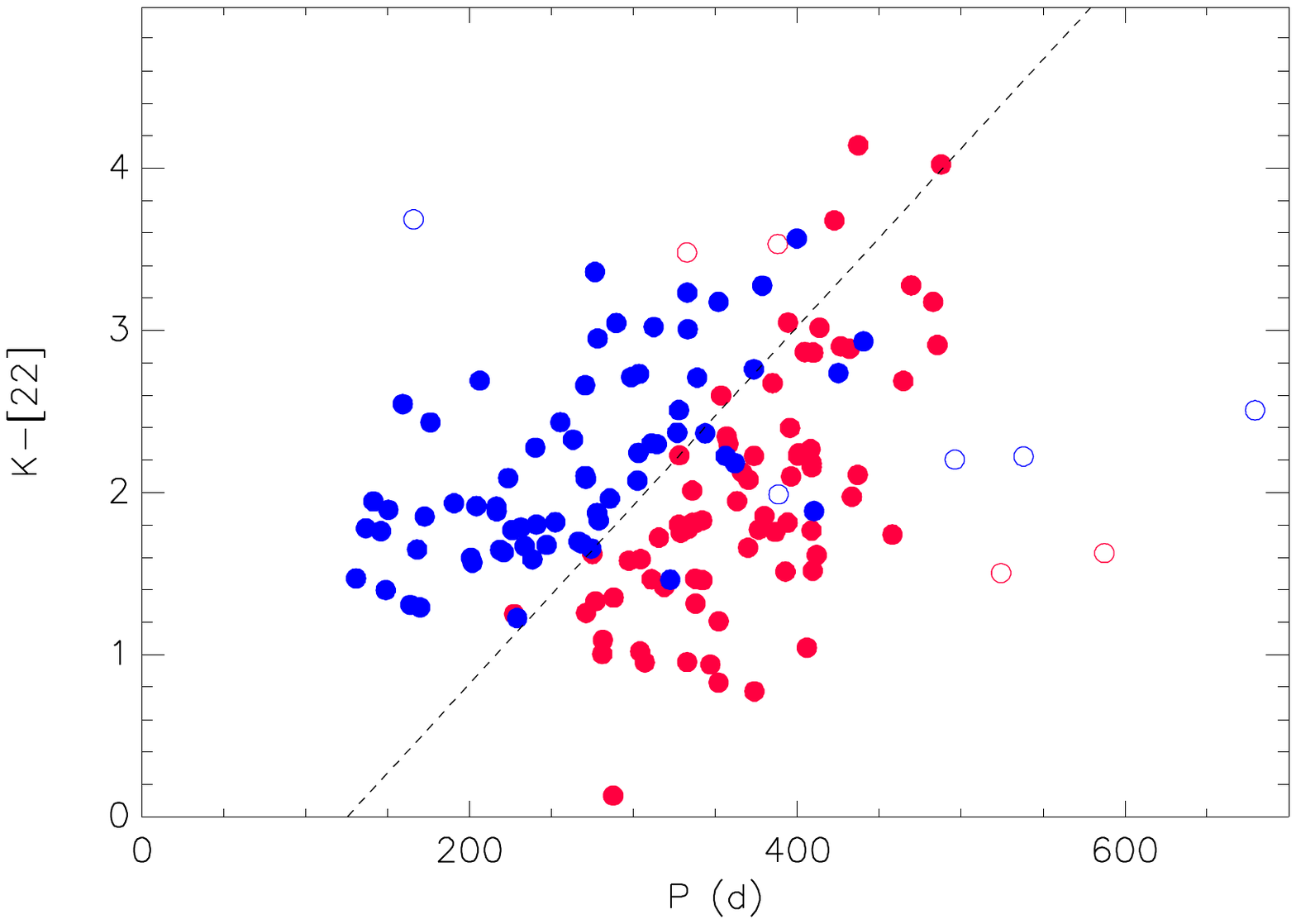}
  \caption{{\it Upper panel:} \mbox{$P$ vs.\ $K-[22]$} diagram with linear
  least-squares fits to different sub-samples of Miras. Blue dashed line: Fit
  to the Tc-poor Miras (filled blue circles), excluding the identified outliers
  (open circles, see text). Red dashed line: Same for the Tc-rich Miras (filled
  and open red circles, respectively). Black solid line: Fit to all Miras.
  {\it Lower panel:} \mbox{$P$ vs.\ $K-[22]$} diagram with the linear relation
  that best separates Tc-poor from Tc-rich Miras (Equ.~\ref{divlin_equ}).}\label{linfit}
\end{figure}

In the upper panel of Fig.~\ref{linfit} we show linear least-squares fits
to the sub-samples of Tc-poor (filled blue circles) and Tc-rich (filled red
circles) Miras, respectively. In this plot, the stars are not differentiated by
their spectral types. This figure demonstrates very clearly how well the two
groups are separated. The outliers mentioned above were excluded from
the fits (open red and blue circles, respectively). The linear fit to the Tc-rich
Miras has a much steeper slope than that to the Tc-poor ones. The two fits
formally cross at $P\approx505$\,d, but the longest-period Tc-poor Mira that is not
a hot bottom burner is at $P\approx440$\,d. It is also noteworthy that the linear
fit to all Mira stars has a quite shallow slope. Thus, without making a distinction
for the Tc content, the correlation between pulsation period and dust mass-loss
rate seems weak. However, for the two separate groups, the correlation becomes very
strong. Table~\ref{coeffs} summarises the slopes, intercepts, and correlation
coefficients $r$ of the three linear fits plotted in the upper panel of
Fig.~\ref{linfit}.

\begin{table}
\caption{Slopes, intercepts, and correlation coefficients ($r$) of the three linear
least-squares fits shown in the upper panel of Fig.~\ref{linfit}.}\label{coeffs}
\centering
\begin{tabular}{lcccc}
\hline\hline
 Sample & Slope & Intercept & $r$ \\
\hline
All Miras & $0.00218\pm0.00056$ & $+1.369\pm0.180$ & 0.298 \\
Tc-poor   & $0.00463\pm0.00073$ & $+0.946\pm0.187$ & 0.612 \\
Tc-rich   & $0.00935\pm0.00117$ & $-1.500\pm0.417$ & 0.696 \\
\hline
\end{tabular}
\end{table}

Tc-poor and Tc-rich Miras mainly occupy two different regions in the
\mbox{$P$ vs.\ $K-[22]$} diagram, and we note that the transition is very sharp.
We can thus construct a straight line that best separates the two groups. An amoeba
routine was used to find slope and intercept of the line at which the accuracy,
as defined for binary tests (i.e. the probability that a randomly chosen instance
will be correct), becomes maximal. The visually defined line of Paper~I was used as
starting point. The optimisation routine yields the relation

\begin{equation}
K - [22] = 0.011 \times P - 1.380.
\label{divlin_equ}
\end{equation}

Out of the 73 Tc-poor Miras, 62 are above this line, and out of the 78 Tc-rich ones,
70 are below the line. This yields an accuracy of $(62+70)/(73+78)\approx0.87$
(outliers discussed above were not excluded). Of the stars that are
on the "wrong" side of the line, many miss the line only by a very narrow margin.
The lower panel of Fig.~\ref{linfit} illustrates the regions that are occupied
by the two groups of Miras and that are separated by the (dashed) line of
Eq.~\ref{divlin_equ}. In Sect.~\ref{bulgesec} we use this separating line to
investigate how common 3DUP is among Galactic bulge Miras.

\subsection{Other combinations of filters}\label{other_col}

The behaviour of Tc-poor and Tc-rich Miras could be unique to the $K-[22]$ colour
if the $[22]$ band covered a particular mid-IR dust feature that is very
sensitive to stellar atmospheric composition, for example, or to other parameters that could be
influenced by 3DUP. The $[22]$ filter of {\it WISE} has its peak sensitivity between
$\approx20-25$\,$\mu{\rm m}$ \citep{Wri10}.

A few dust features fall in this range. One of them is the 23\,$\mu{\rm m}$ feature
of crystalline silicates in oxygen-rich stars, but this reaches only an
intensity of a few percent of the local continuum flux \citep{Jones12}. Another
feature is that of FeO at 20\,$\mu{\rm m}$, which may be present in metal-poor,
oxygen-rich stars \citep{McD10,Posch02}. Finally, the so-called 21\,$\mu{\rm m}$
feature, which peaks at $\sim20.1\,\mu{\rm m}$ \citep{VKH99}, falls in
the $[22]$ filter. This feature has been associated with TiC as carrier
\citep{vH00}. However, it has been observed only in C-rich post-AGB stars and is
thus unlikely to play an important role in our stars, many of which are O rich.
Nevertheless, we also inspected other combinations of near- to mid-IR colours
that were collected in Table~\ref{taba1}.

\begin{figure}
  \centering
  \includegraphics[width=\linewidth,bb=20 36 500 340, clip]{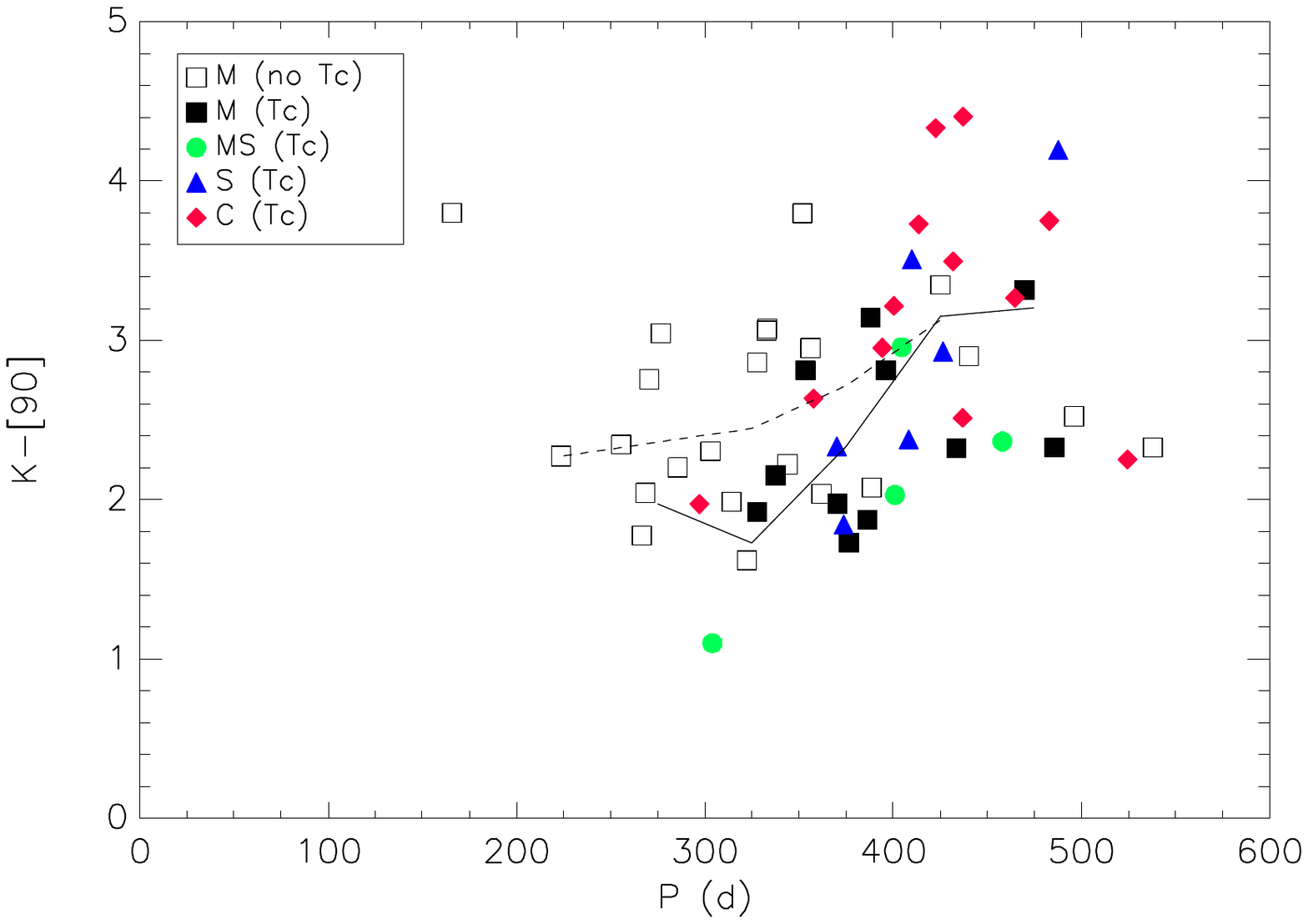}
  \includegraphics[width=\linewidth,bb=20  2 500 333, clip]{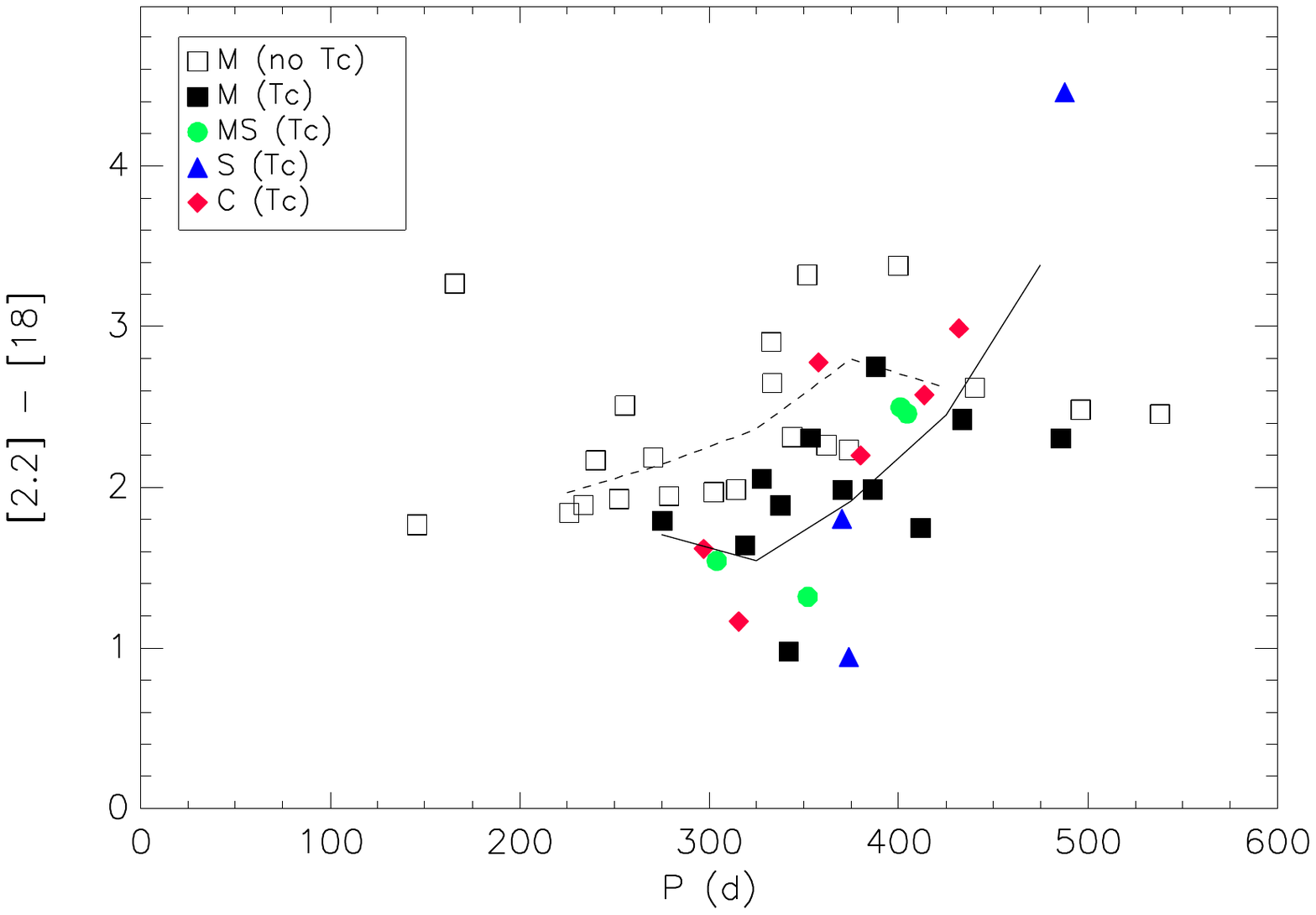}
  \caption{Same as Fig.~\ref{K-22}, but for $K-[90]$ ({\it top panel}) and
    $[2.2]-[18]$ ({\it bottom panel}).}
  \label{othercol}
\end{figure}

Two examples are shown in Fig.~\ref{othercol}. The upper panel of this figure shows
the \mbox{$P$ vs.\ $K-[90]$} diagram, with $[90]$ being the 90\,$\mu{\rm m}$ band
of the {\it Akari} satellite. The lower panel shows the $[2.2]-[18]$ colour as a
function of period. Here, $[2.2]$ is the time-averaged flux measured in the
2.2\,$\mu{\rm m}$ band of the Diffuse Infrared Background Experiment ({\it DIRBE})
on board the {\it Cosmic Background Explorer} \citep[{\it COBE},][]{Pri10}, while
the 18\,$\mu{\rm m}$ flux comes from {\it Akari}. The number of stars in this plot
is reduced because on the one hand, only a few of our sample stars are included
in the data set of \citet{Pri10}, and on the other hand, very bright stars are
saturated in {\it Akari}. Still, the $[2.2]$ band data are expected to be
of high quality because many individual measurements distributed over a number
of pulsation cycles within the 3.6 years time of DIRBE observations have been
averaged. Both panels broadly reproduce what is found in the $K-[22]$ colour
(Fig.~\ref{K-22}): At a given pulsation period, Tc-poor Miras tend to have a
higher near- to mid-IR colour index than Tc-rich Miras, even among a modest
number of stars.

We thus can establish that Tc-poor Miras have a higher near- to mid-IR excess
than Tc-rich Miras over a wide range of the mid-IR region. This means that the
observed separation is not related to a particular mid-IR dust feature, but
rather related to a general difference in the intensity of the mid-IR emission.

\subsection{ISO dust spectra}\label{iso_specs}

\subsubsection{General dust emission}

The conclusion of the previous section can be further substantiated by inspecting
broad-band dust spectra of stars in the two groups. The dust spectra of nearby
AGB stars obtained with the Short Wavelength Spectrometer (SWS) on board {\it ISO}
cover the range $2.4-45.4\,\mu{\rm m}$. These spectra, processed and renormalized
in a uniform manner, have been published by
\citet{Slo03b}\footnote{\tt https://users.physics.unc.edu/$\sim$gcsloan/library/sws\-atlas/atlas.html}.
Unfortunately, there is little overlap between the {\it ISO} sample and the sample
of AGB stars with information on Tc content discussed here. Nevertheless,
interesting conclusions can be drawn already from this limited overlap.

\begin{figure}
  \centering
  \includegraphics[width=\linewidth]{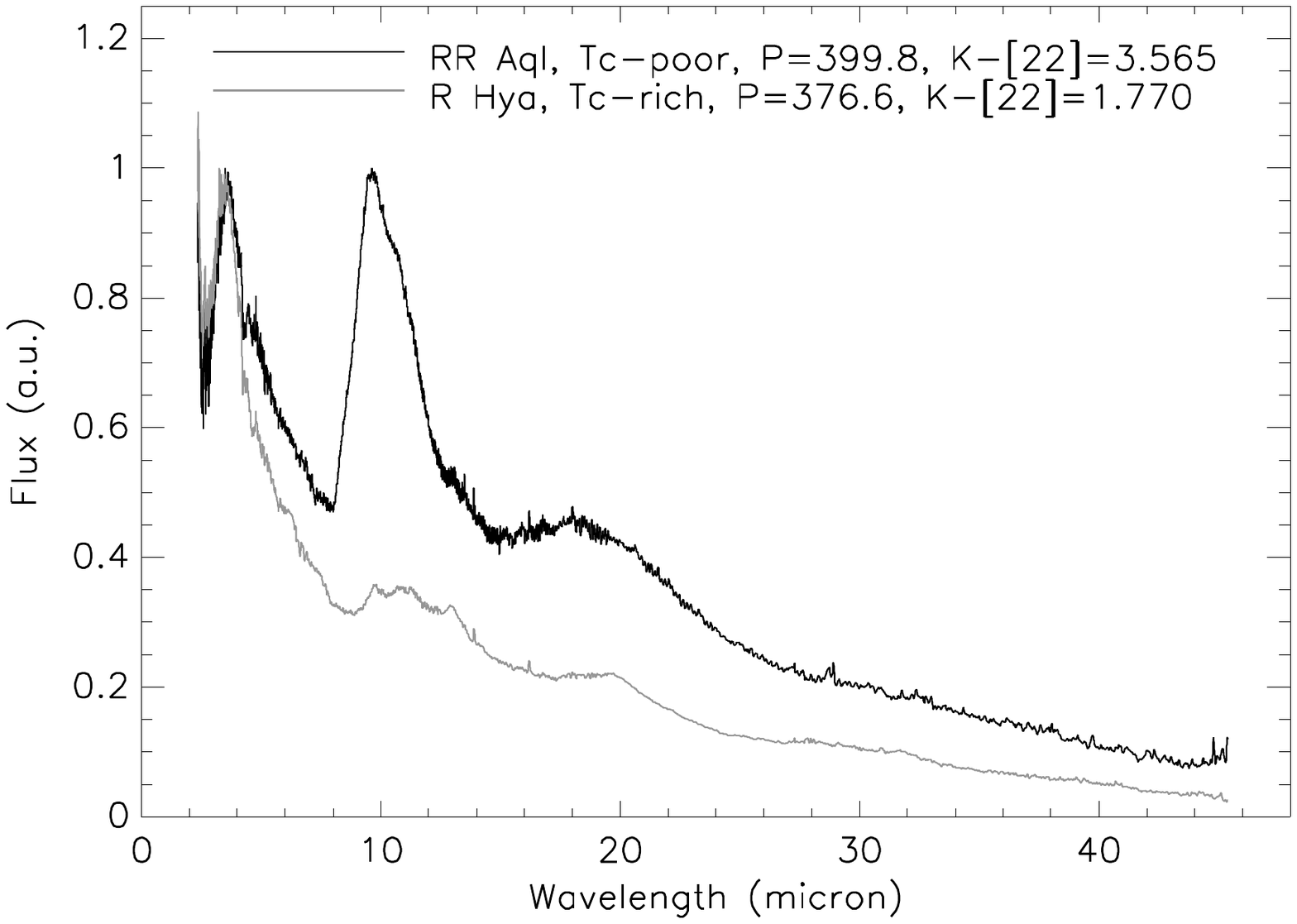}
  \caption{{\it ISO} dust spectra of RR~Aql (Tc-poor, black graph) and R~Hya
    (Tc-rich, grey graph). The spectra are normalised to a maximum flux of 1.0
    between $3-4\,\mu{\rm m}$. The stellar periods and $K-[22]$ colours are
    indicated in the legend.}
  \label{ISO_fig}
\end{figure}

In particular, it is insightful to compare {\it ISO} dust spectra of Tc-poor and
Tc-rich Miras with similar pulsation periods. As an example, this is done for
one such pair, RR~Aql and R~Hya, in Fig.~\ref{ISO_fig}. The short-wavelength end of
the spectra is longward of the K band, therefore no direct scaling of the spectra
to the same K-band flux is possible. Therefore, the flux maxima of the spectra
between 3 and 4\,$\mu{\rm m}$ were normalised to 1.0 for comparison purposes
(the flux shortward of 3\,$\mu{\rm m}$ is reduced by the CO first-overtone band).
The aim here is just to illustrate the difference in flux between near- and mid-IR
wavelengths, and the chosen normalisation should be sufficiently precise for this
purpose. As indicated in the figure legend, the two stars have similar pulsation
periods, but very different $K-[22]$ colours. This fact is reflected in the appearance
of the {\it ISO} spectra: The Tc-poor Mira RR~Aql has more flux than the Tc-rich Mira
R~Hya (relative to the maximum at $3-4\,\mu{\rm m}$) not only in the wavelength
range in which the WISE [22] bandpass is most sensitive ($\sim20-25\,\mu{\rm m}$),
but basically at all wavelengths longward of $\sim5\,\mu{\rm m}$. In particular,
the silicate features at 10 and 18\,$\mu{\rm m}$ are very prominent in RR~Aql, but
weak in R~Hya. Comparisons with other Tc-rich stars, for instance, T~Cep ($P=386.6$\,d,
$K-[22]=1.756$) and S~Vir ($P=370.4$\,d, $K-[22]=2.077$), yield similar results.
Although the sample is small, the Tc-poor and Tc-rich Miras distinctly differ in the
{\it ISO} dust spectra, with the Tc-rich stars having much lower mid-IR excess than
the Tc-poor ones over a wide wavelength range. This confirms the results found in
Sect.~\ref{Miras} from the combinations of various near- and mid-IR filters.

\subsubsection{ 13\,$\mu{\rm m}$ feature}\label{13mufeat}

On the other hand, inspection of the {\it ISO} spectra shows that, among other
features, the 13\,$\mu{\rm m}$ feature is clearly present in R~Hya, but absent
or lost in the wing of the silicate feature in RR~Aql. The 13\,$\mu{\rm m}$ is
thought to be emitted by grains of alumina dust \citep[Al$_2$O$_3$,][]{Slo03a}.
The presence of the feature seems to be related neither to the presence of Tc
nor to the pulsation period. It is clearly related to the $K-[22]$
excess, however, thus to the (dust) mass-loss rate, confirming earlier results by
\citet{Leb06} based on observations of globular cluster stars. This is
demonstrated in Table~\ref{13micron}, in which the stars in common between the
Tc and the {\it ISO} SWS sample are sorted by increasing $K-[22]$ colour. The
reddest star showing the 13\,$\mu{\rm m}$ feature is R~Cen with $K-[22]=2.221$,
the bluest one without is NO~Aur at $K-[22]=2.110$.
Hence, the threshold at which the 13\,$\mu{\rm m}$ feature disappears is
$K-[22]\simeq2.17$. Using the relation between $K-[22]$ and the gas
mass-loss rate that we derive in Section~\ref{gmlr}
(Eq.~\ref{Mgas_K22_rel}), we find that this threshold corresponds to a
mass-loss rate of $\dot{M}_{\rm g}=(2.6\pm0.5)\times10^{-7}\,M_{\sun}\,yr^{-1}$.
For the error estimate, the formal variation of the coefficients in
Equ.~\ref{Mgas_K22_rel}, as well as the uncertainty in the $K-[22]$ threshold,
have been taken into account. We note, however, that the total variation in gas
mass-loss rate at the $K-[22]$ threshold value is large, up to an order of
magnitude either side of the mean value (see Fig.~\ref{Mdotgas_K22}).
Current dynamic AGB atmosphere models producing mass loss, such as
those of \citet{Hoef16}, propose Al$_2$O$_3$ grains as seed particles for
larger silicate grains. Our results suggest that at a mass-loss rate of
$\sim2.6\times10^{-7}\,M_{\sun}\,yr^{-1}$ the density in the outflow is high
enough for the formation of significant silicate mantles around the alumina
seed grains so that the alumina 13\,$\mu{\rm m}$ feature disappears. Models
of AGB winds can be tested by reproducing the observed threshold.

\begin{table}
\caption{Occurrence of the 13\,$\mu{\rm m}$ feature in the sample stars
  observed by {\it ISO}. The presence of the 13\,$\mu{\rm m}$ feature reported
  in the last column is taken from \citet{Slo03b}. The stars are ordered by
  increasing $K-[22]$ colour.}
\label{13micron}
\centering
\begin{tabular}{lccrccc}
\hline\hline
 Object & Tc & Sp.\ & Var.\ & Period & $K-[22]$ & 13\,$\mu{\rm m}$ \\
 name   &    & type & type  & (d)    &          & feature         \\
\hline
R Dor   & 0 & M  & SRV   & 327.1 & 1.290 & y \\
RX Lac  & 1 & MS & SRV   & 331.4 & 1.295 & y \\
T Cet   & 1 & MS & SRV   & 159.3 & 1.444 & y \\
T Cep   & 1 & M  & Mira  & 386.6 & 1.756 & y \\
R Hya   & 1 & M  & Mira  & 376.6 & 1.770 & y \\
W Hya   & 0 & M  & Mira? & 388.6 & 1.987 & y \\
ST Her  & 1 & MS & SRV   & 147.0 & 1.999 & y \\
S Vir   & 1 & M  & Mira  & 370.4 & 2.077 & y \\
NO Aur  & 1 & MS & SRV   & 102.1 & 2.110 & n \\
R Cen   & 0 & M  & Mira  & 538.1 & 2.221 & y \\
$\pi_1$ Gru & 1 & S  & SRV   & 198.8 & 2.253 & n \\
$\chi$ Cyg  & 1 & S  & Mira  & 408.2 & 2.266 & n \\
R Aql   & 0 & M  & Mira  & 270.6 & 2.660 & n \\
R And   & 1 & S  & Mira  & 409.9 & 2.861 & n \\
$o$ Cet & 1 & M  & Mira  & 332.7 & 3.479 & n \\
R Aqr   & 1 & M  & Mira  & 388.2 & 3.531 & n \\
RR Aql  & 0 & M  & Mira  & 399.8 & 3.565 & n \\
\hline
\end{tabular}
\end{table}

\subsection{Semi-regular variables}\label{SRVs_sec}

\begin{figure}
  \centering
  \includegraphics[width=\linewidth,bb=4 38 500 360,clip]{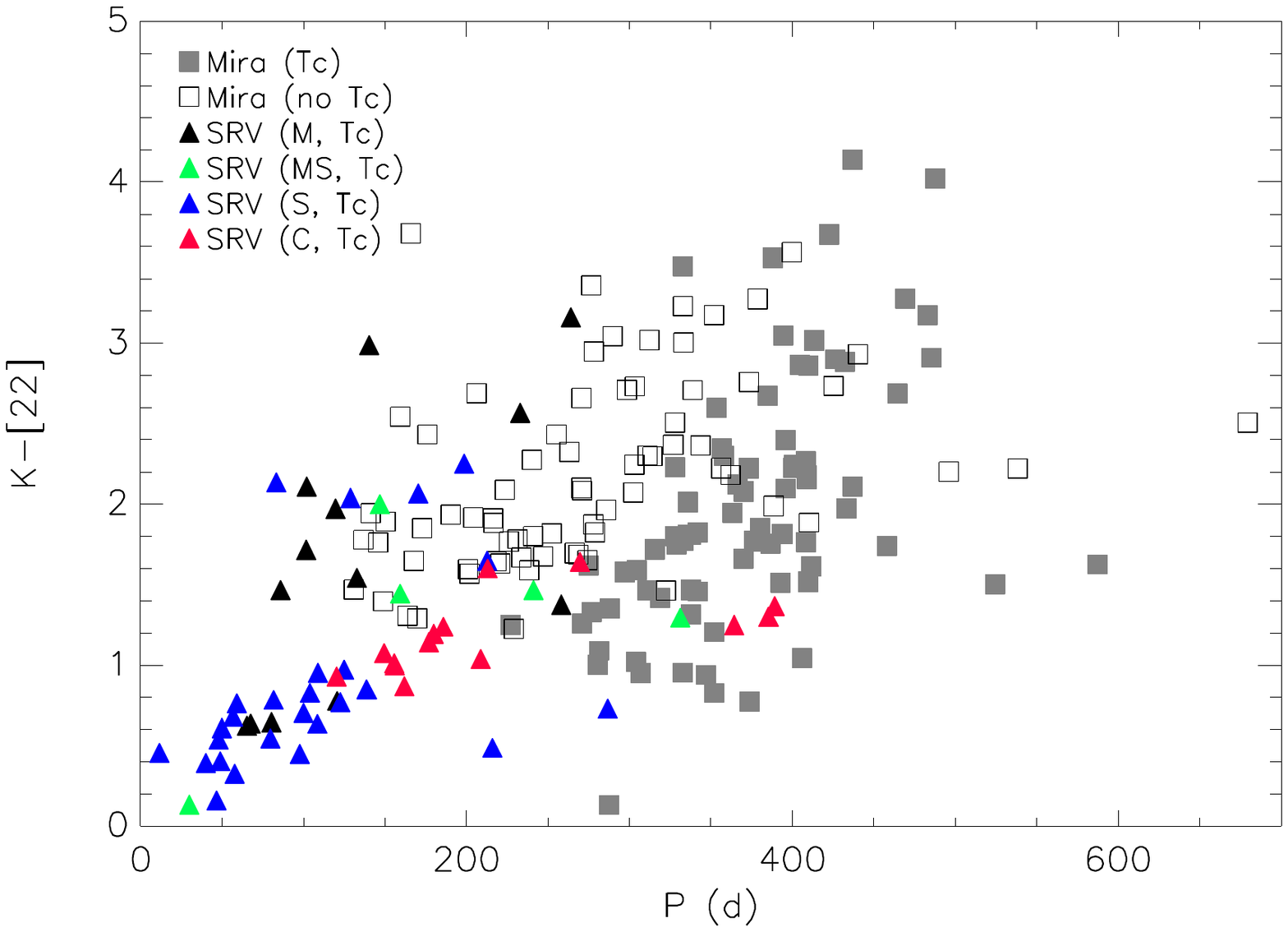}
  \includegraphics[width=\linewidth,bb=4  2 500 352,clip]{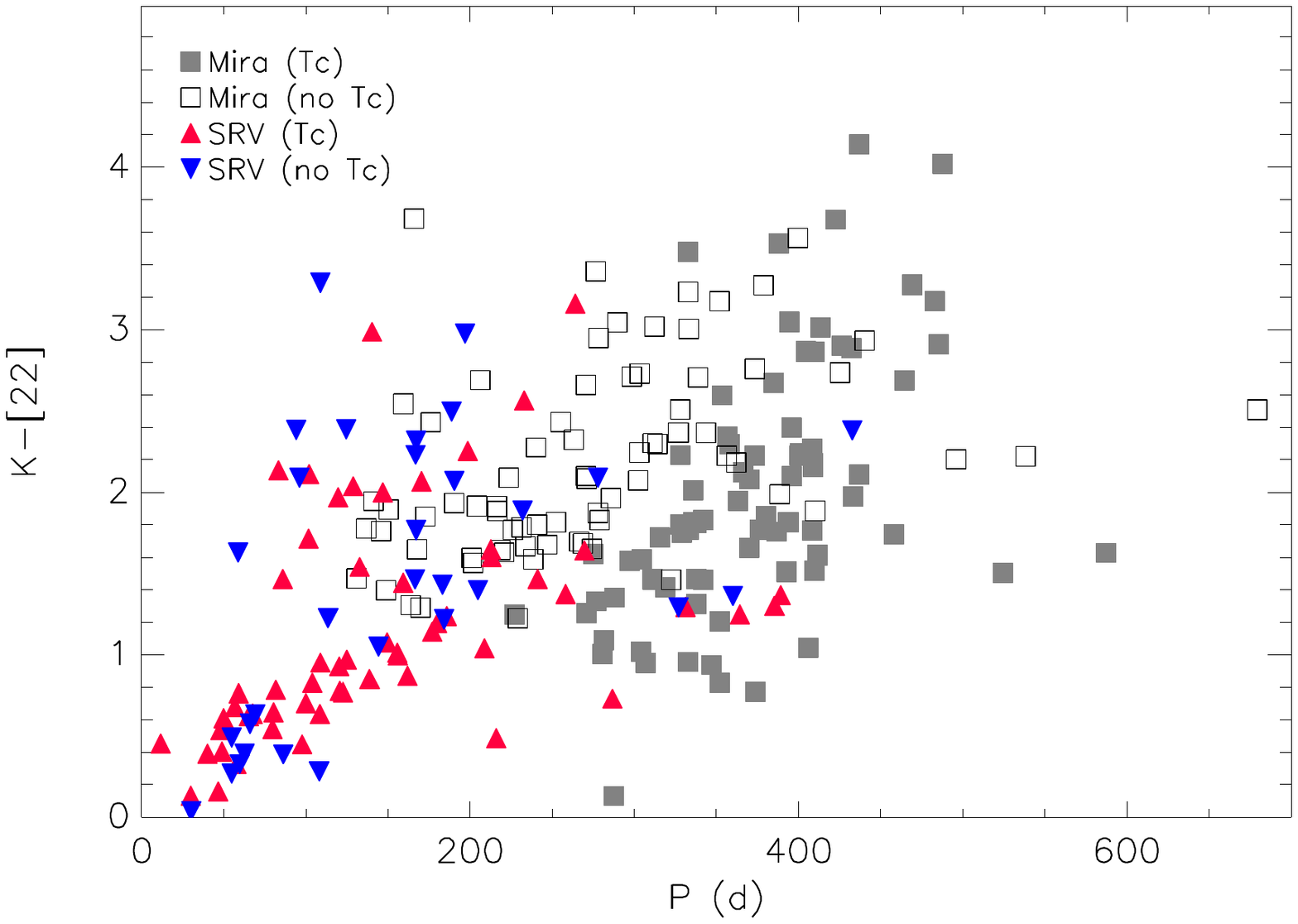}
  \caption{{\it Upper panel:} \mbox{$P$ vs.\ $K-[22]$} diagram including Tc-rich
    SRVs (triangles). Different spectral types are colour-coded, see legend.
    {\it Lower panel:} Same as upper panel, but showing M-type Tc-poor
    (downward-pointing blue triangles) and Tc-rich (upward-pointing red triangles)
    SRVs. Mira stars are shown in both panels as open and filled grey squares.}\label{SRVs}
\end{figure}

Investigating the behaviour of SRVs may give interesting insights because they differ
from the Miras mainly in that they tend to pulsate in one of the overtones, not in
the fundamental mode. The upper panel in Fig.~\ref{SRVs} thus shows how Tc-rich SRVs
distribute in the well-known \mbox{$P$ vs.\ $K-[22]$} diagram. Three different groups
may be distinguished among them.

First, there is a relatively tight sequence of S and C stars with $K-[22]$ colour
increasing with pulsation period. This group extends the sequence formed by the
Miras to shorter periods. This sequence has already been noted in Paper~I.
Interestingly, a number of C-type SRVs are near the long-period end of the sequence
that is formed mostly by S-type stars, which are thought to precede the C-type
stars in the evolution. Because there are very few Miras with $K-[22]<1$, it seems
likely that SRVs switch to fundamental mode (Mira phase) only from the upper end
of this sequence.

Next, there are a few SRVs that have longer periods and fall among the Tc-rich Miras
(filled grey squares). These are likely SRa variables, that is,\ fundamental mode pulsators.
First-overtone pulsators indeed only reach periods up to $\sim240$\,d; already above
200\,d of period, the number of first-overtone pulsators is very low \citep{Rie10}.
Interestingly, most of these long-period, Tc-rich SRVs are in a region of the diagram
where Tc-rich Miras are found as well, namely at $K-[22]$ colours that are bluer than
those of Tc-poor Miras, supporting the finding from that group of stars.

Nevertheless, there are Tc-rich SRVs that are scattered around at relatively red $K-[22]$.
There are 15 stars with $K-[22]\ge1.3$ and $P\le240$\,d. Among these,
L$_2$~Pup \citep[e.g.][]{Ker15} and $\pi_1$~Gru \citep[e.g.][]{May14} are confirmed binary
stars. They have the largest and third-largest $K-[22]$ excess among the Tc-rich SRVs,
respectively. Among the remaining stars in this area of the diagram, T~Cet, W~Cyg, and
ST~Her are listed by \citet{OG16} as binary star candidates with main-sequence companions
based on {\it GALEX} UV photometry. Thus, it seems likely that also among SRVs the
$K-[22]$ colour, hence the dust-production rate, is enhanced by the presence of a
binary companion, just like among Miras (see Sect.~\ref{Miras}).

The Tc-poor SRVs in the sample do not show much structure in the \mbox{$P$ vs.\ $K-[22]$}
diagram, see the blue triangles in the lower panel of Fig.~\ref{SRVs}. There are a few stars
at the base of the sequence that is formed by the Tc-rich SRVs, which might be direct
precursors of the latter. Again, some of the longer-period stars could be SRa.
Some of these stars have relatively red $K-[22]$ colours compared to Tc-rich
SRVs at similar periods, as might be expected. Furthermore, binary stars with enhanced
dust-production rate might be found among the Tc-poor SRVs.
In conclusion, SRVs evidently confirm the findings from the Mira stars.

\subsection{Gas mass-loss rate}\label{gmlr}

\begin{figure}
  \centering
  \includegraphics[width=\linewidth,bb=30 10 484 334, clip]{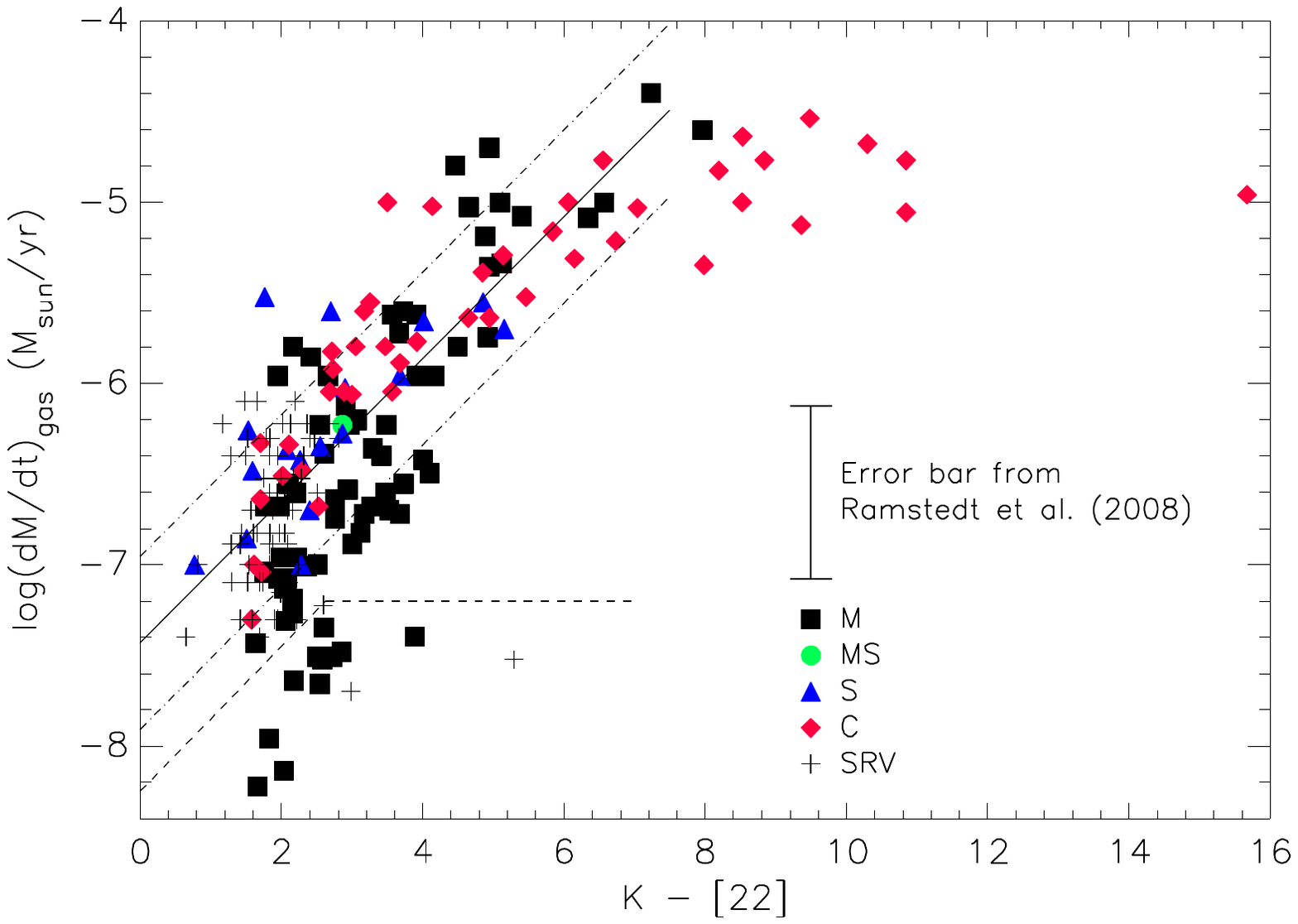}
  \caption{Gas mass-loss rate $\log(\dot{M}_{\rm g})$ as a function of $K-[22]$ of the
    collection of AGB stars in Table~\ref{taba2}. The data of Mira stars are plotted as
    different symbols according to their spectral type, see the legend in the lower right corner of
    the diagram. SRVs from \citet{Olo02} are plotted as black crosses; they are all of
    spectral type M. Stars to the lower right of the black dashed line as well as stars
    redder than $K-[22]=7.5$ have been excluded from the ordinary least-squares fit that is
    shown as the black solid line. The dash-dotted lines show the fit moved up and down by the
    estimated factor-of-3 uncertainty \citep{Ram08}.}\label{Mdotgas_K22}
\end{figure}

As outlined in the introduction, the main motivation of this paper is to expand the
analysis to the gas mass-loss rate $\dot{M}_{\rm g}$ of the stars.
In order to investigate the \mbox{$P$ -- $\log(\dot{M}_{\rm g})$} diagram, we collected
from the literature the data outlined in Sect.~\ref{co_data} that are presented in
Table~\ref{taba2} in the appendix. First, we examine the distribution of
stars in the $\log(\dot{M}_{\rm g})$ vs.\ $K-[22]$ diagram, see Fig.~\ref{Mdotgas_K22}.
Generally, $\log(\dot{M}_{\rm g})$ increases with increasing $K-[22]$ excess. The regions
of the diagram occupied by \mbox{M-,} \mbox{S-,} and C-type Miras, as well as the (M-type)
SRVs, largely overlap (there is only one MS-type star in the sample). As expected, the
carbon Miras extend to larger $K-[22]$ colours than the other stars. The increase in
$\log(\dot{M}_{\rm g})$ seems to flatten off at $K-[22]\gtrsim7.5$.
At the other end of the distribution, several M-type stars have
fairly low gas mass-loss rates, even though some of them have a considerable $K-[22]$
excess. We note that this tail of low mass-loss rate Miras comes from the
works of \citet{You95} and \citet{Groen99} and is separated from the remaining sample
by a dashed line in Fig.~\ref{Mdotgas_K22}. (A comparison of mass-loss rates of
stars in common between these two and more recent works reveals that i) there seems
to be no systematic difference between \citet{You95} and more recent works;
ii) $\log(\dot{M}_{\rm g})$ could be systematically underestimated by \citet{Groen99},
but not by much; and iii) both works contain a number of stars whose $\log(\dot{M}_{\rm g})$
is severely lower than what more recent works find. We thus conclude that the mass-loss
rates of the stars from \citet{You95} and \citet{Groen99} not separated out by the dashed
line in Fig.~\ref{Mdotgas_K22} should be fairly reliable.) Between these regions, the
relationship between $K-[22]$ and $\log(\dot{M}_{\rm g})$ appears to be fairly linear. An
ordinary linear least-squares fit to the data yields the relation

\begin{equation}
\log(\dot{M}_{\rm g})=(0.392\pm0.019)\times(K-[22]) - (7.431\pm0.064).
\label{Mgas_K22_rel}
\end{equation}

This relation, plotted in Fig.~\ref{Mdotgas_K22} as a solid line, can be used to
estimate the gas mass-loss rate of AGB stars and should be applicable roughly in
the range $1.5\lesssim K-[22]\lesssim 7.5$.

We also plot in Fig.~\ref{Mdotgas_K22} the error bar of a factor of 3 that was
found by \citet{Ram08} to be a lower limit to the uncertainty in current mass-loss
rate estimates. The $\log(\dot{M}_{\rm g})$ values scatter with a standard
deviation of $\sim0.397$ around the linear fit, corresponding to a factor of
$\sim2.5$. From the existence of the relation between $K-[22]$ and $\log(\dot{M}_{\rm g})$,
one might expect that the same two sequences as in the \mbox{$P$ vs.\ $K-[22]$}
diagram might show up in the \mbox{$P$ -- $\log(\dot{M}_{\rm g}$)} plane,
although with increased scatter.

\begin{figure}
  \centering
  \includegraphics[width=\linewidth,bb=35 14 488 333, clip]{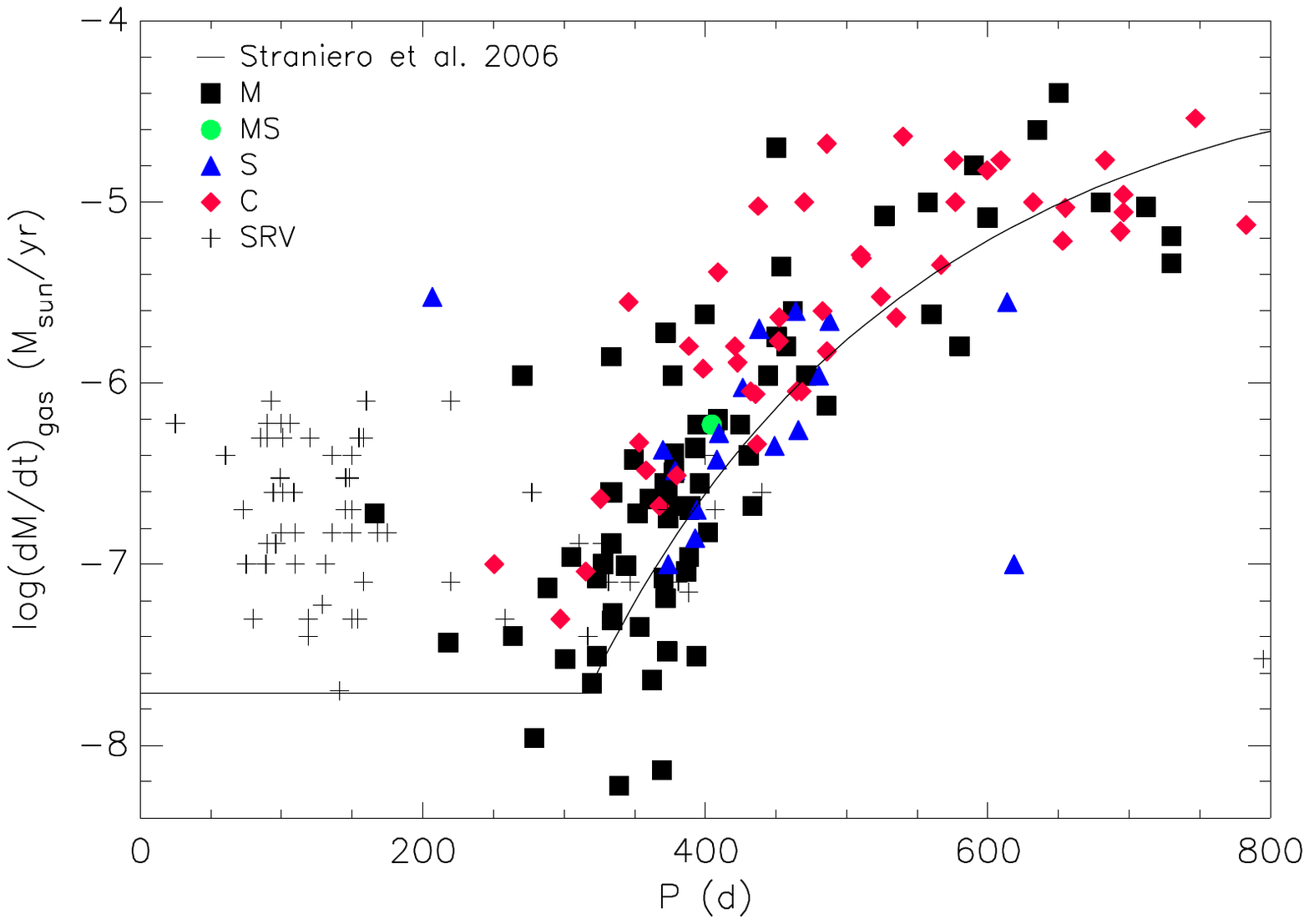}
  \caption{Gas mass-loss rate $\log(\dot{M}_{\rm g})$ as a function of pulsation period
    for Miras in Table~\ref{taba2} and SRVs from \citet{Olo02}. The solid line is the
    empiric period -- mass-loss-rate relation of \citet{SGC06}. Two SRVs with no known
    period are plotted at $P=25$\,d.}\label{Mdotgas_P}
\end{figure}

Next, we study the \mbox{$P$ -- $\log(\dot{M}_{\rm g})$} diagram of all stars for
which the gas mass-loss rates were collected, see Fig.~\ref{Mdotgas_P}. Such diagrams
have been used in the past to construct empirical period -- mass-loss-rate relations
\citep{VW93,SGC06}. The precise functional form of this relation has profound impacts
on evolutionary models of the AGB because the mass-loss rate often greatly exceeds
the nuclear-burning rate. Figure~\ref{Mdotgas_P} is an updated version with almost
double the number of stars as that in \citet{SGC06}. Their relation is over-plotted
as a solid line in our figure. It captures the  distribution of stars very well at the
longer periods, but seems to somewhat underestimate the mass-loss rate at periods
$\lesssim300$\,d. We note, however, that their relation extends to longer periods where
most of the mass loss occurs on the AGB and that is not covered by the present sample.
The overtone SRVs also probably have higher initial (and current) masses than
the short-period Miras \citep{VW93,FWY72}.

Importantly, Miras of all spectral types share more or less the same area in
Fig.~\ref{Mdotgas_P}. This has been shown before by \citet[][their Fig.~6]{GJ98} to
be the case for S- and C-type Miras. Clearly, there is an increase in gas mass-loss
rate between $\sim350$ and $\sim600$\,d. A marked increase in mass-loss in this
period range has been noted by \citet{VW93}, \citet[][their Fig.~10]{Mat05},
and \citet{MZ16}. At longer periods, the mass-loss rate flattens off around
$\log(\dot{M}_{\rm g})\approx-5$ \citep{VW93}. Moreover, the SRa variables with periods
above 200\,d share the same area as the Miras in this period range. This agrees with
the findings of \citet[][their Fig.~6]{GJ98}. The shorter-period SRVs also have gas
mass-loss rates in the same range, despite their shorter periods. However, the Mira
sequences observed in $K-[22]$ colour are not apparent from Fig.~\ref{Mdotgas_P}.

\begin{figure}
  \centering
  \includegraphics[width=\linewidth,bb=35 14 488 333, clip]{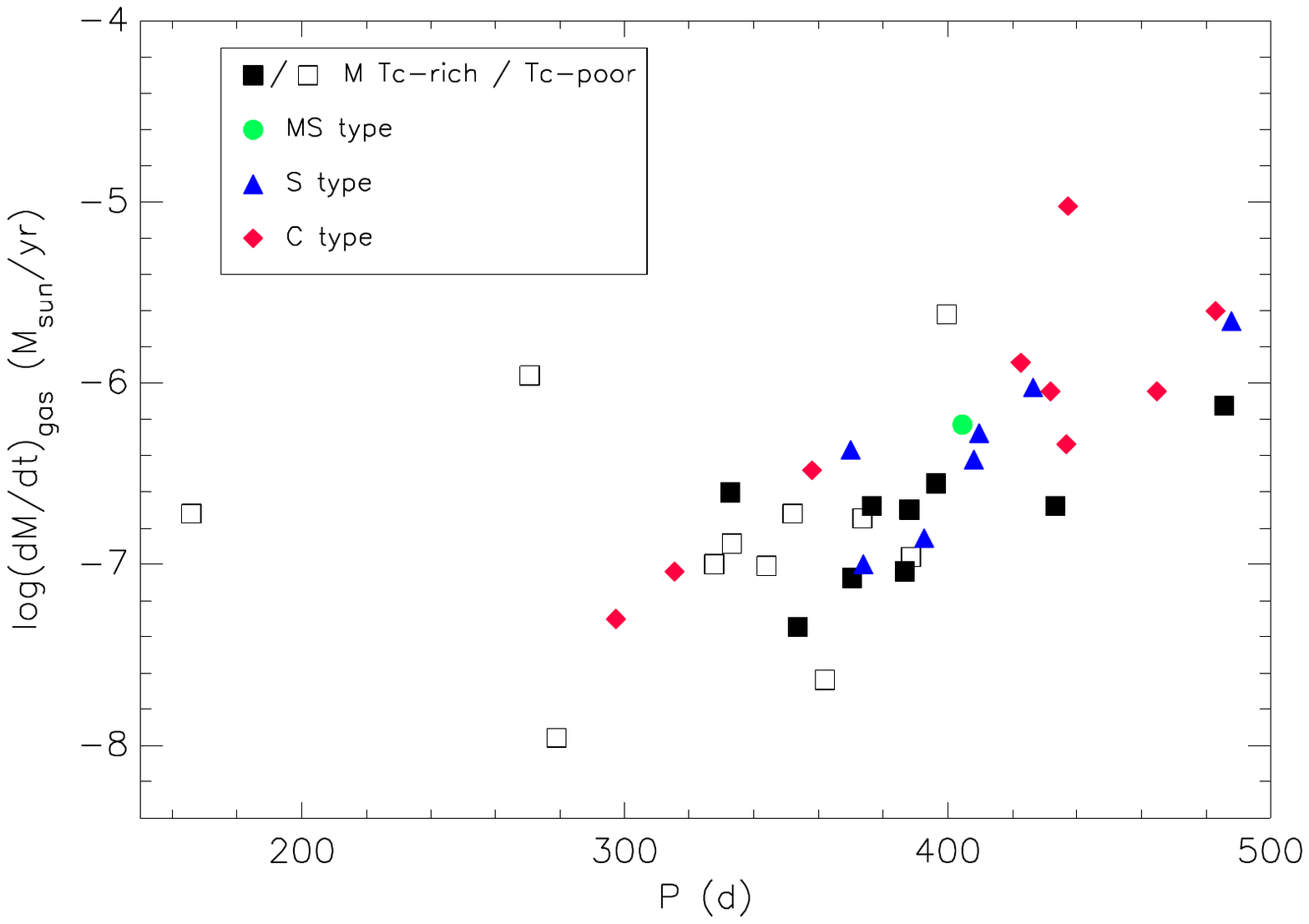}
  \caption{Gas mass-loss rate $\log(\dot{M}_{\rm g})$ as a function of pulsation
    period of Miras with information on their Tc content. Symbols are the same as in
    Fig.~\ref{Mdotgas_K22}, except for Tc-poor and Tc-rich M-type Miras (open
    and filled black squares, see legend). Note the zoom-in on the x-axis
    compared to Fig.~\ref{Mdotgas_P}.}\label{Mdotgas_P_Tc}
\end{figure}

We investigate this in more detail in Fig.~\ref{Mdotgas_P_Tc}, which includes only Miras
with information on their Tc content. Two Tc-poor Miras clearly have higher gas
mass-loss rates than the Tc-rich Miras at similar periods. These are R~Aql and RR~Aql,
at $P\simeq270$\,d and 400\,d. They have higher gas mass-loss rates than all
Tc-rich Miras, except for three carbon Miras and one S-type Mira, all of which
have longer pulsation periods, however. The gas mass-loss rate of R~Cet, at $P\simeq170$\,d and noted in Paper~I to be
very red in $K-[22]$ for its period, is much higher than what
would be expected if the Mira sequence of Fig.~\ref{Mdotgas_P} were extrapolated to its
period. However, many SRVs with similar pulsation periods have gas
mass-loss rates of the same order of magnitude. We also note that the Tc-rich M-type Mira
with the shortest period in Fig.~\ref{Mdotgas_P_Tc} is the binary $o$~Cet. Another binary,
R~Aqr, is included in this diagram. Their stellar wind might therefore deviate
considerably from spherical symmetry \citep[e.g.][]{Ram18}, which is a usual assumption
to derive gas mass-loss rates from radio CO line observations, and their mass-loss rate
might be more uncertain. Finally, we note that the three M-type Miras with the
lowest gas mass-loss rates in this diagram are in the tail of low mass-loss rate
stars that are identified and separated by the dashed line in Fig.~\ref{Mdotgas_K22}.

In conclusion, with the available observations it can neither be confirmed nor excluded
that Tc-poor Miras also have higher {\it \textup{gas}} mass-loss rates than Tc-rich ones at a
similar pulsation period and that there is a real split. Several limitations prevent us
from drawing stronger conclusions. First of all, the same stars as in
Fig.~\ref{Mdotgas_P_Tc} plotted in the \mbox{$P$ vs.\ $K-[22]$} plane do not clearly
exhibit the two sequences either. Second, as shown in Sect.~\ref{Miras}, the two
sequences merge in $K-[22]$ somewhere between 350 and 400\,d of period, but there are
few stars with period below 350\,d in Fig.~\ref{Mdotgas_P_Tc}. It is also important to
take into account that there are larger uncertainties attached to the
$\log(\dot{M}_{\rm g})$ determination than to the $K-[22]$ colour. The determination of
$\log(\dot{M}_{\rm g})$ is a complex undertaking that uses radiative transfer models of the
circumstellar envelope. Uncertainties introduced by necessary assumptions might contribute
to scatter the sequences that are observed in $K-[22]$, which might be a much more direct
and reliable measure of the mass-loss rate. Variability of the stars will not play a large
role in the colour because (as much as possible) averaged $K$-band magnitudes were
used, whereas the variability amplitude in the 22\,$\mu{\rm m}$ band is expected to be
small \citep{GN11}.

\begin{figure}
  \centering
  \includegraphics[width=\linewidth,bb=32 12 488 335, clip]{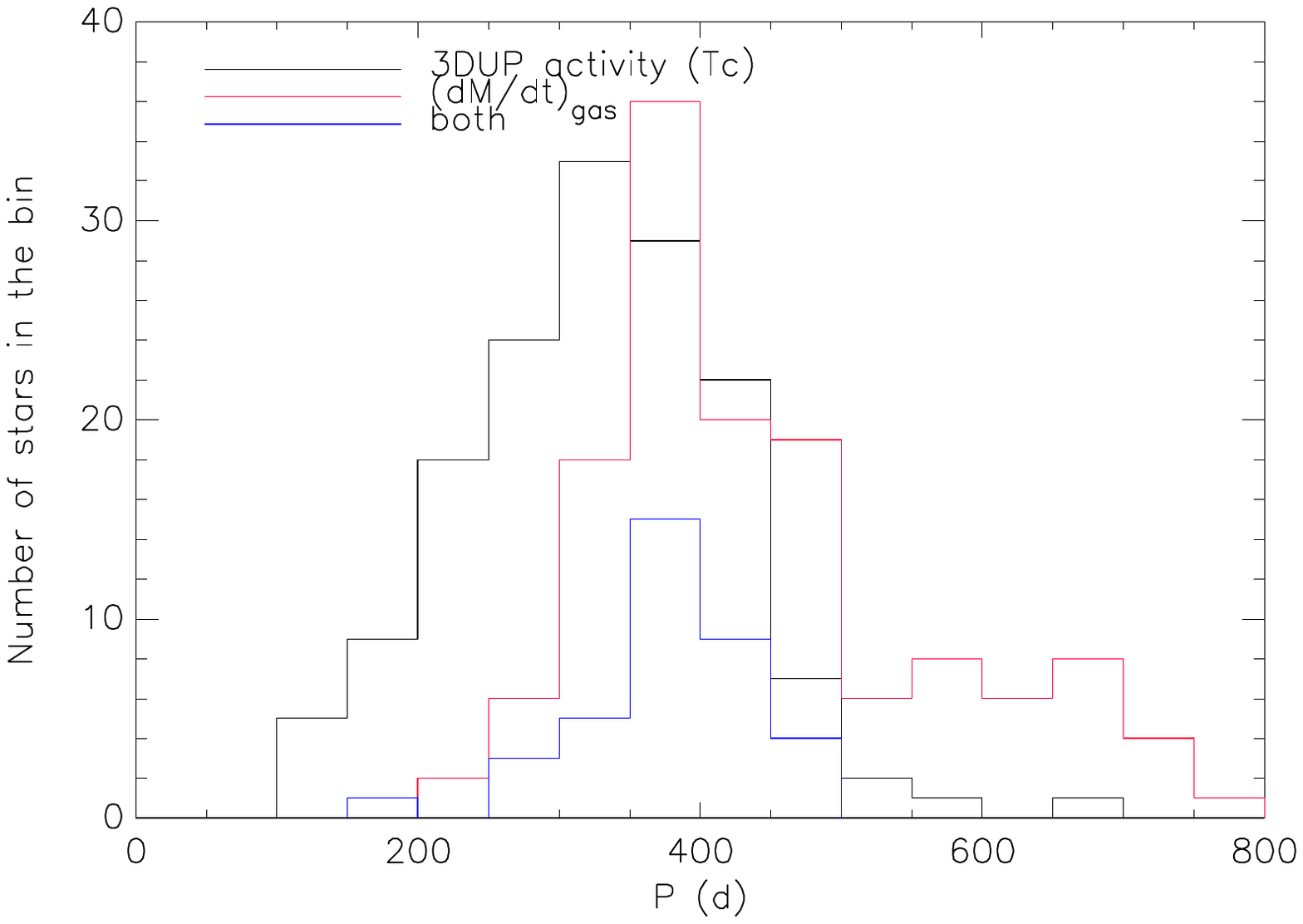}
  \caption{Period distribution of Miras with information on their 3DUP activity
    (black histogram), of Miras with a measured gas mass-loss rate (red),
    and of Miras with information on both (blue). Observational biases are the
    probable cause of the different distributions.}\label{P_distrib}
\end{figure}

An obvious obstacle in the analysis is the small number of stars available, however.
While samples of 135 Miras with a determined gas mass-loss rate and 151 Miras with
information on their 3DUP activity were compiled, only for a mere 37 stars do we have
information on both. As illustrated in Fig.~\ref{P_distrib}, the two samples have
period distributions that differ considerably. The reason for this is most likely an
observational bias: The Tc lines can only be inspected in optically bright Miras that
are not too much obscured by circumstellar absorption, which favours shorter-period
stars as targets. On the other hand, traditionally, the longer-period Miras have been
targeted for radio CO measurements, also due to sensitivity reasons. We also observe
in the literature that the same Miras are targeted over and over again for their
molecular radio emission, while hardly any new objects are added to the sample. With
APEX and ALMA, for example, very sensitive radio facilities are available to researchers,
who are encouraged to use them for observations of shorter-period Miras
($P\lesssim400$\,d). In general, more Miras with information on their Tc content
should be observed to investigate the interplay between pulsation,
mass loss, and 3DUP in more detail.

Finally, we recall that, as already observed in Sect.~\ref{Miras} (see also
Fig.~\ref{linfit}), the transition between the regions occupied by Tc-poor and
the Tc-rich Miras, respectively, is very sharp. This sharp transition could be
accomplished by a mechanism that acts very quickly on the $K-[22]$ colour.
The 22\,$\mu{\rm m}$ band probes relatively warm dust that is thought
to be at a distance of a few or several stellar radii from the photosphere of
the star. Assuming a typical outflow velocity of the wind ($\sim10$\,km\,s$^{-1}$),
it takes only a few decades to fully replace the material in this region of
the CSE. The radio CO lines observed to determine the gas mass-loss rate, on
the other hand, probe a much larger region of the CSE, which is replenished
only within $\sim5000$\,yrs. Thus, the CO lines measure the mass-loss rate
averaged over a time span that is longer by some two orders of magnitude than
that probed by the $K-[22]$ colour. Hence, even if the uncertainties in
determining the gas mass-loss rates from radio CO lines were very small, it
is not granted that a similar separation between Tc-poor and Tc-rich Miras can
be found in the \mbox{$P$ vs.\ $\log(\dot{M}_{\rm g})$} diagram because the
transition that the star undergoes could be much slower in $\log(\dot{M}_{\rm g})$
than in $K-[22]$ (e.g.\ if a single 3DUP episode is sufficient to impact the
$K-[22]$ colour). Nevertheless, we encourage observers to add more Miras to
the sample of stars with known gas mass-loss rate, especially both Tc-poor
and Tc-rich Miras in the period range of overlap.

\subsection{Pulsation amplitudes and mass-loss rates}\label{amplitude}

It has been suggested that pulsation is the main trigger of strong mass loss
\citep[e.g.][]{MZ16}. However, it is not clear if pulsation period or (radial)
pulsation amplitude is the decisive factor for strong mass loss. From the
observational side this is difficult to investigate because the amplitude in the
visual range, which would be readily available for a large sample of stars, is
not a good indicator for the true pulsation amplitude due to its strong
dependence on molecular absorption bands, which in turn are very sensitive to
the chemistry,  temperature, density, and related changes over a pulsation
cycle. Light curves and thus amplitudes in the near-IR range, on the other
hand, would be much better measures of the true pulsation amplitude, but are
available only for relatively few stars. A homogeneous sample of IR
light curves has been published by \citet{Pri10}. We cross-matched our sample
with their data of variable stars and found 48 stars in common, a reasonable
number to analyse the dependence of dust mass-loss rate (i.e.\ the $K-[22]$
colour) on pulsation amplitude.

\begin{figure}
  \centering
  \includegraphics[width=\linewidth,bb=14 3 481 335, clip]{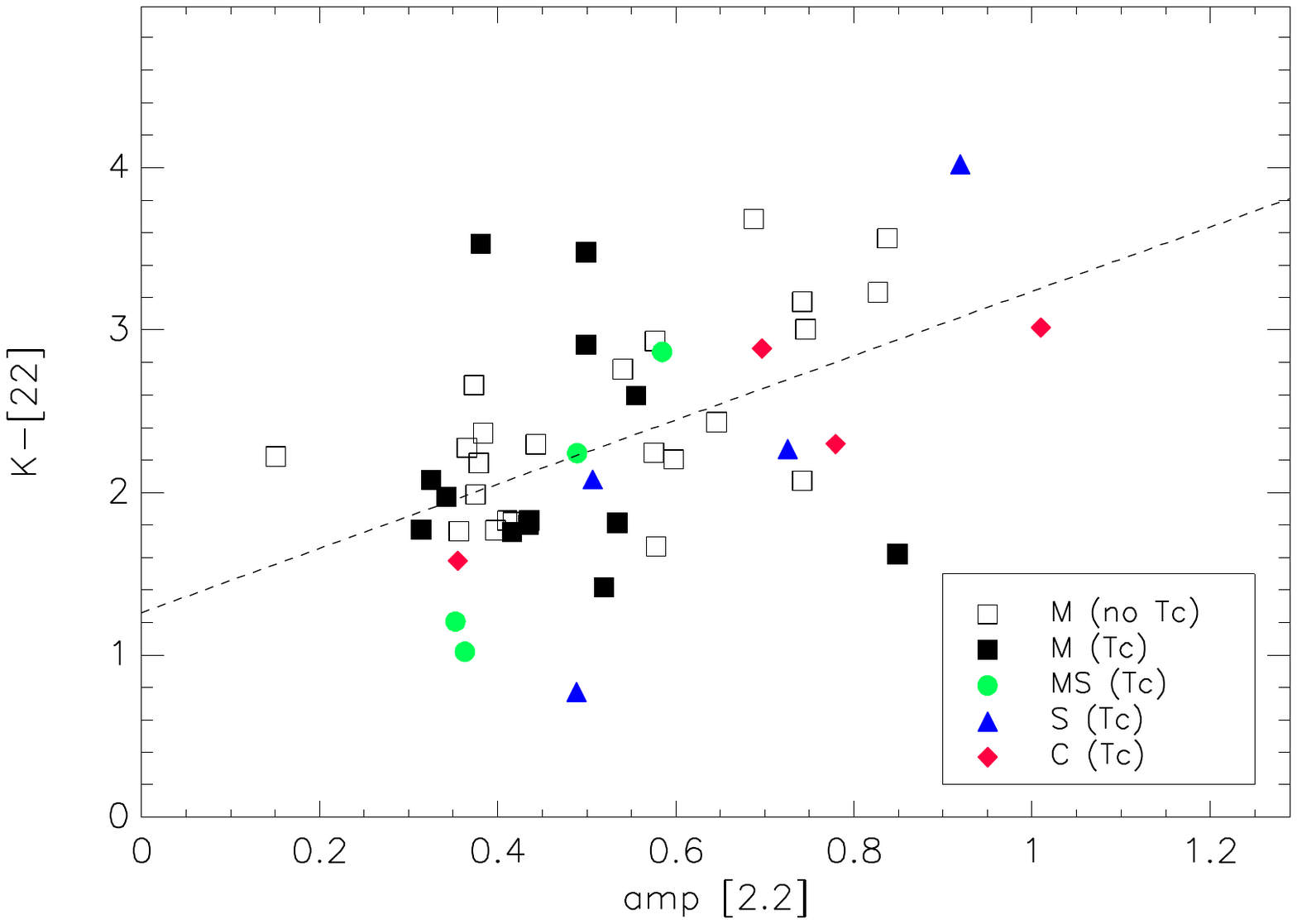}
  \caption{$K-[22]$ colour as a function of amplitude at 2.2\,$\mu{\rm m}$ of the
    sample stars in common with \citet{Pri10}. Chemical types are identified by the
    legend in the lower right corner. The dashed line is a linear least-squares fit
    to the data.}\label{DIRBE_amp22}
\end{figure}

Figure~\ref{DIRBE_amp22} shows a diagram of the $K-[22]$ colour as a function of the
(full) amplitude at 2.2\,$\mu{\rm m}$, which is essentially the same wavelength as the
$K$ band. It becomes clear from this diagram that the dust mass-loss rate increases with
increasing amplitude. This holds also for the amplitude at 1.25\,$\mu{\rm m}$, which is
also reported by \citet{Pri10}, with the main difference that the amplitude is larger at
this shorter wavelength. The correlation also holds for the different chemical subtypes,
except for the Tc-rich M-stars (filled squares), which do not seem to show this
correlation. However, behaviour is skewed by the fact that the two Tc-rich M-type
stars with the reddest $K-[22]$ colour are binaries, and there is only one Tc-rich
M-star with a full amplitude above 0\fm6.

When we exclude the binaries $o$~Cet and R~Aqr, the correlation coefficient is 0.59. This
correlation coefficient is comparable to that between pulsation period and $K-[22]$
colour of the Tc-poor and Tc-rich Miras, respectively, see Table~\ref{coeffs}.
Interestingly, there is hardly any correlation between pulsation period and amplitude.
It seems that the dust mass-loss rate depends both on pulsation period and amplitude,
but that pulsation period and amplitude vary almost independently from one to another.

A correlation between mass-loss rate and near-IR pulsation amplitude has in the
past been shown to exist, for instance,\ by \citet{WPF87}. Recently, \citet{Rie15} have
demonstrated a correlation between dust-production rate and {\it Spitzer} [3.6]
amplitude among C-rich stars in the Magellanic Clouds. However, the correlation
among their O-rich stars is feeble at best. Fig.~\ref{DIRBE_amp22} demonstrates
that such a relation is also present among O-rich stars (type M and S), and that
it is independent from the occurrence of 3DUP.

\section{Hypotheses for the origin of the two sequences}\label{dicuss}

As has been noted in Paper~I, it is counter-intuitive that Tc-poor Miras have a redder
$K-[22]$ colour (i.e.\ dust mass-loss rate) than Tc-rich Miras at the same pulsation
period. (It should be noted here that a difference in temperature could also
cause part of the separation between the two sequences of Miras. Stars of lower
temperature have larger $K-[22]$ colour. We find that black bodies of temperatures
2600\,K and 3600\,K differ in $K-[22]$ by $0\fm48$. Here, 2600\,K is a realistic
effective temperature for many Miras, whereas the hottest Miras could have an effective
temperature close to 3600\,K. Thus, the full difference between the two sequences of
$\sim1\fm0$ (Fig.~\ref{K-22}) cannot be explained by a difference in temperature. Most
of all, it also defies a straightforward explanation why Tc-rich Miras should be hotter
than Tc-poor ones and only shifts the problem.) Na\"ively, one would expect that
an AGB star evolves in the \mbox{$P$ vs.\ $K-[22]$} plane towards longer period (also
by mode switching) and higher (dust) mass-loss rate, when at some point 3DUP sets in,
and the star continues to evolve to higher mass-loss rate without much
disturbance. (We note that the TP preceding a 3DUP event has a profound but
short-term impact on both period and mass loss, but only a fraction of $\sim1\%$ of
Miras is expected to undergo a TP at any one time.) In particular, the relatively
small abundance change imposed by one 3DUP event is not expected to lead to large
changes in the pulsation period or mass-loss rate, or both. Thus, the observed
behaviour of stars is surprising and deserves further attention to understand the
underlying mechanism(s). We developed hypotheses to explain the observations, which
we discuss in the following.

\subsection{Hypothesis 1: 3DUP influences the pulsation period}\label{hyp1}

One hypothesis that was formulated in Paper~I was that 3DUP somehow influences
the pulsation period of a star, that is,\ shifting it to longer periods by
changing its density. In this way, a previously Tc-poor star would be shifted
to a longer period while the $K-[22]$ colour could essentially stay constant,
and the star would appear in the Tc-rich sequence. \citet{SIW14} investigated
the effect of moderate abundance changes that may be induced by 3DUP, such as
of the C/O ratio, on the atmospheric structure and colours of a series of Mira
pulsation models. They found relatively little influence on\ the pulsation
periods of the model stars, for example. Furthermore, one might expect that such
a period shift would also be noticeable in period -- magnitude
(or period -- luminosity) diagrams by a broadening of the fundamental mode
sequence, for instance. This is not observed, however \citep{Whi08,Rie10,Rie15}.
Because hypothesis~1 is supported neither by theoretical nor observational
evidence, we regard it as falsified.

It should also be stated here that the Tc-poor Miras are most likely
fundamental-mode pulsators, not overtone pulsators, just as the Tc-rich Miras.
As already mentioned, first-overtone pulsators only reach periods up to $\sim240$\,d;
already above 200\,d of period, the number of first-overtone pulsators is very low
\citep{Rie10}. Many of the Tc-poor Miras, on the other hand, have pulsation
periods longer than 240\,d, up to 440\,d (Fig.~\ref{linfit}). Thus, the two
sequences in the \mbox{$P$ vs.\ $K-[22]$} plane cannot be explained by Tc-poor
and Tc-rich Miras oscillating in different modes.

\subsection{Hypothesis 2: Tc-poor and Tc-rich Miras are groups of stars with different masses}\label{hyp2}

AGB evolutionary models predict that a threshold mass exists below which a star never
experiences a 3DUP event on the AGB; this threshold mass depends on metallicity.
In Paper~I, the hypothesis was put forward that the two sequences of Miras could be
related to two groups of stars with different masses, the Tc-rich Miras being the more
massive group at a given period. It was speculated that the two groups evolve differently,
thus occupying different areas in the \mbox{$P$ vs.\ $K-[22]$} diagram. Among others, this
hypothesis was given credibility based on the fact that the Tc-rich Miras are (on average)
closer to the Galactic midplane, which indicates a higher average birth mass of the stars.
Here we perform a more careful analysis than in Paper~I of the distance from the Galactic
midplane as a function of period.

As in Paper~I, we derived the distance of the stars from the Sun using the period-magnitude
relation of \citet[][their Sequence~1]{Rie10}, assuming that all sample Miras are
fundamental-mode pulsators and assuming a distance modulus to the LMC of 18\fm50. If some
of the sample stars were overtone pulsators, their distances would be significantly
overestimated. The lack of an absolute luminosity and hence distance calibration means that
the distances assigned to AGB stars are a major uncertainty in this analysis. Nevertheless,
the so determined distances from the Sun were then used to calculate their distance from
the Galactic midplane using their Galactic coordinates.

\begin{figure}
  \centering
  \includegraphics[width=\linewidth,bb=2 4 502 356]{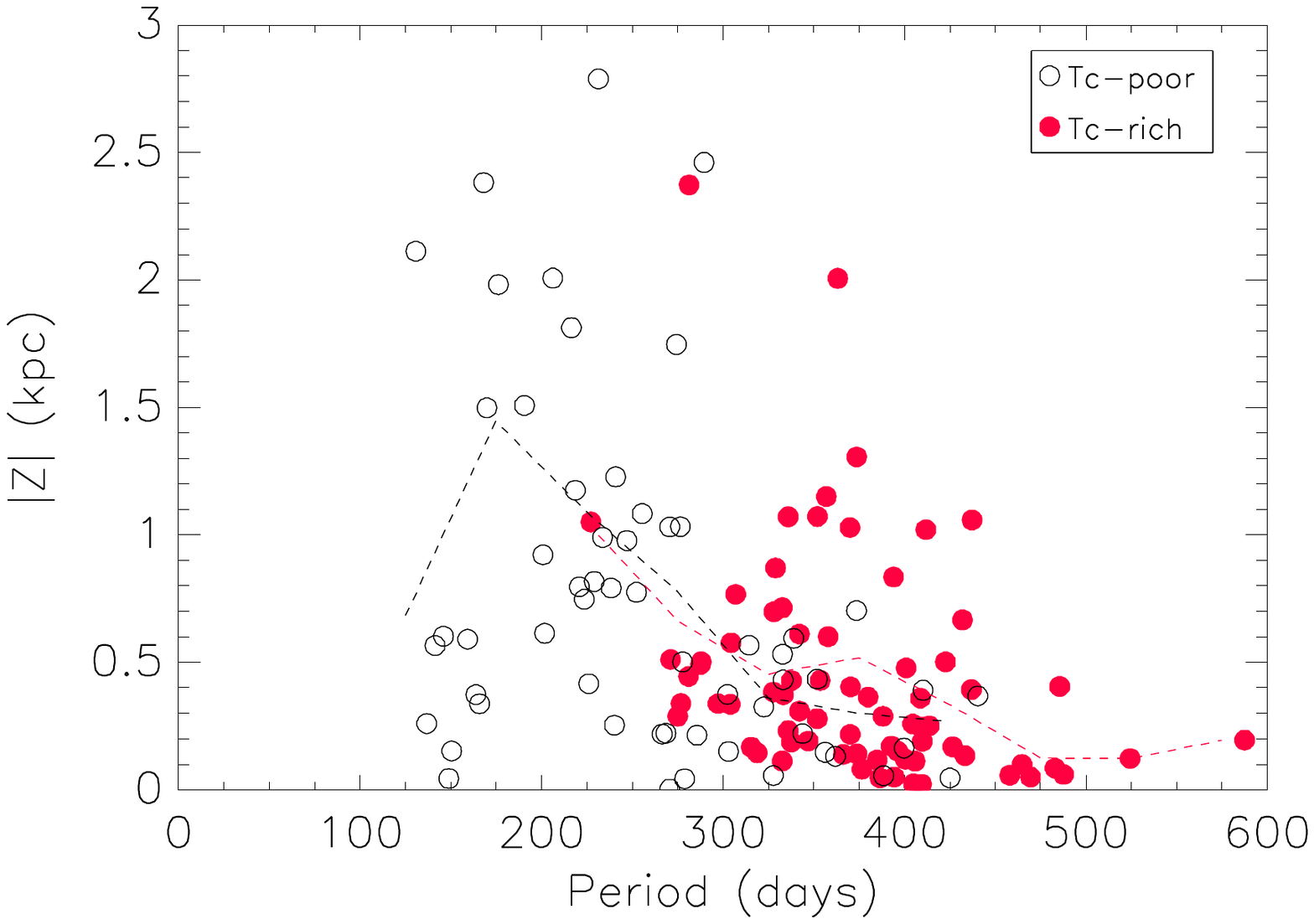}
  \caption{Absolute value of the distance to the Galactic midplane $|Z|$ as a function
    of pulsation period of the Galactic disc Miras in the sample. Open circles
    are Tc-poor Miras, whereas the filled red circles are the Tc-rich ones. The
    dashed lines show the respective run of the mean in 50\,d period bins. The
    Tc-poor Mira U~Psc falls outside the plotting range, at $P=172.6$\,d and
    $|Z|\approx4.2$\,kpc.}\label{PZ}
\end{figure}

Figure~\ref{PZ} shows the absolute value of the distance from the Galactic midplane $|Z|$
as a function of pulsation period of the Galactic disc Miras. Tc-rich Miras (filled red
circles) on average have longer periods than the Tc-poor ones (open circles). However, at
a given period, the two samples have, on average, a similar distance from the Galactic
midplane. This is illustrated by the dashed lines, which show the run of the mean $|Z|$
in 50\,d period bins. The shortest-period objects are surprisingly close to the midplane,
but this can easily be understood as a bias towards close-by, apparently bright stars
that will of course also be relatively close to the midplane. With increasing period, the
mean distance from the midplane decreases in parallel for the two groups. Because the mean
distance from the midplane is a good indicator of the age of a group of stars, hence of
their initial mass,  the mass also seems to increase smoothly with pulsation period,
regardless of the presence of Tc.

The mass-loss rate itself could be more directly related to the stellar mass than Tc-content.
At any period, Miras with a higher mass-loss rate could be the less massive ones, as their
surface gravities may be expected to be lower, hence their atmospheres are more weakly 
bound. Therefore, we also separated the sample into two groups, one of them bluer than the
median $K-[22]=1.884$, and one redder than this, and again inspected $|Z|$ as a function
of pulsation period. However, these two groups show no significant difference either. From
this analysis we do not find evidence that Tc-rich and Tc-poor Miras are groups of stars
of different masses at a given pulsation period.
We caution, however, that the distances to the stars analysed here have significant
uncertainties.

\begin{figure}
  \centering
  \includegraphics[width=\linewidth,bb=4 13 498 349]{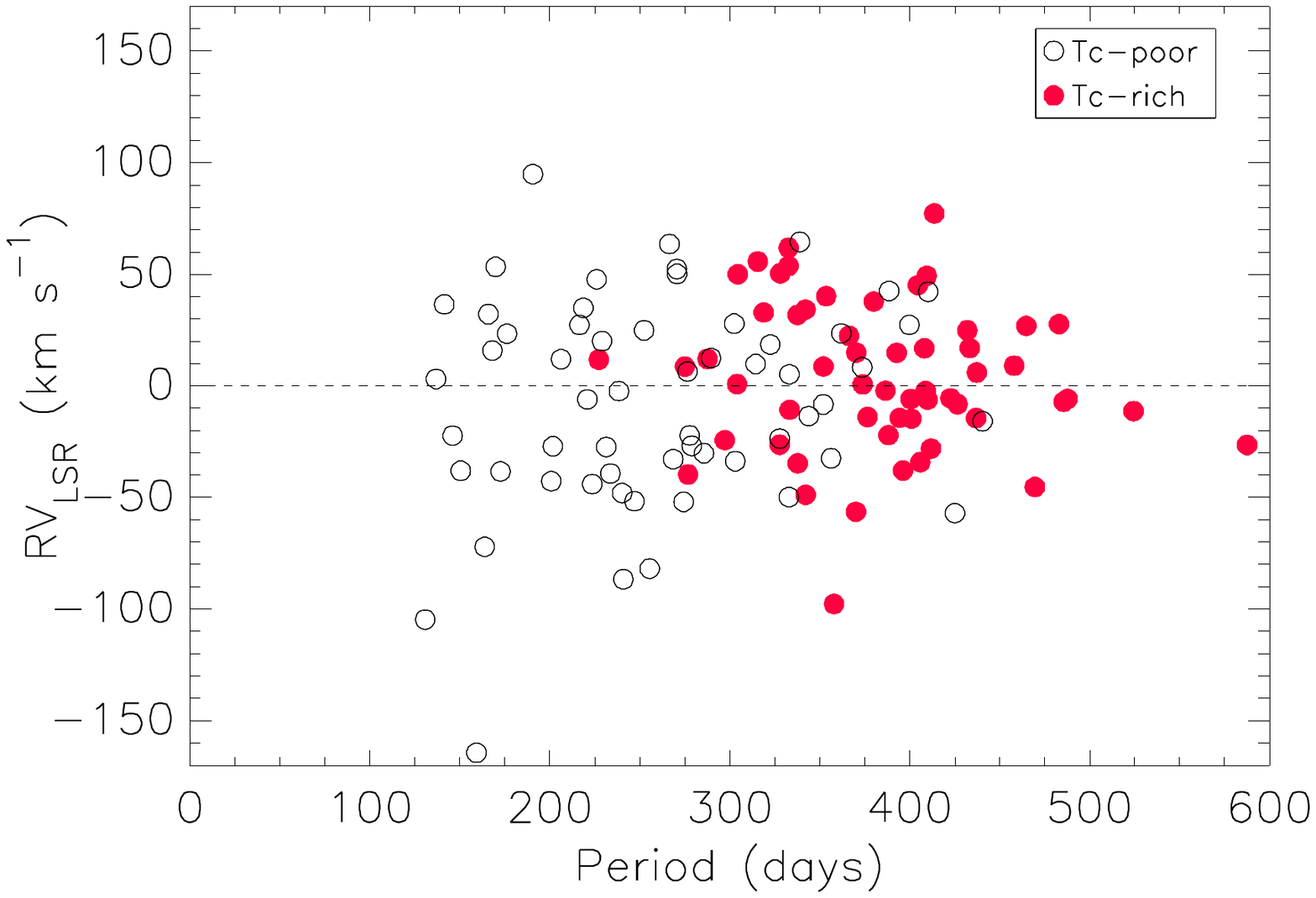}
  \caption{Radial velocities relative to the local standard of rest as a function
  of pulsation period of the Galactic disc Miras. Tc-poor S~Car is outside the plotting
  range at $P=148.9$\,d and $v_{\rm LSR}=+277$\,km\,s$^{-1}$. Symbols are the same as in
  Fig.~\ref{PZ}.}\label{V_LRS}
\end{figure}

One kinematic tracer that can be assessed independently from the knowledge of the
stellar distances is the radial velocity. Just as height above the Galactic midplane,
the radial velocity dispersion of Miras decreases with increasing period, hence with
decreasing age \citep{Feast63}. Heliocentric radial velocity measurements are indeed
available for a substantial fraction of the Galactic disc Miras in our sample (59
Tc-poor, 72 Tc-rich). These were converted into radial velocities with respect to the
local standard of rest ($v_{\rm LSR}$) and are plotted as a function of period in
Fig.~\ref{V_LRS}. The figure seems to confirm the general decrease of velocity
dispersion with increasing period. However, the transition from Tc-poor to Tc-rich
stars is smooth. In the period range of overlap between 227 and 441\,d (longer than
the shortest-period Tc-rich Mira, but shorter than the longest-period Tc-poor Mira),
the standard deviation of the velocities is 39.4\,km\,s$^{-1}$ and 34.9km\,s$^{-1}$
for the 37 Tc-poor and 48 Tc-rich Miras, respectively. The small difference may be
entirely attributed to the general decrease in velocity dispersion with period.
This means that kinematics does not provide evidence either for a difference in mass at a
given period between the two groups of Miras.

\subsection{Hypothesis 3: 3DUP decreases the mass-loss rate}\label{hyp3}

\begin{figure}
  \centering
  \includegraphics[width=\linewidth,bb=38 60 682 532,clip]{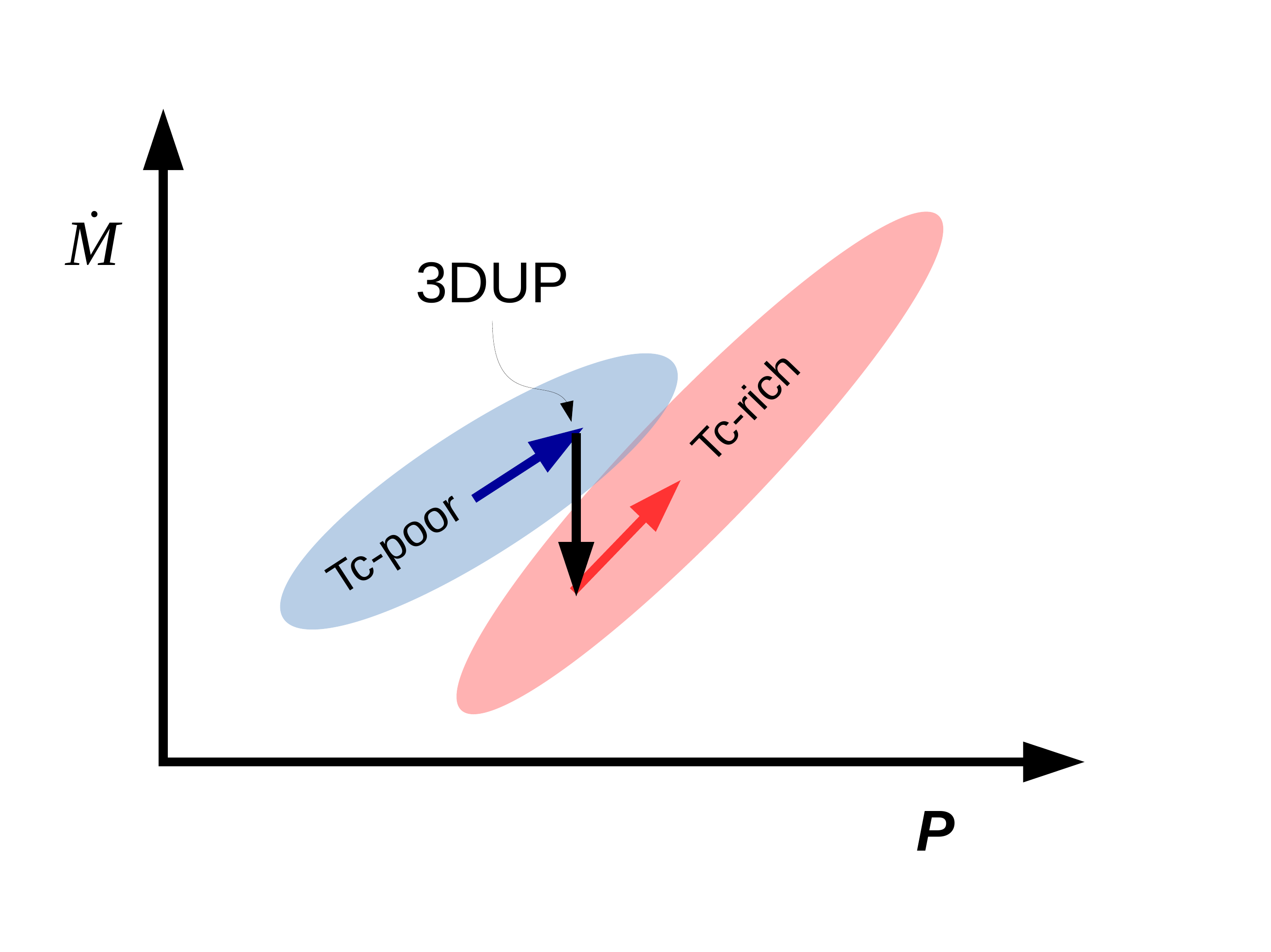}
  \caption{Schematic illustration of the evolution of a Mira in the
  \mbox{$\dot{M}$ vs.\ $P$} diagram under the assumption of hypothesis~3
  (thick arrows): The stellar mass-loss rate decreases upon 3DUP, while
  the period stays constant. The shaded areas indicate the observed
  sequences of Tc-poor and Tc-rich Miras (Fig.~\ref{K-22}).}\label{hyp3_fig}
\end{figure}

The third hypothesis that is put forward assumes that a 3DUP event somehow leads to a
reduction of the (dust) mass-loss rate from a star. In this way, upon a 3DUP event, a
Mira from the Tc-poor sequence would be shifted downwards in the \mbox{$P$ vs.\ $K-[22]$}
diagram at a constant pulsation period. Figure~\ref{hyp3_fig} shows a schematic
representation of the evolution of a star in the \mbox{$\dot{M}$ vs.\ $P$} diagram
under this paradigm. We note that the fractional change to the pulsational period of
a Mira during its lifetime as a Mira can only be rather small \citep{Whi94}. The period
at which 3DUP and thus the reduction in mass loss-rate takes place is mostly a function
of mass and metallicity of the star. Later in its evolution as Tc-rich object, the
star may enter the super-wind phase where its circumstellar envelope (CSE) becomes so
thick that the flux especially in the blue spectral range becomes too low to detect
the Tc lines.

The physical mechanism by which 3DUP could reduce the mass-loss rate is not
obvious, however. If 3DUP somehow reduces the formation of dust, it could lead to a reduction
of the mass-loss rate because the transfer of momentum from the stellar photons to
the dust and gas would become less efficient. The atmospheric composition of
post-3DUP stars differs from that of pre-3DUP stars in a number of aspects, most
notably in the content of He and C, but also in that of heavier elements, including
radioactively unstable isotopes. The ions produced by the decay of unstable isotopes
could influence the formation of dust. For example, these ions could produce so many
seeds for the nucleation of dust that only small grains form in the atmosphere, so
that absorption and most of all scattering \citep{Hoef08} of stellar photons becomes
inefficient. This would reduce the transfer of momentum to the gas, thus reduce
mass-loss rate. Moreover, the ions could strongly influence a key reaction in the
formation of dust grains. Quantum-chemical calculations of dust-forming aluminum
oxide clusters show that the structures, potential energies, and dust formation
routes differ significantly when ions are included, compared to the neutral case
\citep{Des97}.

We made an estimate of the energy input and the number of ions per second produced by
decaying unstable isotopes in a typical AGB atmosphere. For this purpose, we selected
from the {\sc FRUITY} database \citep{Cri09} the model with $M=2.0M_{\sun}$, $Z=0.01$,
zero initial rotation velocity, and a standard $^{13}$C pocket. This should be a
fairly representative model of a typical AGB star in the solar neighbourhood. For the
estimate, we selected from this model the composition after the third 3DUP event.
This is relatively early in the evolution of this model, which undergoes
nine 3DUP events in total. This choice is justified by the fact that already stars with very
little s-element enrichment (spectral type M) but with Tc in their spectra appear to
have reduced $K-[22]$.

Only three isotopes contribute significantly to the total amount of energy released
per second by radioactive decay, namely $^{26}$Al (91.9\%), $^{59}$Ni (4.5\%), and
$^{60}$Fe (2.9\%, taking into account the energy released by the following, fast
decay of $^{60}$Co to $^{60}$Ni). Interestingly, the element Al is part of alumina
dust (see Sect.~\ref{13mufeat}) that is a potential candidate to trigger the onset
of dust formation in oxygen-rich AGB stars \citep[see e.g.][]{Kar13,Dec17}. We also
note that the lithium-rich, intermediate-mass AGB star candidates are comparably
blue in $K-[22]$ for their long pulsation periods, and they are predicted to be
rich in $^{26}$Al.

The decay of $^{26}$Al and other unstable isotopes releases about
189\,MeV\,s$^{-1}$\,g$^{-1}$, where the energy carried away by the antineutrino
is already subtracted. For the ionisation by the electrons and positrons
from the $\beta$-decays, we further assume for the sake of simplicity that the
gas is solely composed of H (76\% by mass) and He (24\%), thus the weighted mean
ionisation potential per gas particle is 14.697\,eV. From this it follows that
about $1.3\times10^{7}$ ions per second per gram of gas are produced. We further
find that the recombination timescale is $\sim3300$\,s.

This can be compared to other sources of ions in the AGB atmosphere. Atmospheric
models that follow the non-equilibrium chemistry in gas shocked periodically by
the stellar pulsations \citep[e.g.][]{Gob16} find ionisation fractions of
$4.83\times10^{-13}$ in post-shock gas at $T=2000$\,K (depending on the
composition, dust grains condense at $T\approx1000$\,K). Assuming the
same composition of H+He as above, this translates into a number of
$2.36\times10^{14}$ ions per gram of gas. Even if this ionisation fraction is
reached only for a small fraction of the pulsation period of a few hundred
days, it is clear that this outweighs the number of ions produced by
radioactive decays by far. Eventually, other sources of ions must also be considered, for
example UV radiation from the stellar chromosphere, far-UV photons from the
interstellar radiation field, and cosmic rays.

The effect of radioactive decays on circumstellar envelopes has previously been
investigated by \citet{Gla95}, who concluded that the ionisation of AGB winds is
likely to be due in part to {\it \textup{in situ}} radioactive decays, at least for stars
such as \object{IRC +10216}. Thus, a mechanism as sketched under hypothesis~3 might
accomplish the sharp separation of Tc-poor and Tc-rich stars observed here (lower
panel of Fig.~\ref{linfit}), but further investigation of this issue is certainly
warranted.


\subsection{Hypothesis 4: 3DUP decreases dust emissivity}\label{hyp4}

A variation of hypothesis~3 is that 3DUP reduces the dust emissivity, and thus the
$K-[22]$ colour, but not necessarily the dust mass-loss rate. The dust emissivity
depends on the grain size distribution, and on the composition, shape, and
structure of the dust grains. All these could be influenced by a change in gas
composition due to 3DUP, as sketched above. Again, the physical mechanism that
could be at work is not immediately obvious. We note that \citet{Jones12} ruled out
changes in silicate composition with period. Nevertheless, radio CO observations
to measure the gas mass-loss rate of Tc-poor and Tc-rich Miras, respectively,
would be very helpful to rule out one or the other hypothesis.


\subsection{Hypothesis 5: Tc lines are unobservable in some stars}\label{hyp5}

Finally, we consider the hypothesis that Tc may become invisible in stars not
because it is underabundant, but because the Tc lines are not excited. This may
be because lines are blanketed (e.g. by TiO, or C$_2$, CH, or CN), because Tc
becomes ionised, or because gas-phase Tc condenses into molecules. In this
way, a separation of Tc-poor and Tc-rich Miras could be mimicked.

Line blanketing of Tc could occur in particularly cool objects, but the
relative opacity at these wavelengths is weak. Ionisation of
\ion{Tc}{i} $\rightarrow$ \ion{Tc}{ii} can occur at $\sim$2500\,K at 10$^{-7}$ bar,
rising to $\sim$3000\,K at 10$^{-5}$ bar (\citet{Gor05}; cf.\ \citet{Speck12} for
temperature--pressure curves in AGB stars). This may become significant,
depending on the stellar effective temperature (as measured around the wavelength
of Tc lines) and the presence of a stellar chromosphere. The chemistry of Tc,
particularly at low pressures, is not well understood. TcO is not well-measured
in the literature, but TcO$_2$ is known to be stable up to high ($\gtrsim1200$\,K)
temperatures \citep{Rar83}.

Figure~\ref{demoTcdetect} shows that the absorption features are
weaker in the Tc-poor star than in the Tc-rich one. This may indicate that higher
atmospheric opacity obscures the lines in the Tc-poor star, for exmaple\ by facilitated
dust production. We compared lines of \ion{Fe}{i} and \ion{V}{i} with similar
excitation potentials as the Tc resonance lines and confirm that all of these
lines are somewhat weaker in the star without Tc than in the star with Tc.
However, other comparisons of Tc-poor and Tc-rich star spectra do not generally
reveal such a trend \citep[e.g.][]{Utt07,Utt11}. Furthermore, the effect of line
weakening has been observed in Mira stars, which varies with phase and even from
cycle to cycle \citep{MDK62}. Thus, any conclusions from such comparisons are not
straightforward.

Without further characterisation of the line formation region in these stars, and
the technetium chemistry involved, it is not possible to state with confidence
whether the factors mentioned above contribute to Tc being absent in some stars.
Including a more realistic and detailed Tc chemistry and line formation in
current model atmospheres would be very helpful to further study this aspect.
However, we note that stars with higher mass-loss rate but shorter period
(possibly smaller and hotter) are preferentially Tc-poor, indicating that
either higher temperature or pressure may be important.

\section{Searching for 3DUP in Galactic bulge Miras}\label{bulgesec}

Third dredge-up and subsequent mass loss are important for the chemical
evolution and global properties of a stellar population. Because the position
of a Mira star in the \mbox{$P$ vs.\ $K-[22]$} diagram is a strong predictor
of the presence of Tc (lower panel of Fig.~\ref{linfit}), we may use this
diagram in the reverse way to assess the occurrence of 3DUP in Miras in a
specific stellar population. In Paper~I, this method of detecting 3DUP has been
applied to a small sample of Galactic bulge Miras. Here we apply the same
approach to a much larger sample, using the best-separating line calculated
above (Eq.~\ref{divlin_equ}).

The Galactic bulge has been found to lack intrinsic carbon stars, that is, luminous stars
on the TP-AGB that owe their carbon enhancement to internal nucleosynthesis and
dredge-up, rather than to mass transfer from a binary companion. It appears that the
implied luminosities of C-stars in the direction of the bulge are too low for them to be
on the TP-AGB \citep{Azz88,BT89,TR91}, except maybe for a very few objects \citep{Mis13,Mat17}.
The central regions of other galaxies also exhibit a lack of carbon stars \citep{Boy13}.
This has been interpreted as a sign that 3DUP does not occur or is not efficient enough
to form carbon stars in these stellar populations. This may be either due to low mass
(old age), high metallicity, or high helium content \citep{Kar14}. A high oxygen
abundance would also prevent the formation of carbon stars, even if the stars experience
3DUP, because the amount of carbon needed to reach C/O$>$1 increases with increasing
oxygen abundance. However, a few AGB stars in the outer Galactic bulge, which are
included in the present sample, have been found to contain Tc \citep{Utt07}. Here we
would like to shed light on the question if 3DUP is not efficient enough in the
Galactic bulge to form carbon stars, or if it is not very common at all. The goal
of this investigation is not to compile a comprehensive list of 3DUP candidates in the
bulge, but rather to check how large a fraction of bulge Miras potentially experiences
3DUP and how common this mixing process is in bulge stars.

\begin{figure}
  \centering
  \includegraphics[width=\linewidth,bb=4 2 500 358]{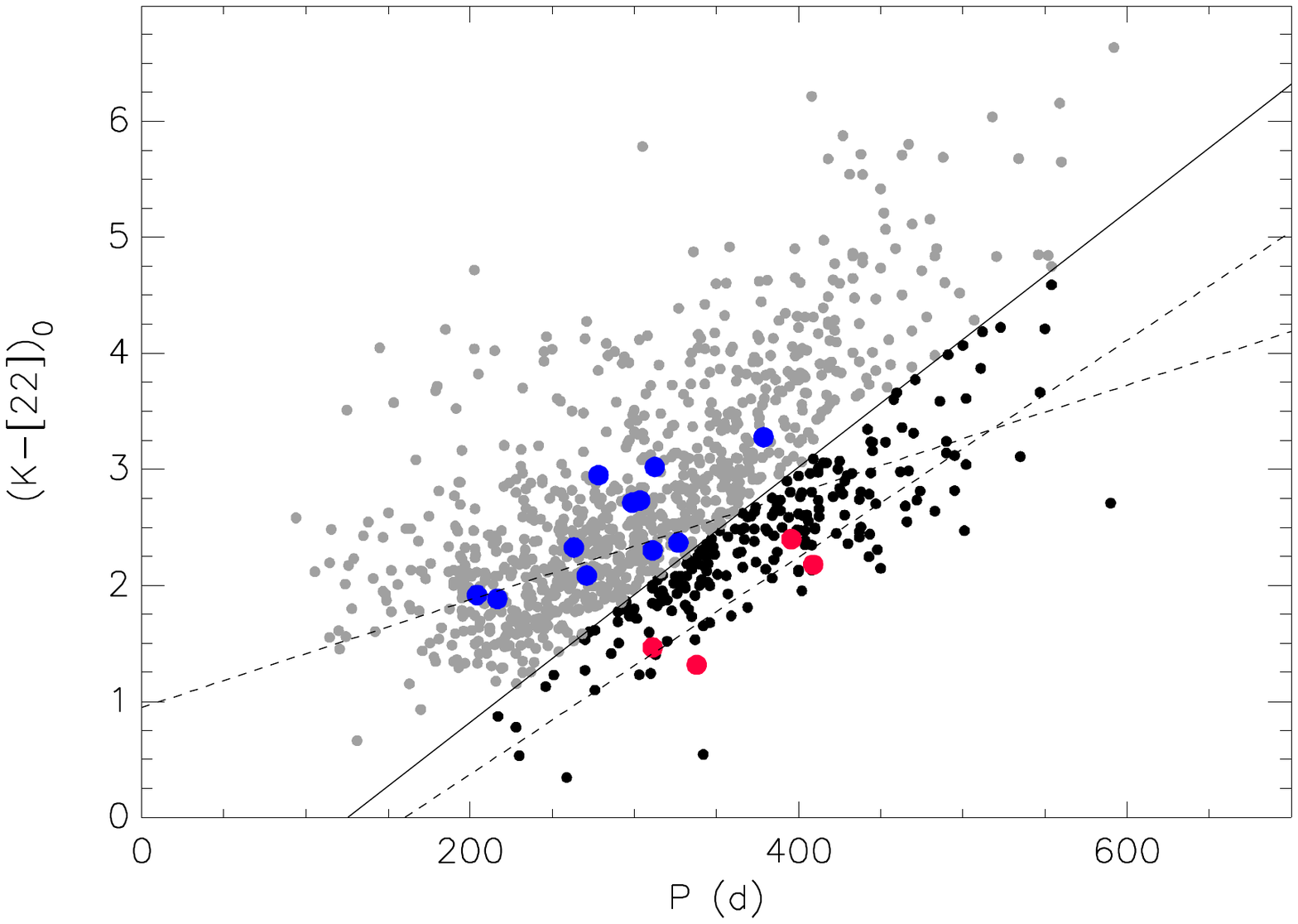}
  \caption{\mbox{P vs.\ $(K_{\rm s}-[22])_0$} diagram of 1091 MACHO Galactic
  bulge Miras from \citet{BH13}. The solid line is the relation that was found
  to best separate Tc-poor from Tc-rich stars (Equ.~\ref{divlin_equ}). MACHO
  Miras falling above this line, i.e. putative Tc-poor stars, are plotted as
  grey symbols, whereas Miras below the line, i.e. putative Tc-rich stars,
  as black symbols. Large symbols represent the Plaut stars analysed for
  their Tc content in \citet{Utt07} (blue: Tc-poor, red: Tc-rich). The dashed
  lines are the linear fits to the Tc-poor and Tc-rich Miras from the upper
  panel of Fig.~\ref{linfit}.}\label{bulgefig}
\end{figure}

The sample considered here are the 1094 Galactic bulge Miras in the Massive
Astrophysical Compact Halo Object (MACHO) survey database identified by
\citet[][and references therein]{BH13}. This sample was chosen because all
light curves and pulsation periods have been checked visually, thus it is expected to
constitute a very clean sample of bulge Miras. Nevertheless, for three of the MACHO
stars, no precise period could be determined, thus reducing the sample to 1091 objects.
$K_{\rm s}$-band magnitudes were extracted from the 2MASS catalogue and 22\,$\mu{\rm m}$
magnitudes from the {\it WISE} catalogue. (Only 2MASS magnitudes are used in this
section, therefore we use the notation $K_{\rm s}$ band here. The sample discussed in
the rest of the paper has also been observed in other photometric systems, therefore,
although the magnitudes have been transformed to the 2MASS system, we use the notation
$K$-band in all other sections or when referring to that sample.)  Interstellar
extinction is non-negligible towards stars  n the Galactic bulge, therefore we
extracted the $K_{\rm s}$-band extinction towards the objects from the BEAM
calculator\footnote{{\tt http://mill.astro.puc.cl/BEAM/calculator.php}} provided
by \citet{Gon11}. The reddening law of \citet{Nis09} was assumed because it is
more appropriate for the bulge than other laws \citep{Nataf16,Qin18}, as well as an
extinction ratio $A_{[22]}/A_{Ks}=0.02/0.093$ to calculate the de-reddened colour
$(K_{\rm s}-[22])_0$. The result of this exercise is shown in Fig.~\ref{bulgefig}.

The whole sample exhibits a clearly increasing $(K_{\rm s}-[22])_0$ colour with
increasing period. We note that the ordinate range in Fig.~\ref{bulgefig} is expanded
with respect to Fig.~\ref{K-22}. While the reddest stars with information on their
Tc content in Table~\ref{taba1} have $K-[22]$ just above 4, some bulge Miras have
$(K_{\rm s}-[22])_0$ greater than 6. Very red Miras are missing in the former
sample because they would be unobservable in the blue spectral range. The Plaut stars are also included
in Fig.~\ref{bulgefig}. They  have been investigated for their Tc
content by spectral observations by \citet{Utt07}. These stars clearly follow the
linear relations defined by all Tc-poor and Tc-rich Miras, respectively, in the
upper panel of Fig.~\ref{linfit}. Furthermore, they exactly follow the separation
into Tc-poor and Tc-rich Miras prescribed by Eq.~\ref{divlin_equ}.

Of the total sample of 1091 MACHO objects, 885 (81\%) fall above the best separating
line, and 206 (19\%) fall below it. To compare these fractions to the disc sample
(Sect.~\ref{Miras}), short-lived intermediate-mass star candidates have to be excluded
because they are not expected to be present in the predominantly old bulge
($>3$\,Gyr). When these are excluded from the disc sample, 87\% of Tc-poor Miras are above
the line. This would mean that in the Galactic bulge a few more stars fall below the
line than would be expected if none of them had 3DUP. This means that a few of the Galactic
bulge Miras may indeed undergo 3DUP.

However, we must also take into account several effects that may broaden the whole
distribution in the $(K_{\rm s}-[22])_0$ colour: i) only one epoch of $K_{\rm s}$-band
observations is available for the MACHO Miras (the minimum peak-to-peak $K$-band
amplitude for Mira classification is 0\fm4.) ; ii) there is an uncertainty attached to
the extinction correction; iii) a high Galactic dust background emission impedes an
accurate measurement of the 22\,$\mu{\rm m}$ flux; and iv) the different metallicities
of these stars may produce a different $(K_{\rm s}-[22])_0$ colour at a given
mass-loss rate. The number of Miras below the line may be increased by these effects.
If the whole sample were shifted upward by 0\fm2 in $(K_{\rm s}-[22])_0$, which may be
comparable to making the distribution narrower by this amount, a fraction of 88\% of
stars would fall above the line. We conclude that, while 3DUP may occur in some
Galactic bulge Miras, we do not find evidence that it is a widespread phenomenon.
Carbon stars in the Galactic bulge may be rare because 3DUP may be a rare event in
first place.

Any bulge AGB star that undergoes 3DUP may have the same origin as the C-rich
bulge Miras identified by \citet{Mat17}. Following their discussion, we suggest that
these could be either merged binaries \citep{RG90}, members of a younger, metal-rich
population \citep{Bens13}, or accreted from a dwarf galaxy such as the Sagittarius
dwarf galaxy \citep{WIC96}.


\section{Summary and conclusions}\label{conclusio}


We presented a follow-up paper on a previous work \citep[][Paper~I]{Utt13}, using a larger
sample and expanding the analysis. The main finding of Paper~I is confirmed here, namely
that Tc-poor and Tc-rich Miras form two separate sequences in the \mbox{$P$ vs.\ $K-[22]$}
diagram, and that at a given period, Tc-poor Miras have a higher $K-[22]$ excess than
Tc-rich Miras. Both sequences show a very strong correlation with pulsation period,
whereas the correlation is weak if no distinction is made with respect to the presence of
Tc. Other combinations of near-to-mid-IR colours and {\it ISO} dust spectra are used to
demonstrate that the higher excess is not confined to the $[22]$ band or a specific dust
feature in this band.

Tc-rich SRVs are found to form another sequence of increasing $K-[22]$ colour with
increasing pulsation period (Fig.~\ref{SRVs}). A mode switch from overtone
to fundamental-mode (Mira) pulsation most probably only occurs near the upper
end of this SRV sequence because most Miras have much redder colours than the SRVs
on that sequence. Fundamental-mode SRa-type pulsators basically confirm the
findings from Miras. In general, it is concluded that red outliers in the
\mbox{$P$ vs.\ $K-[22]$} plane, be they SRVs or Miras, are good candidates to host
a close binary systems.

We also find that there is a sharp threshold at $K-[22]$ above which the
13\,$\mu{\rm m}$ feature disappears. This colour corresponds to a total
mass-loss rate of $\dot{M}\approx2.6\times10^{-7}M_{\sun}yr^{-1}$.

A central aim of this work was to investigate if gas mass-loss rates, determined from
radio CO lines, also show two sequences if a distinction is made with respect to Tc
in the atmosphere of the stars. The gas mass-loss rates $\dot{M}_{\rm g}$ were found
to have a strong correlation with the $K-[22]$ colour. We established a linear
relationship that relates the $K-[22]$ colour to gas mass-loss rate in the range
$1.5\lesssim K-[22]\lesssim7.5$, which may be used to estimate mass-loss rates from
$K-[22]$ colour. Moreover, $\dot{M}_{\rm g}$ clearly increases with pulsation period,
where a strong increase is seen between $\sim350$ and $\sim600$\,d of period.
However, the available data revealed no such sequences in the
\mbox{$\dot{M}_{\rm g}$ vs.\ $P$} diagram. A number of obstacles inhibit our
endeavour, one of them being uncertainties attached to the determined gas mass-loss
rates, but most importantly the low number of Miras with both a determined 
$\dot{M}_{\rm g}$ and information on Tc content in the period range of interest. We
encourage researchers to increase the number of radio CO line observations for gas
mass-loss rate determinations of (optically bright) Miras in particular in the period
range $250\lesssim P/{\rm d}\lesssim400$ to study the interplay between
pulsation, mass loss, and third dredge-up in more detail.

An inspection of near-IR pulsation amplitudes reveals that it is tightly correlated
with $K-[22]$ colour, with a similar correlation coefficient as between period
and $K-[22]$ colour for Tc-poor and Tc-rich Miras, respectively. Thus, the dust
mass-loss rate seems to be dependent on both pulsation period and amplitude.

We also discussed hypotheses to elucidate the process(es) that might be at work to form the
two sequences of Miras. We find that a mechanism that leads to an increase of pulsation
period upon 3DUP appears unlikely. We do not find supportive evidence that the sequences
could be the result of two groups of stars with different masses (at a given pulsation
period) because the implied distance from the Galactic mid-plane as well as the
radial velocity dispersion decrease in parallel and smoothly with increasing period for
both groups. The analysis suffers from uncertainties on the distances, however. Two more
hypotheses suggest that the (dust) mass-loss rate and/or the dust emissivity decrease
upon 3DUP. It is conceivable that dust grains form less efficiently when the atmospheric
composition of the star changes upon 3DUP, or that reactions that are important for dust
formation are strongly influenced by the presence of ions that are produced by the
radioactive decay of unstable isotopes that result from internal nucleosynthesis
(e.g.\ $^{26}$Al). However, the physical mechanism at the microscopic scale that
would be required by these hypotheses are not directly obvious. Finally, the
chemistry and line formation of Tc in cool stellar atmospheres are not well understood,
and we considered the possibility that stars in a certain parameter range appear Tc-poor
not because Tc is not present, but because the Tc lines are not excited. In conclusion,
at this point in time we cannot offer a hypothesis that plausibly explains the available
observational evidence.

We also applied the clear separation of Tc-poor from Tc-rich Miras to a large sample of
Galactic bulge Miras to investigate if 3DUP occurs in this population. Several bulge Miras
were indeed found in the region of the \mbox{$P$ vs.\ $K-[22]$} diagram where
post-3DUP Miras are expected to be found. However, their number is not large, and we
conclude that 3DUP is possible in bulge stars, but probably not a wide-spread
phenomenon. This is in agreement with the finding that intrinsic carbon stars on the
TP-AGB are rare in the bulge.


We encourage further studies of the observational evidence presented in this paper,
which might be a hint at a hitherto unknown, but potentially important process
influencing mass loss on the TP-AGB.

\begin{acknowledgements}
  We thank T.\ Lebzelter for helpful comments and discussions. SU thanks his
  brother Richard for help with installing the GNU Data Language (GDL) that was
  used for the data analysis in this paper. IM acknowledges support from the UK
  Science and Technology Facilities Council under grant ST/L000768/1. We thank
  the XSL team for collecting, reducing, and making available the data used for
  this work, in particular Ana\"{i}s Gonneau for her support with retrieving the XSL data. 
  We acknowledge with thanks the variable star observations from the AAVSO International
  Database contributed by observers worldwide and used in this research. This publication
  makes use of data products from the Two Micron All Sky Survey, which is a joint project
  of the University of Massachusetts and the Infrared Processing and Analysis
  Center/California Institute of Technology, funded by the National Aeronautics and Space
  Administration and the National Science Foundation. This publication makes use of data
  products from the Wide-field Infrared Survey Explorer, which is a joint project of the
  University of California, Los Angeles, and the Jet Propulsion Laboratory/California
  Institute of Technology, funded by the National Aeronautics and Space Administration.
  This research is based on observations with AKARI, a JAXA project with the participation
  of ESA.
\end{acknowledgements}

\begin{appendix}
\longtab[1]{
\begin{landscape}
\begin{longtable}{lcrcrrccccrcccccccccc}
\caption{The data.}\label{taba1}\\
\hline
\hline
Name& Tc & Ref. & P     & VT & ST & $K$ & $\sigma(K)$ & $J-K$& $\sigma(J-K)$ & Ref. & [1.25]& [2.20]& [12] & [25] & [60] & [9]  & [18] & [65] & [90] & [22] \\
    &    &      & (days)&    &    &(mag)&(mag)&(mag)&(mag)& & (mag) & (mag) & (mag)& (mag)& (mag)& (mag)& (mag)& (mag)& (mag)& (mag)\\
 (1)& (2)& (3)  & (4)   & (5)& (6)& (7) & (8)  & (9)   & (10)  & (11) & (12) & (13) & (14) & (15) & (16) & (17) & (18) & (19) & (20) & (21) \\
\hline
\endfirsthead
\caption{Continued.}\\
\hline
\hline
Name& Tc & Ref. & P     & VT & ST & $K$ & $\sigma(K)$& $J-K$& $\sigma(J-K)$ & Ref. & [1.25]& [2.20]& [12] & [25] & [60] &  [9] & [18] & [65] & [90] & [22] \\
    &    &      & (days)&    &    &(mag)&(mag)& (mag)&(mag)& & (mag) & (mag) & (mag)& (mag)& (mag)& (mag)& (mag)& (mag)& (mag)& (mag)\\
 (1)& (2)& (3)  & (4)   & (5)& (6)& (7) & (8)  & (9)   & (10)  & (11) & (12) & (13) & (14) & (15) & (16) & (17) & (18) & (19) & (20) & (21) \\
\hline
\endhead
\hline
\endfoot
\hline
\endlastfoot
\object{R And}       & 1 &  1,3 & 409.9 & Mira &  S & $ +0.031$ & 0.226 & 1.902 & 0.358 & $a,g$ &  &          & $-2.657$ & $-3.493$ & $-3.271$ &          & $-3.145$ & $-3.482$ & $-3.490$ & $-2.843$ \\
\object{W And}       & 1 &  1,2 & 396.3 & Mira &  M & $ +0.376$ & 0.184 & 1.461 & 0.309 & $a,g$ &  &          & $-1.927$ & $-2.575$ & $-2.629$ & $-1.076$ & $-1.980$ & $-2.303$ & $-2.381$ & $-1.667$ \\
\object{Y And}       & 0 &    1 & 220.9 & Mira &  M & $ +4.152$ & 0.036 & 1.101 & 0.040 & $a$ &  &          & $+2.360$ & $+1.828$ &          & $+3.090$ & $+2.277$ & & & $+2.522$ \\
\object{Z Ant}       & 1 &    5 & 128.9 & SRV  &  S & $ +1.532$ & 0.242 & 1.396 & 0.357 & $a$ &  &          & $-0.080$ & $-0.813$ & $-0.519$ & $+0.666$ & $-0.526$ & & $+0.066$ & $-0.505$ \\
\object{R Aql}       & 0 & 3,10 & 270.6 & Mira &  M & $-0.788$ & 0.120 & 1.361 & 0.038 & $c$ & $+0.538$ & $-0.998$ & $-2.881$ & $-3.903$ &          & $-1.861$ & $-3.182$ & $-4.097$ & $-3.581$ & $-3.486$ \\
\object{RR Aql}      & 0 &    7 & 399.8 & Mira &  M &  $+0.626$ & 0.065 & 1.576 & 0.030 & $a,c,f$ & $+1.940$ & $+0.311$ & $-2.585$ & $-3.356$ & $-3.369$ & $-1.857$ & $-3.067$ & $-3.509$ & & $-2.939$ \\
\object{RT Aql}      & 0 &    1 & 327.8 & Mira &  M & $ +1.120$ & 0.113 & 1.331 & 0.031 & $a,c,g$ &          &          & $-1.051$ & $-1.802$ & $-1.865$ & $-0.356$ & $-1.556$ & $-1.497$ & $-1.742$ & $-1.387$ \\
\object{SY Aql}      & 0 &    1 & 356.3 & Mira &  M & $ +2.166$ & 0.200 & 1.436 & 0.074 & $a,c$ &          &          & $-0.903$ & $-1.713$ & $-1.593$ & $+0.329$ & $-1.100$ & $-1.350$ & $-1.507$ & $-0.780$ \\
\object{UV Aql}      & 1 &   12 & 385.5 & SRV  &  C & $ +1.558$ & 0.216 & 1.972 & 0.368 & $a$ & $+3.457$ & $+1.626$ & $+0.100$ & $-0.116$ & $-0.881$ & $+0.587$ & $+0.234$ & & $-0.776$ & $+0.257$ \\
\object{V335 Aql}    & 0 &   15 & 176.2 & Mira &  M & $ +6.786$ & 0.037 & 1.210 & 0.041 & $a,i$ &          &          & $+4.860$ & $+4.182$ &          & $+5.493$ &          & & & $+4.363$ \\
\object{V899 Aql}    & 1 &   11 & 373.7 & Mira &  S & $ +5.560$ &  & 1.440 &  & $o$ &          &          & $+3.189$ & $+2.836$ &          & $+3.909$ & $+3.200$ & & & $+3.336$ \\
\object{V915 Aql}    & 1 &    4 &  81.8 & SRV  &  S & $ +2.090$ & 0.294 & 1.405 & 0.427 & $a$ &          &          & $+1.036$ & $+0.754$ & $+0.420$ & $+1.530$ & $+1.147$ & & & $+1.306$ \\
\object{V 1717Aql}   & 1 &   11 & 357.0 & Mira &  S & $ +5.490$ &  & 1.540 &  & $o$ &          &          & $+3.197$ & $+2.743$ &          & $+3.898$ & $+2.862$ & & & $+3.146$ \\
\object{W Aql}       & 1 &   11 & 487.8 & Mira &  S & $ -0.556$ & 0.358 & 2.090 & 0.439 & $a$ & $+1.835$ & $-0.114$ & $-4.364$ & $-4.995$ &          &          & $-4.573$ & $-5.167$ & $-4.750$ & $-4.578$ \\
\object{R Aqr}       & 1 &    7 & 388.2 & Mira &  M &  $-0.821$ & 0.051 & 1.578 & 0.035 & $a,c,f$ & $+0.420$ & $-1.027$ & $-4.367$ & $-4.769$ & $-4.371$ &          & $-3.777$ & $-3.988$ & $-3.962$ & $-4.352$ \\
\object{T Aqr}       & 0 &    1 & 201.8 & Mira &  M & $ +3.237$ & 0.070 & 1.174 & 0.031 & $a,c,g$ &          &          & $+1.825$ & $+1.438$ & $+1.154$ & $+2.246$ & $+1.583$ & & & $+1.670$ \\
\object{RZ Ari}      & 0 &   15 &  54.8 & SRV  &  M & $ -0.951$ & 0.083 & 1.117 & 0.083 & $a,j$ & $+0.122$ & $-1.012$ & $-1.789$ & $-1.865$ & $-1.796$ & $-1.303$ & $-1.596$ & $-1.420$ & $-1.647$ & $-1.440$ \\
\object{AU Aur}      & 1 &   13 & 400.5 & Mira &  C & $ +2.928$ & 0.264 & 2.276 & 0.267 & $a$         &          & $+0.520$ & $+0.269$ & $-0.058$ & $+0.876$ & $+0.471$ & & $-0.287$ & $+0.702$ \\
\object{AZ Aur}      & 1 &   13 & 413.7 & Mira &  C & $ +2.883$ & 0.252 & 2.125 & 0.311 & $a$ & $+4.086$ & $+2.327$ & $-0.609$ & $-0.818$ & $-1.133$ & $+0.622$ & $-0.246$ & $-0.744$ & $-0.837$ & $-0.124$ \\
\object{NO Aur}      & 1 &   14 & 102.1 & SRV  &  M & $ +0.971$ & 0.196 & 1.151 & 0.377 & $a$ & $+2.564$ & $+0.990$ & $-0.467$ & $-1.330$ & $-1.584$ & $+0.199$ &          & $-1.462$ & $-1.523$ & $-1.139$ \\
\object{RU Aur}      & 1 &    7 & 469.6 & Mira &  M &  $+1.764$ & 0.322 & 2.073 & 0.477 & $a$ &          &          & $-1.710$ & $-2.701$ & $-2.472$ &          & $-1.897$ & $-2.102$ & $-2.349$ & $-2.309$ \\
\object{SZ Aur}      & 1 &   10 & 458.1 & Mira & MS & $ +0.760$ & 0.222 & 1.393 & 0.340 & $a$ &          &          & $-0.841$ & $-1.407$ & $-1.563$ & $-0.132$ & $-1.263$ & $-1.391$ & $-1.605$ & $-0.980$ \\
\object{UV Aur}      & 1 &    1 & 394.4 & Mira &  C & $ +2.129$ & 0.220 & 1.900 & 0.307 & $a$ &          &          & $-0.974$ & $-1.215$ & $-1.130$ &          & $-0.746$ & & $-0.824$ & $-0.918$ \\
\object{VX Aur}      & 0 &    1 & 322.6 & Mira &  M & $ +1.460$ & 0.202 & 1.211 & 0.271 & $a,g$ &          &          & $+0.130$ & $-0.363$ & $-0.315$ & $+0.715$ & $-0.140$ & & $-0.092$ & $+0.066$ \\
\object{R Boo}       & 0 &    1 & 223.6 & Mira &  M & $ +2.0436$ & 0.228 & 1.203 & 0.350 & $a,g$ &          &          & $+0.491$ & $-0.162$ & $-0.104$ & $+0.746$ & $-0.106$ & & $-0.168$ & $+0.018$ \\
\object{V Boo}       & 1 &    4 & 258.2 & SRV  &  M & $ +1.035$ & 0.210 & 1.133 & 0.376 & $a$ & $+2.248$ & $+1.009$ & $-0.421$ & $-0.826$ & $-0.647$ & $+0.228$ & $-0.471$ & & $-0.382$ & $-0.339$ \\
\object{T Cae}       & 1 &   15 & 156.0 & SRV  &  C & $ +2.238$ & 0.250 & 1.519 & 0.362 & $a$ & $+3.737$ & $+2.401$ & $+0.812$ & $+0.838$ & $+0.658$ & $+1.526$ & $+1.106$ & & & $+1.241$ \\
\object{T Cam}       & 1 &    1 & 373.9 & Mira &  S & $ +0.814$ & 0.194 & 1.741 & 0.310 & $a$ & $+1.977$ & $+0.675$ & $-0.410$ & $-0.619$ & $-1.211$ & $+0.158$ & $-0.269$ & $-0.700$ & $-1.029$ & $+0.042$ \\
\object{T Cap}       & 0 &    7 & 270.7 & Mira &  M &  $+3.297$ & 0.048 & 1.295 & 0.027 & $a,f$ &          &          & $+1.349$ & $+0.849$ & $+1.032$ & $+2.023$ & $+1.350$ & & & $+1.196$ \\
\object{AU Car}      & 1 &   11 & 347.0 & Mira &  S & $ +4.270$ &  & 1.620 &  & $o$ &          &          & $+3.320$ & $+3.070$ &          & $+3.743$ & $+3.113$ & & & $+3.332$ \\
\object{CM Car}      & 0 &   15 & 339.0 & Mira &  M & $ +4.497$ & 0.020 & 1.448 & 0.039 & $a$ &          &          & $+2.011$ & $+1.402$ & $+1.237$ & $+2.986$ & $+2.014$ & & & $+1.789$ \\
\object{KN Car}      & 1 &    5 & 352.0 & Mira &  S & $ +2.516$ & 0.248 & 1.509 & 0.371 & $a$ &          &          & $+1.668$ & $+1.507$ & $+1.329$ & $+2.159$ & $+1.707$ & & & $+1.690$ \\
\object{RX Car}      & 1 &   11 & 336.0 & Mira &  S & $ +4.900$ &  & 1.640 &  & $o$ &          &          & $+3.066$ & $+2.763$ &          & $+3.779$ & $+3.043$ & & & $+3.089$ \\
\object{S Car}       & 0 &    1 & 148.9 & Mira &  M & $ +1.859$ & 0.278 & 1.354 & 0.392 & $a$ &          &          & $+0.288$ & $+0.005$ & $-0.153$ & $+0.844$ & $+0.416$ & & & $+0.462$ \\
\object{U Cas}       & 1 &    1 & 276.9 & Mira &  S & $ +2.878$ & 0.306 & 1.451 & 0.400 & $a$ &          &          & $+1.321$ & $+1.080$ & $+0.872$ & $+1.887$ & $+1.516$ & & & $+1.550$ \\
\object{W Cas}       & 1 &   13 & 405.9 & Mira &  C & $ +2.218$ & 0.196 & 1.720 & 0.289 & $a$ &          &          & $+1.220$ & $+0.874$ & $+0.222$ & $+1.951$ & $+1.364$ & & & $+1.501$ \\
\object{R Cen}       & 0 &    1 & 538.1 & Mira &  M & $ -0.644$ & 0.027 & 1.286 & 0.019 & $a,c,g$ & $+0.711$ & $-0.555$ &          & $-3.677$ & $-3.271$ & $-1.744$ & $-3.010$ & $-3.068$ & $-2.973$ & $-2.865$ \\
\object{V744 Cen}    & 0 &  4,5 & 167.3 & SRV  &  M & $ -0.809$ & 0.434 & 1.271 & 0.507 & $a$ & $+0.443$ & $-0.700$ & $-2.236$ & $-3.082$ & $-2.868$ & $-1.412$ & $-2.510$ & $-2.680$ & $-2.797$ & $-2.575$ \\
\object{V763 Cen}    & 0 &   15 &  60.0 & SRV  &  M & $ +0.807$ & 0.278 & 1.112 & 0.410 & $a$ & $+1.699$ & $+0.675$ & $+0.152$ & $+0.100$ & $-0.199$ & $+0.607$ & $+0.374$ & & $+0.145$ & $+0.481$ \\
\object{VX Cen}      & 1 &    5 & 109.1 & SRV  &  S & $ +0.504$ & 0.252 & 1.413 & 0.404 & $a$ &          &          & $-0.538$ & $-0.808$ &          & $-0.023$ & $-0.509$ & & & $-0.448$ \\
\object{T Cep}       & 1 &1,3,10& 386.6 & Mira &  M &  $-1.824$ & 0.182 & 1.328 & 0.308 & $a$ & $-0.385$ & $-1.751$ & $-3.563$ & $-3.996$ & $-3.859$ &          & $-3.738$ & $-3.680$ & $-3.698$ & $-3.580$ \\
\object{$o$ Cet}     & 1 &  1,3 & 332.7 & Mira &  M & $ -2.451$ & 0.076 & 1.296 & 0.037 & $a,c,d,g$ & $-1.368$ & $-2.662$ & $-5.592$ & $-6.315$ & $-6.008$ &          &          & & & $-5.899$ \\
\object{R Cet}       & 0 &    1 & 165.9 & Mira &  M & $ +2.538$ & 0.066 & 1.240 & 0.022 & $a,c,d,g$ & $+4.028$ & $+2.548$ & $-0.030$ & $-1.251$ & $-1.261$ & $+0.411$ & $-0.718$ & & $-1.247$ & $-1.130$ \\
\object{T Cet}       & 1 &   14 & 159.3 & SRV  & MS & $ -0.808$ & 0.276 & 1.304 & 0.337 & $a$ & $+0.405$ & $-0.738$ & $-2.112$ & $-2.297$ & $-2.707$ & $-1.336$ & $-2.282$ & $-2.336$ & $-2.452$ & $-2.252$ \\
\object{U Cet}       & 0 &  7,9 & 233.7 & Mira &  M & $ +2.770$ & 0.176 & 1.220 & 0.060 & $f$ & $+4.146$ & $+2.782$ & $+1.136$ & $+0.716$ & $+0.722$ &          & $+0.892$ & & & $+1.062$ \\
\object{W Cet}       & 1 &  1,9 & 352.1 & Mira & MS & $ +2.128$ & 0.051 & 1.245 & 0.021 & $a,c$ & $+3.409$ & $+2.181$ & $+0.820$ & $+0.576$ & $+0.538$ & $+1.396$ & $+0.863$ & & & $+0.935$ \\
\object{R Cha}       & 1 &   15 & 337.8 & Mira &  M & $ +2.313$ & 0.258 & 1.483 & 0.354 & $a$ & $+3.491$ & $+2.181$ & $+0.522$ & $+0.058$ & $+0.085$ & $+1.061$ & $+0.296$ & & $+0.161$ & $+0.502$ \\
\object{CR Cir}      & 0 &    6 & 108.2 & SRV  &  S & $ +2.064$ & 0.242 & 1.253 & 0.341 & $a$ &          &          & $+1.477$ & $+1.385$ &          & $+1.909$ & $+1.659$ & & & $+1.783$ \\
\object{BZ CMa}      & 1 &   11 & 332.8 & Mira &  S & $ +4.960$ &  & 1.630 &  & $o$ &          &          & $+3.942$ & $+3.642$ &          & $+4.092$ & $+3.659$ & & & $+4.006$ \\
\object{V355 CMa}    & 1 &   11 & 370.0 & Mira &  S & $ +4.760$ &  & 1.460 &  & $o$ &          &          & $+3.353$ & $+2.862$ &          &          & $+3.263$ & & & $+3.101$ \\
\object{R CMi}       & 1 &    1 & 337.7 & Mira &  C & $ +2.661$ & 0.079 & 1.369 & 0.034 & $a,b$ &          &          & $+0.689$ & $+0.701$ & $+0.601$ & $+1.510$ & $+1.189$ & & & $+1.024$ \\
\object{U CMi}       & 0 &    1 & 410.3 & Mira &  M & $ +2.608$ & 0.035 & 1.265 & 0.032 & $a,c,g$ &          &          & $+1.006$ & $+0.171$ & $+0.278$ & $+1.692$ & $+0.628$ & & & $+0.697$ \\
\object{V CMi}       & 1 &    1 & 366.3 & Mira &  M & $ +1.965$ & 0.061 & 1.249 & 0.021 & $a,c$ &          &          & $+0.135$ & $-0.462$ & $-0.558$ & $+0.938$ & $+0.081$ & & & $-0.158$ \\
\object{XY CMi}      & 1 &   11 & 271.2 & Mira &  S & $ +5.370$ &  & 1.420 &  & $o$ &          &          & $+3.898$ &         &          & $+4.557$ & $+4.011$ & & & $+4.114$ \\
\object{R Cnc}       & 0 &    1 & 362.0 & Mira &  M & $ -0.622$ & 0.059 & 1.298 & 0.044 & $a,c,d,g$ & $+1.008$ & $-0.491$ & $-2.534$ & $-3.024$ & $-2.961$ & $-1.540$ & $-2.753$ & $-2.661$ & $-2.657$ & $-2.802$ \\
\object{RS Cnc}      & 1 & 4,14 & 241.2 & SRV  & MS & $ -1.873$ & 0.158 & 1.163 & 0.228 & $a$ & $-0.448$ & $-1.644$ & $-3.074$ & $-3.730$ & $-3.594$ &          & $-3.597$ & $-3.558$ & $-3.581$ & $-3.338$ \\
\object{R Col}       & 1 &    1 & 328.1 & Mira &  M & $ +3.066$ & 0.260 & 1.296 & 0.363 & $a$ &          &          & $+0.959$ & $+0.158$ & $+0.370$ & $+1.400$ & $+0.578$ & & & $+0.838$ \\
\object{T Col}       & 0 &    1 & 226.1 & Mira &  M & $ +1.957$ & 0.063 & 1.151 & 0.016 & $a,c,g$ & $+3.108$ & $+1.868$ & $+0.039$ & $-0.451$ & $-0.215$ & $+0.611$ & $+0.027$ & & & $+0.190$ \\
\object{RY CrA}      & 0 &   15 & 206.3 & Mira &  M & $ +7.271$ & 0.034 & 1.243 & 0.042 & $a$ &          &          & $+4.863$ & $+4.311$ &          & $+5.613$ &          & & & $+4.582$ \\
\object{V CrB}       & 1 & 12,13& 358.0 & Mira &  C & $ +1.321$ & 0.276 & 2.153 & 0.388 & $a$ & $+3.737$ & $+1.814$ & $-1.413$ & $-1.700$ & $-1.806$ & $-0.581$ & $-0.962$ & $-1.399$ & $-1.333$ & $-0.997$ \\
\object{X CrB}       & 0 &    1 & 240.9 & Mira &  M & $ +3.420$ & 0.258 & 1.388 & 0.333 & $a,g$ &          &          & $+1.738$ & $+1.239$ &          & $+2.447$ & $+1.435$ & & & $+1.619$ \\
\object{U Crt}       & 0 &   15 & 168.1 & Mira &  M & $ +5.492$ & 0.016 & 1.205 & 0.024 & $a$ &          &          & $+4.126$ &          &          & $+4.573$ & $+3.895$ & & & $+3.844$ \\
\object{BH Cru}      & 1 &   10 & 524.4 & Mira &  C & $ +1.448$ & 0.028 & 1.632 & 0.029 & $a,b$ &          &          & $+0.329$ & $+0.028$ & $-0.335$ & $+0.384$ & $-0.028$ & & $-0.691$ & $+0.058$ \\
\object{AA Cyg}      & 1 &   14 & 212.7 & SRV  &  S & $ +0.625$ & 0.224 & 1.441 & 0.365 & $a$ &          &          & $-0.368$ & $-0.906$ & $-1.616$ & $+0.090$ & $-0.589$ & $-1.550$ & $-2.017$ & $-1.024$ \\
\object{AW Cyg}      & 1 &   12 & 213.0 & SRV  &  C & $ +2.114$ & 0.302 & 1.947 & 0.423 & $a$ & $+1.696$ & $+0.416$ & $+0.254$ & $+0.036$ & $-0.692$ & $+0.745$ & $+0.363$ & & $-0.520$ & $+0.514$ \\
\object{$\chi$ Cyg}  & 1 &  1,3 & 408.2 & Mira &  S & $ -1.902$ & 0.156 & 1.863 & 0.240 & $a,g$ & $-0.448$ & $-1.971$ & $-4.440$ & $-4.584$ & $-4.578$ &          &          & $-4.605$ & $-4.280$ & $-4.168$ \\
\object{LX Cyg}      & 1 &   10 & 587.5 & Mira &  C & $ +2.929$ & 0.272 & 2.199 & 0.358 & $a$ &          &          & $+1.693$ & $+1.528$ &          & $+1.389$ & $+0.900$ & & & $+1.304$ \\
\object{R Cyg}       & 1 &    1 & 426.6 & Mira &  S & $ +0.861$ & 0.246 & 1.390 & 0.399 & $a$ &          &          & $-1.424$ & $-2.224$ & $-2.500$ & $-0.547$ & $-1.447$ & $-1.999$ & $-2.068$ & $-2.039$ \\
\object{RU Cyg}      & 1 &    4 & 233.2 & SRV  &  M & $ -0.128$ & 0.208 & 1.445 & 0.324 & $a$ & $+1.118$ & $-0.335$ & $-2.067$ & $-2.983$ & $-2.961$ & $-1.332$ & $-2.762$ & $-2.733$ & $-2.989$ & $-2.694$ \\
\object{U Cyg}       & 1 &   13 & 464.8 & Mira &  C & $ +0.456$ & 0.190 & 1.732 & 0.295 & $a$ &          &          & $-1.494$ & $-1.815$ & $-2.132$ & $-0.752$ & $-1.445$ & $-1.571$ & $-2.094$ & $-1.512$ \\
\object{V441 Cyg}    & 1 &    9 & 286.6 & SRV  &  S & $ +1.704$ & 0.068 & 1.516 & 0.079 & $a,e$ &          &          & $+0.861$ & $+0.579$ &          & $+1.318$ & $+0.834$ & & & $+0.960$ \\
\object{V460 Cyg}    & 1 &   12 & 180.0 & SRV  &  C & $ +0.270$ & 0.184 & 1.532 & 0.287 & $a$ & $+1.750$ & $+0.291$ & $-1.073$ & $-1.251$ & $-2.219$ & $-0.544$ & $-0.962$ & $-1.758$ & $-2.143$ & $-0.923$ \\
\object{W Cyg}       & 1 & 3,4,7& 132.8 & SRV  &  M &  $-1.444$ & 0.040 & 1.193 & 0.016 & $a,g,h$ & $-0.229$ & $-1.402$ & $-2.728$ & $-3.311$ & $-3.147$ & $-1.959$ & $-2.983$ & $-2.972$ & $-3.052$ & $-2.965$ \\
\object{WX Cyg}      & 1 &   13 & 409.5 & Mira &  C & $ +2.376$ & 0.322 & 2.121 & 0.414 & $a$ &          &          & $+1.066$ & $+0.698$ &          & $+1.590$ & $+1.020$ & & & $+0.868$ \\
\object{R Del}       & 0 &    1 & 285.7 & Mira &  M & $ +1.932$ & 0.262 & 1.526 & 0.404 & $a,g$ &          &          & $-0.102$ & $-0.619$ & $-0.467$ & $+0.723$ & $-0.246$ & & $-0.271$ & $-0.030$ \\
\object{U Del}       & 1 &  4,7 & 119.8 & SRV  &  M & $ -0.353$ & 0.141 & 1.223 & 0.016 & $a,h$ & $+1.009$ & $-0.161$ & $-1.751$ & $-2.662$ & $-2.395$ & $-1.047$ & $-2.185$ & $-1.978$ & $-2.174$ & $-2.324$ \\
\object{Z Del}       & 1 &    1 & 304.5 & Mira &  S & $ +4.135$ & 0.036 & 1.154 & 0.042 & $a$ &          &          & $+2.398$ & $+2.102$ &          & $+2.897$ & $+2.338$ & & & $+2.547$ \\
\object{R Dor}       & 0 &    4 & 327.1 & SRV  &  M & $ -4.227$ & 0.206 & 1.575 & 0.230 & $a$ & $-2.656$ & $-4.046$ & $-5.652$ & $-5.933$ & $-5.780$ &          &          & $-5.821$ & $-5.583$ & $-5.517$ \\
\object{T Dra}       & 1 &   13 & 422.7 & Mira &  C & $ +1.365$ & 0.212 & 2.698 & 0.319 & $a$ &          &          & $-2.107$ & $-2.480$ & $-2.808$ & $-1.622$ & $-2.257$ & $-2.737$ & $-2.695$ & $-2.039$ \\
\object{UX Dra}      & 1 & 12,13& 177.0 & SRV  &  C & $ +0.374$ & 0.194 & 1.640 & 0.374 & $a$ & $+1.747$ & $+0.278$ & $-0.926$ & $-1.149$ & $-1.501$ & $-0.414$ & $-0.726$ & & & $-0.768$ \\
\object{W Dra}       & 0 &   10 & 289.6 & Mira &  M & $ +5.688$ & 0.016 & 1.340 & 0.029 & $a$ &          &          & $+3.264$ & $+3.022$ & $+2.528$ & $+3.362$ & $+2.666$ & & & $+2.644$ \\
\object{WZ Dra}      & 1 &    7 & 411.9 & Mira &  M & $ +2.764$ & 0.018 & 1.379 & 0.102 & $a,e$ & $+4.028$ & $+2.849$ & $+1.249$ & $+0.647$ & $+0.836$ & $+2.172$ & $+1.103$ & & & $+1.151$ \\
\object{T Eri}       & 0 &    1 & 252.4 & Mira &  M & $ +2.418$ & 0.043 & 1.194 & 0.008 & $a,c,g$ & $+4.028$ & $+2.621$ & $+0.749$ & $+0.325$ & $+0.608$ & $+1.361$ & $+0.694$ & & & $+0.628$ \\
\object{U Eri}       & 0 &    1 & 274.4 & Mira &  M & $ +4.047$ & 0.031 & 1.236 & 0.037 & $a$ &          &          & $+2.588$ & $+2.187$ &          & $+3.183$ & $+2.638$ & & & $+2.394$ \\
\object{W Eri}       & 0 &    9 & 373.6 & Mira &  M & $ +1.776$ & 0.085 & 1.366 & 0.093 & $a,c$ & $+2.680$ & $+1.250$ & $-1.340$ & $-1.996$ & $-1.744$ & $-0.179$ & $-0.982$ & & & $-0.979$ \\
\object{NZ Gem}      & 0 &    2 &  30.1 & SRV  &  S & $ +0.557$ & 0.184 & 1.031 & 0.279 & $a$ &          &          & $+0.126$ & $+0.038$ & $+0.076$ & $+0.551$ & $+0.315$ & & $+0.223$ & $+0.521$ \\
\object{R Gem}       & 1 &    1 & 370.1 & Mira &  S & $ +1.664$ & 0.226 & 1.071 & 0.333 & $a,g$ & $+2.817$ & $+1.634$ & $+0.293$ & $-0.121$ & $-0.734$ & $+0.489$ & $-0.171$ & & $-0.671$ & $-0.415$ \\
\object{T Gem}       & 1 &    1 & 287.7 & Mira &  S & $ +2.705$ & 0.270 & 1.412 & 0.356 & $a$ &          &          & $+2.200$ & $+2.049$ &          & $+2.668$ & $+2.280$ & & & $+2.575$ \\
\object{V Gem}       & 1 &    1 & 275.1 & Mira &  M & $ +2.808$ & 0.304 & 1.318 & 0.405 & $a,g$ & $+4.313$ & $+2.946$ & $+1.215$ & $+0.770$ & $+1.292$ & $+1.876$ & $+1.154$ & & & $+1.188$ \\
\object{VX Gem}      & 1 & 12,13& 380.0 & Mira &  C & $ +3.112$ & 0.077 & 1.801 & 0.071 & $a,g$ & $+5.148$ & $+3.565$ & $+1.046$ & $+0.964$ & $+0.448$ & $+2.013$ & $+1.367$ & & & $+1.275$ \\
\object{ZZ Gem}      & 1 &   13 & 315.6 & Mira &  C & $ +3.233$ & 0.032 & 2.037 & 0.059 & $a,b$ & $+4.700$ & $+2.679$ & $+1.308$ & $+1.161$ & $+0.637$ & $+1.846$ & $+1.515$ & & & $+1.516$ \\
\object{$\pi$1 Gru}  & 1 &   14 & 198.8 & SRV  &  S & $ -2.351$ & 0.278 & 1.638 & 0.329 & $a$ & $-0.772$ & $-2.089$ & $-3.766$ & $-4.531$ & $-4.532$ &          &          & $-4.553$ & $-4.367$ & $-4.604$ \\
\object{S Gru}       & 1 &  7,9 & 401.0 & Mira & MS & $ +0.526$ & 0.107 & 1.239 & 0.019 & $a,c$ & $+2.106$ & $+0.691$ & $-1.639$ & $-2.331$ & $-2.157$ & $-0.667$ & $-1.805$ & $-1.766$ & $-1.488$ & $-1.701$ \\
\object{FR Her}      & 0 &   15 & 130.8 & Mira &  M & $ +6.879$ & 0.020 & 1.024 & 0.028 & $a$ &          &          & $+5.541$ &          &          & $+5.718$ &          & & & $+5.410$ \\
\object{OP Her}      & 1 & 4,14 & 120.5 & SRV  &  M & $ +0.198$ & 0.164 & 1.261 & 0.309 & $a$ & $+1.242$ & $+0.150$ & $-0.704$ & $-1.012$ & $-1.124$ & $-0.179$ & $-0.649$ & & $-0.732$ & $-0.579$ \\
\object{RU Her}      & 1 &    7 & 485.6 & Mira &  M & $ +0.192$ & 0.142 & 1.618 & 0.061 & $a,f$ & $+1.652$ & $+0.165$ & $-1.966$ & $-2.662$ & $-2.453$ & $-1.077$ & $-2.138$ & $-2.202$ & $-2.138$ & $-2.717$ \\
\object{S Her}       & 1 &   10 & 304.1 & Mira & MS &  $+1.215$ & 0.080 & 1.166 & 0.025 & $a,c$ & $+2.680$ & $+1.422$ & $-0.209$ & $-0.582$ & $-0.519$ & $+0.521$ & $-0.120$ & & $-0.055$ & $+0.026$ \\
\object{ST Her}      & 1 & 4,14 & 147.0 & SRV  & MS & $ -0.542$ & 0.206 & 1.285 & 0.318 & $a$ & $+0.516$ & $-0.704$ & $-2.118$ & $-2.898$ & $-2.868$ & $-1.397$ & $-2.492$ & $-2.570$ & $-2.580$ & $-2.541$ \\
\object{SV Her}      & 0 &    1 & 238.4 & Mira &  M & $ +4.707$ & 0.029 & 1.131 & 0.035 & $a$ &          &          & $+2.711$ & $+2.343$ &          & $+3.463$ & $+2.661$ & & & $+3.120$ \\
\object{T Her}       & 0 &   10 & 163.9 & Mira &  M &  $+2.988$ & 0.268 & 1.222 & 0.340 & $a$ &          &          & $+1.529$ & $+1.045$ & $+0.993$ & $+2.227$ & $+1.553$ & & & $+1.682$ \\
\object{XZ Her}      & 0 &   15 & 169.9 & Mira &  M & $ +6.489$ & 0.018 & 1.210 & 0.055 & $a$ &          &          &          &          &          & $+5.624$ &          & & & $+5.199$ \\
\object{R Hor}       & 1 &  7,9 & 404.6 & Mira & MS & $ -0.862$ & 0.069 & 1.256 & 0.019 & $a,c$ & $+0.378$ & $-1.044$ & $-3.526$ & $-4.162$ & $-4.134$ & $-1.567$ & $-3.502$ & $-3.951$ & $-3.836$ & $-3.745$ \\
\object{T Hor}       & 0 &  7,9 & 218.7 & Mira &  M & $ +3.363$ & 0.054 & 1.141 & 0.016 & $a,c$ &          &          & $+1.784$ & $+1.373$ & $+1.668$ & $+2.325$ & $+1.601$ & & & $+1.695$ \\
\object{TW Hor}      & 1 &  7,9 & 269.6 & SRV  &  C & $ +0.151$ & 0.026 & 1.497 & 0.064 & $a,e$ & $+1.493$ & $+0.163$ & $-1.303$ & $-1.836$ & $-1.962$ & $-0.627$ & $-1.347$ & $-1.694$ & $-1.639$ & $-1.451$ \\
\object{HV 12667}    & 0 &   15 & 679.4 & Mira &  M & $ +8.717$ & 0.188 & 1.413 & 0.020 & $a,k,l,p$ &          &          &          &          &          & $+7.799$ &          & & & $+6.316$ \\
\object{CZ Hya}      & 1 &   13 & 432.0 & Mira &  C &  $+2.598$ & 0.101 & 2.263 & 0.061 & $a,b$ & $+4.996$ & $+2.621$ & $-0.667$ & $-1.166$ & $-1.255$ & $-0.000$ & $-0.367$ & $-1.077$ & $-1.061$ & $-0.451$ \\
\object{R Hya}       & 1 & 7,10 & 376.6 & Mira &  M &  $-2.503$ & 0.032 & 1.261 & 0.025 & $a,c,d$ & $-1.135$ & $-2.477$ & $-4.374$ & $-4.850$ & $-4.698$ &          &          & $-4.540$ & $-4.391$ & $-4.433$ \\
\object{RR Hya}      & 1 &    1 & 342.1 & Mira &  M & $ +3.008$ & 0.019 & 1.253 & 0.026 & $a,c$ & $+3.561$ & $+2.074$ & $+1.195$ & $+0.764$ & $+0.939$ & $+1.750$ & $+1.101$ & & & $+1.182$ \\
\object{RU Hya}      & 0 &  7,9 & 333.2 & Mira &  M & $ +1.598$ & 0.040 & 1.196 & 0.006 & $c$ & $+3.182$ & $+1.770$ & $-1.091$ & $-1.908$ & $-1.929$ & $-0.005$ & $-0.878$ & $-1.134$ & $-1.475$ & $-1.406$ \\
\object{U Hya}       & 1 & 12,13& 389.4 & SRV  &  C & $ -0.716$ & 0.362 & 1.519 & 0.439 & $a$ & $+0.805$ & $-0.596$ & $-1.894$ & $-2.626$ & $-3.851$ & $-1.614$ & $-2.152$ & $-2.528$ & $-2.689$ & $-2.082$ \\
\object{W Hya}       & 0 & 4,10 & 388.6 & Mira &  M &  $-3.038$ & 0.038 & 1.459 & 0.033 & $a,c,e$ & $-1.747$ & $-3.218$ & $-5.429$ & $-5.619$ & $-5.536$ &          &          & $-5.595$ & $-5.291$ & $-5.202$ \\
\object{T Ind}       & 1 &    1 & 162.0 & SRV  &  C & $ +0.564$ & 0.316 & 1.417 & 0.402 & $a$ & $+2.051$ & $+0.723$ & $-0.544$ & $-0.756$ & $-1.599$ & $+0.029$ & $-0.387$ & $-1.339$ & $-1.459$ & $-0.305$ \\
\object{RX Lac}      & 1 &    7 & 331.4 & SRV  & MS &  $-0.124$ & 0.101 & 1.324 & 0.037 & $a,h$ & $+1.027$ & $-0.215$ & $-1.356$ & $-1.716$ & $-2.121$ & $-0.732$ & $-1.347$ & $-1.867$ & $-1.969$ & $-1.419$ \\
\object{S Lac}       & 0 &    1 & 240.2 & Mira &  M & $ +2.418$ & 0.258 & 1.305 & 0.347 & $a,g$ & $+3.781$ & $+2.481$ & $+0.309$ & $-0.169$ & $+0.047$ & $+1.103$ & $+0.314$ & & & $+0.143$ \\
\object{RR Lib}      & 0 &    1 & 277.8 & Mira &  M & $ +2.513$ & 0.061 & 1.303 & 0.020 & $a,c$ &          &          & $+0.804$ & $+0.369$ & $+0.590$ & $+1.416$ &          & & & $+0.640$ \\
\object{VX Lib}      & 0 &   15 &  58.8 & SRV  &  M & $ +6.883$ & 0.024 & 1.290 & 0.036 & $a,i$ &          &          &          &          &          & $+6.048$ &          & & & $+5.263$ \\
\object{Y Lib}       & 0 &  7,9 & 276.6 & Mira &  M & $ +3.150$ & 0.082 & 1.228 & 0.022 & $a,c$ &          &          & $+0.997$ & $+0.171$ & $+0.318$ & $+1.339$ & $+0.183$ & & $+0.115$ & $-0.199$ \\
\object{S Lup}       & 1 &    1 & 342.2 & Mira &  S & $ +2.924$ & 0.070 & 1.430 & 0.035 & $a,c$ &          &          & $+1.402$ & $+1.204$ &          & $+2.318$ & $+1.778$ & & & $+1.466$ \\
\object{RT Lyn}      & 1 &    7 & 394.1 & Mira &  M & $ +2.718$ & 0.282 & 1.613 & 0.400 & $a$ &          &          & $+0.978$ & $+0.551$ & $+0.601$ & $+1.719$ & $+0.874$ & & & $+0.905$ \\
\object{U Lyn}       & 0 &    7 & 440.5 & Mira &  M & $ +1.533$ & 0.230 & 1.478 & 0.311 & $a$ & $+2.953$ & $+1.362$ & $-1.474$ & $-2.110$ & $-1.825$ & $-0.166$ & $-1.256$ & $-1.359$ & $-1.369$ & $-1.398$ \\
\object{HK Lyr}      & 1 & 12,13& 186.0 & SRV  &  C & $ +1.689$ & 0.286 & 2.054 & 0.389 & $a$ & $+3.221$ & $+1.526$ & $+0.239$ & $-0.035$ & $-0.575$ & $+0.724$ & $+0.326$ & & $-0.463$ & $+0.451$ \\
\object{U Mic}       & 0 &  7,9 & 332.9 & Mira &  M & $ +1.799$ & 0.060 & 1.248 & 0.021 & $a,c$ & $+3.695$ & $+2.040$ & $-0.154$ & $-1.025$ & $-1.275$ & $-0.012$ & $-0.864$ & $-1.441$ & $-1.260$ & $-1.431$ \\
\object{CL Mon}      & 1 &   13 & 483.0 & Mira &  C & $ +1.844$ & 0.072 & 2.718 & 0.047 & $a,b$ &          &          & $-1.503$ & $-1.623$ & $-1.996$ & $-0.848$ & $-1.290$ & $-1.831$ & $-1.809$ & $-1.234$ \\
\object{SY Mon}      & 0 &    7 & 425.1 & Mira &  M &  $+0.848$ & 0.332 & 1.485 & 0.477 & $a$ &          &          & $-1.796$ & $-2.561$ & $-2.562$ & $-0.792$ & $-2.025$ & $-2.281$ & $-2.499$ & $-1.887$ \\
\object{R Nor}       & 0 &   10 & 496.2 & Mira &  M &  $+1.281$ & 0.049 & 1.310 & 0.023 & $a,c,d$ & $+2.747$ & $+1.422$ & $-0.747$ & $-1.641$ & $-1.695$ & $+0.128$ & $-1.061$ & $-1.361$ & $-1.719$ & $-1.398$ \\
\object{U Oct}       & 0 &    1 & 302.5 & Mira &  M & $ +2.163$ & 0.083 & 1.215 & 0.135 & $a,c$ & $+3.561$ & $+2.327$ & $+0.606$ & $+0.079$ & $+0.190$ & $+1.133$ & $+0.357$ & & & $+0.091$ \\
\object{R Oph}       & 0 &    1 & 303.1 & Mira &  M & $ +1.019$ & 0.074 & 1.366 & 0.021 & $a,c,d$ & $+2.633$ & $+1.013$ & $-0.911$ & $-1.339$ & $-1.314$ & $-0.357$ &          & & $-1.284$ & $-1.225$ \\
\object{RY Oph}      & 0 &    1 & 150.5 & Mira &  M & $ +2.958$ & 0.061 & 1.240 & 0.038 & $a,c,g$ &          &          & $+1.396$ & $+0.918$ & $+0.617$ & $+2.154$ & $+1.368$ & & & $+1.071$ \\
\object{V Oph}       & 1 & 12,13& 297.3 & Mira &  C & $ +1.642$ & 0.051 & 1.975 & 0.075 & $a,b$ & $+4.146$ & $+1.916$ & $-0.027$ & $-0.234$ & $-0.558$ &          & $+0.300$ & & $-0.282$ & $+0.111$ \\
\object{$o$1 Ori}    & 1 &1,4,14&  30.0 & SRV  & MS & $ -0.662$ & 0.164 & 1.145 & 0.249 & $a$ & $+0.630$ & $-0.471$ & $-1.198$ & $-1.256$ & $-1.461$ & $-0.713$ & $-0.985$ & & $-1.165$ & $-0.793$ \\
\object{S Ori}       & 1 &   10 & 433.4 & Mira &  M &  $-0.193$ & 0.039 & 1.352 & 0.022 & $a,c,d$ & $+1.326$ & $-0.037$ & $-1.818$ & $-2.442$ & $-2.536$ & $-1.397$ & $-2.458$ & & $-2.822$ & $-2.473$ \\
\object{V1365 Ori}   & 0 &    1 &  62.8 & SRV  &  M & $ +0.713$ & 0.330 & 1.151 & 0.423 & $a$ &          &          & $-0.063$ & $-0.109$ & $-0.237$ &          & $+0.222$ & & $+0.070$ & $+0.324$ \\
\object{W Ori}       & 0 & 12,13& 204.9 & SRV  &  C & $ -0.470$ & 0.402 & 1.752 & 0.492 & $a$ & $+1.303$ & $-0.299$ & $-2.033$ & $-2.214$ & $-2.692$ & $-1.447$ & $-1.950$ & $-2.537$ & $-2.620$ & $-1.868$ \\
\object{NU Pav}      & 0 &    7 &  86.5 & SRV  &  M & $ -1.526$ & 0.248 & 1.313 & 0.298 & $a$ & $-0.263$ & $-1.469$ & $-2.280$ & $-2.352$ & $-2.322$ & $-1.751$ & $-2.106$ & $-2.113$ & $-2.092$ & $-1.911$ \\
\object{SY Pav}      & 0 &   15 & 190.6 & Mira &  M & $ +6.404$ & 0.156 & 1.146 & 0.021 & $a,i$ &          &          & $+4.754$ &          &          & $+5.710$ &          & & & $+4.627$ \\
\object{HR Peg}      & 1 &1,4,14&  50.0 & SRV  &  S & $ +1.041$ & 0.206 & 1.265 & 0.340 & $a$ &          & $+0.890$ & $+0.027$ & $-0.066$ & $+0.000$ & $+0.599$ & $+0.307$ & & $+0.192$ & $+0.434$ \\
\object{RZ Peg}      & 1 &    1 & 436.9 & Mira &  C & $ +2.127$ & 0.302 & 1.807 & 0.434 & $a$ &          &          & $+0.640$ & $-0.047$ & $-0.222$ & $+1.223$ & $+0.342$ & & $-0.386$ & $+0.019$ \\
\object{S Peg}       & 0 &    7 & 314.4 & Mira &  M &  $+1.478$ & 0.063 & 1.280 & 0.025 & $a,f$ & $+2.790$ & $+1.397$ & $-0.370$ & $-0.811$ & $-0.701$ & $+0.214$ & $-0.587$ & & $-0.507$ & $-0.819$ \\
\object{TW Peg}      & 0 &    7 &  94.1 & SRV  &  M &  $-0.539$ & 0.095 & 1.304 & 0.077 & $a,h$ & $+0.656$ & $-0.538$ & $-2.416$ & $-3.385$ & $-3.147$ & $-1.535$ & $-2.734$ & $-2.689$ & $-2.839$ & $-2.920$ \\
\object{TX Peg}      & 1 &    4 &  85.9 & SRV  &  M & $ +1.436$ & 0.206 & 1.218 & 0.340 & $a$ & $+2.489$ & $+1.304$ &          &          &          & $+0.662$ & $-0.065$ & & $-0.012$ & $-0.029$ \\
\object{UW Peg}      & 1 &    4 & 101.7 & SRV  &  M & $ +2.348$ & 0.270 & 1.322 & 0.414 & $a$ & $+3.597$ & $+2.299$ & $+0.836$ & $+0.156$ & $+0.157$ & $+1.555$ & $+0.567$ & & & $+0.633$ \\
\object{W Peg}       & 0 &    7 & 343.9 & Mira &  M & $ +0.008$ & 0.077 & 1.417 & 0.049 & $a,f$ & $+1.479$ & $-0.024$ & $-2.217$ & $-2.863$ & $-2.604$ & $-1.244$ & $-2.337$ & $-2.150$ & $-2.212$ & $-2.357$ \\
\object{X Peg}       & 0 &    1 & 200.9 & Mira &  M & $ +4.468$ & 0.080 & 1.136 & 0.022 & $a,c$ &          &          & $+2.933$ & $+2.503$ &          & $+3.556$ & $+2.811$ & & & $+2.872$ \\
\object{Z Peg}       & 1 &    1 & 327.8 & Mira &  M & $ +1.088$ & 0.188 & 1.310 & 0.351 & $a,g$ & $+2.496$ & $+1.135$ & $-0.697$ & $-1.315$ & $-1.117$ & $+0.078$ & $-0.918$ & & $-0.836$ & $-0.712$ \\
\object{UZ Per}      & 0 &    7 & 926.7 & SRV  &  M &  $+0.894$ & 0.234 & 1.461 & 0.377 & $a$ &          &          & $-0.888$ & $-1.962$ & $-1.873$ & $-0.029$ & $-1.381$ & & $-1.696$ & $-1.538$ \\
\object{RX Psc}      & 1 &   11 & 281.4 & Mira &  S & $ +5.020$ &       & 1.290 &       & $o$ &          &          & $+3.913$ &          &          & $+4.409$ & $+3.840$ & & & $+3.931$ \\
\object{U Psc}       & 0 &   15 & 172.6 & Mira &  M & $ +6.642$ & 0.016 & 0.904 & 0.026 & $a$ &          &          & $+4.794$ &          &          & $+5.782$ &          & & &  $+4.791$ \\
\object{Z Psc}       & 1 & 12,13& 155.8 & SRV  &  C & $ +0.865$ & 0.202 & 1.515 & 0.372 & $a$ & $+2.326$ & $+0.885$ & $-0.180$ & $-0.553$ & $-1.104$ & $+0.148$ & $-0.218$ & & $-1.086$ & $-0.146$ \\
\object{L2 Pup}      & 1 &  1,4 & 140.5 & SRV  &  M & $ -2.308$ & 0.170 & 1.245 & 0.204 & $a$ & $-1.004$ & $-2.377$ & $-4.830$ & $-5.247$ & $-4.745$ &          &          & $-4.867$ & $-4.576$ & $-5.295$ \\
\object{NQ Pup}      & 1 &   15 &  11.5 & SRV  &  S & $ +2.306$ & 0.274 & 1.236 & 0.400 & $a$ & $+3.348$ & $+2.272$ & $+1.514$ & $+1.408$ & $+1.063$ & $+2.039$ & $+1.710$ & & & $+1.853$ \\
\object{CO Pyx}      & 1 &   11 & 329.0 & Mira &  S & $ +5.400$ &       & 1.330 &       & $o$ &          &          & $+3.890$ & $+3.554$ &          & $+4.394$ & $+3.429$ & & & $+3.649$ \\
\object{Y Scl}       & 0 &    9 &  96.0 & SRV  &  M & $ +0.308$ & 0.012 & 1.285 & 0.032 & $a,d$ &          &          & $-1.494$ & $-2.113$ & $-1.783$ & $-0.448$ & $-1.706$ & $-1.539$ & $-1.493$ & $-1.738$ \\
\object{RR Sco}      & 0 &    1 & 278.9 & Mira &  M & $ -0.305$ & 0.071 & 1.207 & 0.028 & $a,c,d,g$ & $+1.128$ & $-0.314$ & $-2.062$ & $-2.547$ & $-2.509$ & $-1.360$ & $-2.258$ & & & $-2.131$ \\
\object{RZ Sco}      & 0 &    1 & 159.3 & Mira &  M & $ +4.152$ & 0.061 & 1.106 & 0.008 & $a,c,g$ &          &          & $+1.582$ & $+1.161$ & $+1.203$ & $+2.745$ & $+1.888$ & & & $+1.608$ \\
\object{V635 Sco}    & 1 &    6 &  58.0 & SRV  &  S & $ +1.198$ & 0.244 & 1.382 & 0.330 & $a$ & $+2.886$ & $+1.538$ & $+0.697$ & $+0.469$ &          & $+1.180$ &          & & & $+0.872$ \\
\object{S Sct}       & 1 &   13 & 149.7 & SRV  &  C & $ +0.578$ & 0.008 & 1.750 & 0.013 & $a,b$ &          &          & $-0.908$ & $-1.025$ & $-2.230$ & $-0.330$ & $-0.680$ & $-1.389$ & $-1.871$ & $-0.505$ \\
\object{DX Ser}      & 0 &    7 & 360.0 & SRV  &  M &  $+1.471$ & 0.085 & 1.364 & 0.027 & $a,e$ & $+2.790$ & $+1.492$ & $+0.100$ & $-0.451$ & $-0.531$ & $+0.830$ & $+0.062$ & & $-0.470$ & $+0.112$ \\
\object{R Ser}       & 1 &  1,3 & 353.6 & Mira &  M & $ +0.630$ & 0.138 & 1.323 & 0.022 & $a,c,g$ & $+1.730$ & $+0.375$ & $-2.073$ & $-2.560$ & $-2.453$ & $-0.986$ & $-1.931$ & $-2.156$ & $-2.078$ & $-1.864$ \\
\object{Y Ser}       & 0 &    7 & 432.7 & SRV  &  M & $ +2.249$ & 0.038 & 1.392 & 0.015 & $a,e$  &          &          & $+0.340$ & $-0.563$ & $+0.056$ & $+1.330$ & $+0.013$ & & $+0.138$ & $-0.130$ \\
\object{R Sgr}       & 0 &    1 & 268.6 & Mira &  M & $ +2.057$ & 0.075 & 1.199 & 0.016 & $a,c,g$ &          &          & $+0.273$ & $-0.058$ & $+0.136$ &          & $+0.159$ & & $+0.038$ & $+0.396$ \\
\object{RV Sgr}      & 1 &    1 & 318.9 & Mira &  M & $ +1.618$ & 0.042 & 1.213 & 0.012 & $a,c,g$ & $+3.208$ & $+1.596$ & $-0.023$ & $-0.451$ & $-0.355$ & $+0.829$ & $-0.042$ & & & $+0.208$ \\
\object{RX Sgr}      & 1 &    1 & 333.3 & Mira &  M & $ +2.999$ & 0.067 & 1.335 & 0.031 & $a,c$ &          &          & $+1.282$ & $+0.852$ &          & $+1.870$ &          & & & $+1.226$ \\
\object{T Sgr}       & 1 &  1,3 & 392.8 & Mira &  S & $ +1.134$ & 0.047 & 1.542 & 0.044 & $a,c$ &          &          & $-0.384$ & $-0.818$ & $-1.327$ & $+0.315$ & $-0.340$ & & & $-0.377$ \\
\object{V781 Sgr}    & 1 & 12,13& 208.7 & SRV  &  C & $ +1.666$ & 0.234 & 1.834 & 0.327 & $a$ & $+3.334$ & $+1.757$ & $+0.473$ & $+0.195$ &          & $+1.109$ & $+0.617$ & & & $+0.628$ \\
\object{SHV 0520}    & 1 &   15 & 264.0 & SRV  &  M & $+12.012$ & 0.019 & 0.854 & 0.021 & $a,k,l$ &          &          &          &          &          &          &          & & & $+8.853$ \\
\object{V1139 Tau}   & 1 &   14 &  80.5 & SRV  &  M & $ +0.978$ & 0.176 & 1.300 & 0.288 & $a$   & $+2.292$ & $+1.074$ & $+0.161$ & $-0.088$ & $-0.861$ & $+0.617$ & $+0.221$ & & $-0.888$ & $+0.333$ \\
\object{BH Tel}      & 0 &   15 & 216.5 & Mira &  M & $ +6.199$ & 0.038 & 1.415 & 0.183 & $a,i$ &          &          & $+4.470$ &          &          & $+4.732$ & $+3.873$ & & & $+4.297$ \\
\object{X TrA}       & 1 &    1 & 364.2 & SRV  &  C & $ -0.613$ & 0.258 & 1.785 & 0.363 & $a$   & $+1.050$ & $-0.522$ & $-2.129$ & $-2.322$ & $-2.737$ & $-1.461$ & $-2.016$ & $-2.516$ & $-2.630$ & $-1.862$ \\
\object{R Tri}       & 0 &    1 & 266.5 & Mira &  M & $ +0.966$ & 0.208 & 1.181 & 0.341 & $a,g$ &          &          & $-0.798$ & $-1.193$ & $-1.133$ & $-0.228$ & $-0.943$ & & $-0.809$ & $-0.730$ \\
\object{T Tuc}       & 0 &  7,9 & 247.1 & Mira &  M & $ +3.245$ & 0.122 & 1.157 & 0.021 & $a,f$ &          &          & $+1.245$ & $+0.764$ & $+0.632$ & $+2.261$ & $+1.392$ & & & $+1.285$ \\
\object{RR UMa}      & 0 &    1 & 231.4 & Mira &  M & $ +5.182$ & 0.015 & 1.259 & 0.027 & $a$   &          &          & $+3.211$ & $+2.904$ &          & $+3.717$ & $+2.963$ & & & $+3.399$ \\
\object{S UMa}       & 1 &  1,2 & 227.3 & Mira &  S & $ +3.041$ & 0.246 & 1.441 & 0.326 & $a,g$ &          &          & $+2.064$ & $+1.705$ & $+1.525$ & $+2.724$ & $+2.071$ & & & $+1.792$ \\
\object{T UMa}       & 0 &    1 & 255.5 & Mira &  M & $ +2.739$ & 0.344 & 1.350 & 0.460 & $a,g$ & $+4.177$ & $+2.896$ & $+0.836$ & $+0.235$ & $+0.047$ & $+1.137$ & $+0.389$ & & $+0.393$ & $+0.307$ \\
\object{VW UMa}      & 0 &    7 &  66.0 & SRV  &  M & $ +1.899$ & 0.260 & 1.219 & 0.350 & $a$   & $+2.817$ & $+1.720$ & $+1.016$ & $+0.794$ & $+0.647$ & $+1.630$ & $+1.165$ & & & $+1.324$ \\
\object{VY UMa}      & 1 & 12,13& 120.4 & SRV  &  C & $ +0.484$ & 0.184 & 1.466 & 0.320 & $a$   & $+1.977$ & $+0.627$ & $-0.658$ & $-0.795$ & $-1.512$ & $-0.159$ & $-0.484$ & $-1.178$ & $-1.009$ & $-0.444$ \\
\object{T UMi}       & 0 &   10 & 229.1 & Mira &  M & $ +2.894$ & 0.318 & 1.553 & 0.411 & $a$   &          &          & $+0.734$ & $+0.239$ & $+0.596$ & $+2.409$ & $+1.489$ & & & $+1.670$ \\
\object{T UMi}       & 0 &   10 & 113.6 & SRV  &  M & $ +2.894$ & 0.318 & 1.553 & 0.411 & $a$   &          &          & $+0.734$ & $+0.239$ & $+0.596$ & $+2.409$ & $+1.489$ & & & $+1.670$ \\
\object{CI Vel}      & 0 &   15 & 141.4 & Mira &  M & $ +6.676$ & 0.020 & 1.167 & 0.030 & $a$   &          &          & $+4.937$ &          &          & $+5.293$ & $+5.355$ & & & $+4.732$ \\
\object{HP Vel}      & 1 &    5 & 104.1 & SRV  &  S & $ +1.893$ & 0.278 & 1.363 & 0.374 & $a$   &          &          & $+1.063$ & $+0.842$ & $+0.105$ & $+1.591$ & $+1.118$ & & $-0.022$ & $+1.067$ \\
\object{ER Vir}      & 0 &  7,9 &  55.0 & SRV  &  M & $ +1.400$ & 0.004 & 1.124 & 0.014 & $a,h$ & $+2.537$ & $+1.479$ & $+0.820$ & $+0.670$ & $+0.635$ & $+1.254$ & $+1.025$ & & & $+1.170$ \\
\object{EV Vir}      & 0 &  7,9 &  69.2 & SRV  &  M & $ +1.517$ & 0.011 & 1.190 & 0.003 & $a,h$ &          &          & $+0.711$ & $+0.497$ & $+0.687$ & $+1.177$ & $+0.803$ & & $+0.452$ & $+0.920$ \\
\object{R Vir}       & 0 &    1 & 146.0 & Mira &  M & $ +2.071$ & 0.076 & 1.117 & 0.025 & $a,c$ & $+3.425$ & $+2.193$ & $+0.522$ & $+0.098$ & $+0.126$ & $+1.012$ & $+0.427$ & & & $+0.311$ \\
\object{RS Vir}      & 0 & 1,7,9& 352.0 & Mira &  M & $ +1.120$ & 0.112 & 1.340 & 0.059 & $a,c$ & $+2.672$ & $+1.135$ & $-1.464$ & $-2.466$ & $-2.500$ & $-0.708$ & $-2.187$ & $-2.533$ & $-2.647$ & $-2.024$ \\
\object{RU Vir}      & 1 &   13 & 437.3 & Mira &  C & $ +1.792$ & 0.057 & 2.865 & 0.037 & $a,b$ &          &          & $-2.275$ & $-2.530$ & $-2.653$ & $-1.605$ & $-2.169$ & $-2.441$ & $-2.522$ & $-2.258$ \\
\object{S Vir}       & 1 &  7,9 & 370.4 & Mira &  M & $ +0.293$ & 0.060 & 1.277 & 0.019 & $a,c$ & $+1.928$ & $+0.513$ & $-1.696$ & $-2.277$ & $-2.102$ & $-0.576$ & $-1.469$ & & $-1.674$ & $-1.777$ \\
\object{X Vol}       & 1 &    5 & 288.0 & Mira &  S & $ +3.243$ & 0.278 & 1.497 & 0.369 & $a$   &          &          & $+1.790$ & $+1.456$ & $+1.992$ & $+2.552$ & $+1.778$ & & & $+1.892$ \\
\object{R Vul}       & 0 &    1 & 136.8 & Mira &  M & $ +3.241$ & 0.248 & 1.462 & 0.336 & $a,g$ &          &          & $+1.551$ & $+1.129$ &          & $+2.270$ & $+1.548$ & & & $+1.463$ \\
\object{RU Vul}      & 0 &   10 & 108.8 & SRV  &  M & $ +4.704$ & 0.020 & 1.223 & 0.028 & $a$   &          &          & $+2.433$ & $+1.890$ & $+1.828$ & $+2.696$ & $+1.541$ & & & $+1.416$ \\
\object{BD+044356}   & 1 &    5 &  49.0 & SRV  &  S & $ +3.735$ & 0.224 & 1.230 & 0.227 & $a$   & $+3.966$ & $+2.904$ & $+3.003$ & $+2.776$ &          & $+3.610$ & $+3.269$ & & & $+3.332$ \\
\object{CSS 344}     & 1 &    5 &  79.9 & SRV  &  S & $ +4.304$ & 0.026 & 1.290 & 0.032 & $a$   &          &          & $+3.584$ & $+3.379$ &          & $+4.073$ & $+3.669$ & & & $+3.759$ \\
\object{CSS 466}     & 1 &   11 & 385.0 & Mira &  S & $ +4.600$ &  & 2.240 &  & $o$ &          &          & $+2.643$ & $+1.836$ &          & $+3.378$ & $+2.242$ & & & $+1.927$ \\
\object{CSS 739}     & 1 &   11 & 336.0 & Mira &  S & $ +5.500$ &  & 1.310 &  & $o$ &          &          & $+3.288$ & $+2.751$ &          & $+4.004$ & $+3.202$ & & & $+3.489$ \\
\object{Hen 4-8}     & 1 &    5 & 170.5 & SRV  &  S & $ +5.355$ & 0.031 & 1.151 & 0.042 & $a$ &          &          & $+3.536$ & $+2.780$ &          & $+4.513$ & $+3.402$ & & & $+3.289$ \\
\object{Hen 4-19}    & 1 &    5 &  48.2 & SRV  &  S & $ +5.433$ & 0.029 & 1.176 & 0.036 & $a$ &          &          & $+4.709$ &          &          & $+5.129$ & $+4.746$ & & & $+4.898$ \\
\object{Hen 4-20}    & 1 &    5 & 138.9 & SRV  &  S & $ +4.277$ & 0.036 & 1.238 & 0.041 & $a$ &          &          & $+3.370$ & $+3.150$ &          & $+3.943$ & $+3.372$ & & & $+3.427$ \\
\object{Hen 4-27}    & 1 &   11 & 363.3 & Mira &  S & $ +5.160$ &  & 1.800 &  & $o$ &          &          & $+3.288$ & $+2.803$ &          & $+4.127$ & $+3.508$ & & & $+3.214$ \\
\object{Hen 4-33}    & 1 &   11 & 405.0 & Mira &  S & $ +5.480$ &  & 1.520 &  & $o$ &          &          & $+3.133$ & $+2.506$ &          & $+3.991$ & $+3.078$ & & & $+3.255$ \\
\object{Hen 4-36}    & 1 &    5 &  56.8 & SRV  &  S & $ +2.653$ & 0.254 & 1.397 & 0.380 & $a$ &          &          & $+1.727$ & $+1.601$ & $+1.360$ & $+2.282$ & $+1.864$ & & & $+1.974$ \\
\object{Hen 4-37}    & 1 &    5 & 108.6 & SRV  &  S & $ +4.273$ & 0.276 & 1.223 & 0.277 & $a$ &          &          & $+3.361$ & $+3.092$ &          & $+3.896$ & $+3.460$ & & & $+3.638$ \\
\object{Hen 4-39}    & 1 &    5 &  46.6 & SRV  &  S & $ +3.209$ & 0.262 & 1.327 & 0.343 & $a$ &          &          & $+2.768$ & $+2.700$ &          & $+3.305$ & $+2.940$ & & & $+3.051$ \\
\object{Hen 4-41}    & 1 &    5 & 125.1 & SRV  &  S & $ +2.727$ & 0.280 & 1.626 & 0.400 & $a$ &          &          & $+1.514$ & $+1.199$ & $+1.157$ & $+2.160$ & $+1.688$ & & & $+1.756$ \\
\object{Hen 4-57}    & 1 &    6 & 122.5 & SRV  &  S & $ +3.848$ & 0.228 & 1.294 & 0.229 & $a$ &          &          & $+3.027$ & $+2.813$ &          & $+3.544$ & $+3.096$ & & & $+3.079$ \\
\object{Hen 4-64}    & 1 &    5 &  59.1 & SRV  &  S & $ +3.495$ & 0.298 & 1.487 & 0.389 & $a$ &          &          & $+2.584$ & $+2.299$ &          & $+3.046$ & $+2.627$ & & & $+2.733$ \\
\object{Hen 4-66}    & 1 &    5 & 216.0 & SRV  &  S & $ +4.277$ & 0.036 & 1.328 & 0.044 & $a$ &          &          & $+3.638$ & $+3.395$ &          & $+4.108$ & $+3.614$ & & & $+3.791$ \\
\object{Hen 4-81}    & 1 &   11 & 408.8 & Mira &  S & $ +4.990$ &  & 1.790 &  & $o$ &          &          & $+3.640$ & $+3.141$ &          & $+4.169$ & $+3.504$ & & & $+3.226$ \\
\object{Hen 4-84}    & 1 &   11 & 408.9 & Mira &  S & $ +4.630$ &  & 1.860 &  & $o$ &          &          & $+2.508$ & $+1.909$ &          & $+3.704$ & $+2.764$ & & & $+2.475$ \\
\object{Hen 4-88}    & 1 &    6 &  97.8 & SRV  &  S & $ +4.235$ & 0.016 & 1.207 & 0.023 & $a$ &          &          &          &          &          & $+4.055$ & $+3.574$ & & & $+3.786$ \\
\object{Hen 4-89}    & 1 &    6 & 100.1 & SRV  &  S & $ +3.708$ & 0.254 & 1.499 & 0.389 & $a$ &          &          &          &          &          & $+3.349$ & $+2.961$ & & & $+3.005$ \\
\object{Hen 4-95}    & 1 &    5 &  83.3 & SRV  &  S & $ +3.798$ & 0.254 & 1.084 & 0.256 & $a$ &          &          & $+2.272$ & $+1.285$ &          & $+2.927$ & $+1.934$ & & & $+1.663$ \\
\object{Hen 4-104}   & 1 &    5 &  40.3 & SRV  &  S & $ +4.160$ & 0.036 & 1.252 & 0.041 & $a$ &          &          & $+3.526$ & $+3.150$ &          & $+4.045$ & $+3.677$ & & & $+3.769$ \\
\object{Hen 4-109}   & 1 &   11 & 281.1 & Mira &  S & $ +4.830$ &  & 1.370 &  & $o$ &          &          & $+3.788$ & $+3.606$ &          & $+4.345$ & $+3.681$ & & & $+3.827$ \\
\object{Hen 4-122}   & 1 &   11 & 307.0 & Mira &  S & $ +5.260$ &  & 1.620 &  & $o$ &          &          & $+4.031$ &          &          & $+4.596$ & $+4.038$ & & & $+4.309$ \\
\object{Hen 4-162}   & 1 &    6 &  65.4 & SRV  &  M & $ +2.864$ & 0.256 & 1.410 & 0.382 & $a$ & $+4.498$ & $+3.074$ & $+2.087$ & $+1.929$ &          & $+2.600$ & $+2.152$ & & & $+2.239$ \\
\object{Hen 4-177}   & 1 &    6 &  67.9 & SRV  &  M & $ +3.173$ & 0.280 & 1.532 & 0.383 & $a$ &          &          & $+2.197$ &          &          & $+2.795$ & $+2.474$ & & & $+2.536$ \\
\object{Plaut 3-45}  & 0 &    8 & 271.0 & Mira &  M & $ +6.650$ & 0.332 & 1.450 & 0.081 & $a,i,m$ &          &          &          &          &          & $+5.580$ &          & & & $+4.567$ \\
\object{Plaut 3-70}  & 0 &    8 & 166.5 & SRV  &  M & $ +7.150$ & 0.177 & 1.290 & 0.052 & $a,n$   &          &          &          &          &          & $+6.346$ &          & & & $+5.688$ \\
\object{Plaut 3-100} & 0 &    8 & 298.7 & Mira &  M & $ +6.370$ & 0.171 & 1.290 & 0.019 & $a,m$   &          &          & $+3.879$ & $+2.244$ &          & $+4.958$ & $+3.236$ & & & $+3.660$ \\
\object{Plaut 3-143} & 0 &    8 & 204.2 & Mira &  M & $ +7.960$ & 0.109 & 1.250 & 0.025 & $a,i,m$ &          &          &          &          &          & $+6.817$ &          & & & $+6.045$ \\
\object{Plaut 3-195} & 0 &    8 & 216.6 & Mira &  M & $ +7.580$ & 0.063 & 1.210 & 0.065 & $a,i,m$ &          &          &          &          &          & $+6.102$ &          & & & $+5.697$ \\
\object{Plaut 3-277} & 0 &    8 & 263.2 & Mira &  M & $ +7.110$ & 0.276 & 1.320 & 0.029 & $a,m$   &          &          &          &          &          & $+5.835$ &          & & & $+4.785$ \\
\object{Plaut 3-315} & 0 &    8 & 326.8 & Mira &  M & $ +6.540$ & 0.049 & 1.370 & 0.029 & $a,m$   &          &          & $+4.788$ &          &          & $+5.142$ & $+4.032$ & & & $+4.172$ \\
\object{Plaut 3-328} & 0 &    8 & 166.8 & SRV  &  M & $ +7.750$ & 0.006 & 1.200 & 0.003 & $a,n$   &          &          &          &          &          & $+6.462$ &          & & & $+5.522$ \\
\object{Plaut 3-331} & 0 &    8 & 311.1 & Mira &  M & $ +6.600$ & 0.135 & 1.460 & 0.074 & $a,m$   &          &          & $+4.455$ &          &          & $+5.212$ & $+4.143$ & & & $+4.299$ \\
\object{Plaut 3-626} & 1 &    8 & 311.1 & Mira & MS & $ +7.190$ & 0.127 & 1.260 & 0.050 & $a,i,m$ &          &          &          &          &          & $+6.027$ &          & & & $+5.726$ \\
\object{Plaut 3-639} & 0 &    8 & 167.3 & SRV  &  M & $ +7.450$ & 0.087 & 1.260 & 0.027 & $a,i,n$ &          &          &          &          &          & $+6.165$ &          & & & $+5.136$ \\
\object{Plaut 3-719} & 0 &    8 & 277.8 & SRV  &  M & $ +6.610$ & 0.198 & 1.300 & 0.026 & $a,i,n$ &          &          &          &          &          & $+5.382$ & $+4.162$ & & & $+4.523$ \\
\object{Plaut 3-794} & 0 &    8 & 303.5 & Mira &  M & $ +6.040$ & 0.150 & 1.350 & 0.072 & $a,m$   &          &          &          &          &          & $+4.613$ & $+3.600$ & & & $+3.310$ \\
\object{Plaut 3-942} & 1 &    8 & 338.0 & Mira & MS & $ +6.550$ & 0.102 & 1.340 & 0.005 & $a,i,n$ &          &          &          &          &          & $+5.888$ &          & & & $+5.236$ \\
\object{Plaut 3-1002}& 0 &    8 & 190.5 & SRV  &  M & $ +7.080$ & 0.111 & 1.220 & 0.049 & $a,i,n$ &          &          &          &          &          &          &          & & & $+5.012$ \\
\object{Plaut 3-1008}& 0 &    8 & 232.1 & SRV  &  M & $ +6.720$ & 0.066 & 1.250 & 0.018 & $a,i,m$ &          &          &          &          &          & $+5.844$ &          & & & $+4.835$ \\
\object{Plaut 3-1059}& 0 &    8 & 144.1 & SRV  &  M & $ +7.690$ & 0.037 & 1.230 & 0.044 & $a,n$   &          &          &          &          &          &          &          & & & $+6.641$ \\
\object{Plaut 3-1147}& 1 &    8 & 395.6 & Mira &  M & $ +5.770$ & 0.042 & 1.500 & 0.064 & $a,m$   &          &          & $+3.405$ & $+2.763$ &          &          &          & & & $+3.373$ \\
\object{Plaut 3-1176}& 0 &    8 & 184.1 & SRV  &  M & $ +6.940$ & 0.072 & 1.330 & 0.028 & $a,i,n$ &          &          &          &          &          &          &          & & & $+5.724$ \\
\object{Plaut 3-1179}& 0 &    8 & 278.2 & Mira &  M & $ +7.200$ & 0.104 & 1.360 & 0.033 & $a,i,m$ &          &          & $+3.881$ & $+2.961$ &          & $+5.599$ & $+4.253$ & & & $+4.251$ \\
\object{Plaut 3-1204}& 0 &    8 & 197.0 & SRV  &  M & $ +7.310$ & 0.034 & 1.330 & 0.010 & $a,m$   &          &          & $+4.807$ &          &          & $+5.519$ & $+4.236$ & & & $+4.334$ \\
\object{Plaut 3-1287}& 0 &    8 & 312.5 & Mira &  M & $ +6.640$ & 0.275 & 1.310 & 0.031 & $a,m$   &          &          &          &          &          & $+4.828$ &          & & & $+3.620$ \\
\object{Plaut 3-1313}& 0 &    8 & 378.7 & Mira &  M & $ +6.251$ & 0.021 & 1.470 & 0.036 & $a$     &          &          & $+3.884$ & $+2.797$ &          & $+4.656$ & $+3.570$ & & & $+2.974$ \\
\object{Plaut 3-1347}& 1 &    8 & 408.9 & Mira & MS & $ +5.990$ & 0.111 & 1.480 & 0.078 & $a,m$   &          &          & $+3.983$ & $+3.109$ &          & $+4.429$ & $+3.517$ & & & $+3.812$ \\
\object{Plaut 3-1470}& 0 &    8 & 183.2 & SRV  &  M & $ +7.340$ & 0.022 & 1.210 & 0.011 & $a,n$   &          &          &          &          &          & $+6.570$ &          & & & $+5.911$ \\
\object{Plaut 3-1517}& 0 &    8 & 188.8 & SRV  &  M & $ +7.680$ & 0.087 & 1.260 & 0.028 & $a,i,n$ &          &          &          &          &          & $+6.536$ &          & & & $+5.185$ \\
\object{Plaut 3-1991}& 0 &    8 & 124.7 & SRV  &  M & $ +8.020$ & 0.220 & 1.150 & 0.048 & $a,i,n$ &          &          &          &          &          &          &          & & & $+5.633$ \\
\end{longtable}
\tablefoot{Meaning of the columns: (1): name of the star; (2): Tc content (0=Tc-poor, 1=Tc-rich);
    (3): reference for Tc content: 1: \citet{Lit87}, 2: \citet{SL88}, 3: \citet{Van91}, 4: \citet{LH99},
    5: \citet{VJ99}, 6: \citet{Van00}, 7: \citet{LH03}, 8: \citet{Utt07}, 9: \citet{UL10}, 10: \citet{Utt11},
    11: \citet{Smo12}, 12: \citet{Abia02}, 13: \citet{BM93}, 14: \citet{Van98}, 15: this work (XSL);
    (4): pulsation period; (5): variability type; (6): spectral type; (7): K-band magnitude;
    (8): uncertainty of the K-band magnitude; (9): $J-K$ colour; (10): uncertainty of the $J-K$ colour;
    (11): reference for the sources of J- and K-band magnitudes and uncertainties: $a$: \citet{2MASS},
    $b$: \citet{Whi06}, $c$: \citet{Cat79}, $d$: \citet{Fou92}, $e$: \citet{KH94}, $f$: \citet{Whi00},
    $g$: \citet{Whi08}, $h$: \citet{Ker95}; $i$: \citet{DENIS}; $j$: \citet{Tab09}; $k$: \citet{Cio00};
    $l$: \citet{Mac15}; $m$: \citet{Blo92}; $n$: \citet{Schulthe98}; $o$: \citet{Smo12}; $p$: \citet{2MASS-6X};
    (If there is only one measurement, e.g.\ from 2MASS, the uncertainty is that of the individual measurement
    [K-band] or the quadratically combined uncertainty [$J-K$ colour]; if there are several measurements, the
    uncertainty is calculated as the standard error of the mean.) (12): COBE/DIRBE 1.25 and 2.20\,$\mu$m
    magnitude \citep{Pri10}; (13), (14), and (15): IRAS 12, 25, and 60\,$\mu$m magnitude \citep{Neu84};
    (16), (17), (18), and (19): Akari 9, 18, 65, and 90\,$\mu$m magnitude \citep{Mur07};
    (20): WISE 22\,$\mu$m magnitude \citep{Wri10}. The following zero-magnitude fluxes were used to convert
    fluxes to magnitudes: 1593.7 and 648.3\,Jy for COBE/DIRBE 1.25 and 2.20\,$\mu$m; 28.3, 6.73, and 1.19\,Jy
    for IRAS~12, 25, and 60; 58.08, 10.77, 0.9650, and 0.6276\,Jy for Akari~9, 18, 65, and 90, respectively.
    We refer to the original catalogues for uncertainties and other limitations of the photometric data.
    T~UMi is listed twice because it recently underwent a marked period decrease and switched from fundamental
    mode (Mira) to first overtone (SRV) pulsation \citep{Utt11}. SHV~0520 stands for \object{SHV 0520261-693826}.}
\end{landscape}
}

\longtab[2]{
\begin{longtable}{lrccrrrr}
\caption{Gas mass-loss rates of Mira stars adopted from the literature. Information on Tc content 
(column~2), pulsation period (column~3), spectral type (column~4), $K$-band magnitude (column~5) and
WISE 22\,$\mu{\rm m}$-band magnitude (column~6) are as in Table~\ref{taba1}.}\label{taba2}\\
\hline
\hline
Name	  & Tc & $P$   & SpType & $K$    & [22]   & $\dot{M}_{\rm g}$ &  Ref. \\
          &    & days  &        & mag    & mag    & $M_{\sun}yr^{-1}$ &       \\
\hline
\endfirsthead
\caption{Continued.}\\
\hline
\hline
Name	  & Tc & $P$   & SpType & $K$    & [22]   & $\dot{M}_{\rm g}$ &  Ref. \\
          &    & days  &        & mag    & mag    & $M_{\sun}yr^{-1}$ &       \\
\hline
\endhead
\hline
\endfoot
\hline
\endlastfoot
KU And	  &    & 712.0 &  M  & $+1.115$ & $-3.539$ & 9.4$\times10^{-6}$ & 17  \\
LP And    &    & 599.3 &  C  & $+3.859$ & $-4.335$ & 1.5$\times10^{-5}$ & 15  \\
R And	  &  1 & 409.9 &  S  & $+0.431$ & $-1.667$ & 2.8$\times10^{-7}$ & 17  \\
W And	  &  1 & 396.3 &  M  & $+0.431$ & $-1.667$ & 2.8$\times10^{-7}$ & 17  \\
R Aql	  &  0 & 270.6 &  M  & $-0.826$ & $-3.486$ & 1.1$\times10^{-6}$ & 6   \\
RR Aql	  &  0 & 399.8 &  M  & $+0.626$ & $-2.939$ & 2.4$\times10^{-6}$ & 17  \\
RT Aql 	  &  0 & 327.8 &  M  & $+1.120$ & $-1.387$ & 1.0$\times10^{-7}$ & 6   \\
V1300 Aql &    & 680.0 &  M  & $+2.059$ & $-4.510$ & 1.0$\times10^{-5}$ & 17  \\
V1420 Aql &    & 694.0 &  C  & $+2.080$ & $-3.763$ & 6.9$\times10^{-6}$ & 8   \\
VX Aql	  &    & 618.6 &  S  & $+3.060$ & $+0.779$ & 1.0$\times10^{-7}$ & 13  \\
W Aql	  &  1 & 487.8 &  S  & $-0.556$ & $-4.578$ & 2.2$\times10^{-6}$ & 15  \\
R Aqr	  &  1 & 388.2 &  M  & $-0.821$ & $-4.352$ & 2.0$\times10^{-7}$ & 18  \\
RV Aqr	  &    & 452.6 &  C  & $+1.390$ & $-3.251$ & 2.3$\times10^{-6}$ & 17  \\
T Ari	  &    & 339.0 &  M  & $+0.171$ & $-1.496$ & 6.0$\times10^{-9}$ & 6   \\
U Ari	  &    & 372.0 &  M  & $+1.001$ & $-1.159$ & 6.5$\times10^{-8}$ & 6   \\
NV Aur	  &    & 635.0 &  M  & $+3.861$ & $-4.099$ & 2.5$\times10^{-5}$ & 17  \\
R Aur	  &    & 457.4 &  M  & $-0.690$ & $-2.859$ & 1.6$\times10^{-6}$ & 6   \\
U Aur	  &    & 408.9 &  M  & $+1.021$ & $-2.044$ & 6.3$\times10^{-7}$ & 6   \\
V370 Aur  &    & 683.0 &  C  & $+5.672$ & $-3.167$ & 1.7$\times10^{-5}$ & 3   \\
V373 Aur  &    & 590.0 &  M  & $+2.704$ & $-1.756$ & 1.6$\times10^{-5}$ & 3   \\
V Aur	  &    & 353.1 &  C  & $+3.065$ & $+1.357$ & 4.7$\times10^{-7}$ & 8   \\
BX Cam	  &    & 454.0 &  M  & $+0.908$ & $-4.039$ & 4.4$\times10^{-6}$ & 17  \\
T Cam	  &  1 & 373.9 &  S  & $+0.814$ & $+0.042$ & 1.0$\times10^{-7}$ & 13  \\
TX Cam	  &    & 557.4 &  M  & $-0.013$ & $-5.110$ & 1.0$\times10^{-5}$ & 15  \\
R Cap	  &    & 345.7 &  C  & $+3.537$ & $+0.278$ & 2.8$\times10^{-6}$ & 8   \\
RU Cap	  &    & 349.6 &  M  & $+2.983$ & $-1.023$ & 3.8$\times10^{-7}$ & 8   \\
HV Cas	  &    & 467.9 &  C  & $+2.466$ & $-1.105$ & 9.0$\times10^{-7}$ & 9   \\
R Cas	  &    & 430.5 &  M  & $-1.404$ & $-4.825$ & 4.0$\times10^{-7}$ & 14  \\
S Cas	  &    & 613.5 &  S  & $+1.760$ & $-3.099$ & 2.8$\times10^{-6}$ & 17  \\
T Cas	  &    & 444.7 &  M  & $-1.040$ & $-2.989$ & 1.1$\times10^{-6}$ & 6   \\
V701 Cas  &    & 567.0 &  C  & $+4.975$ & $-3.006$ & 4.5$\times10^{-6}$ & 17  \\
WY Cas	  &    & 480.6 &  S  & $+1.864$ & $-1.815$ & 1.1$\times10^{-6}$ & 15  \\
AQ Cen	  &    & 377.7 &  M  & $+1.756$ & $-2.342$ & 3.2$\times10^{-7}$ & 8   \\
TT Cen	  &    & 464.0 &  S  & $+2.430$ & $-0.273$ & 2.5$\times10^{-6}$ & 15  \\
AX Cep	  &    & 398.4 &  C  & $+1.581$ & $-1.145$ & 1.2$\times10^{-6}$ & 8   \\
S Cep	  &    & 486.0 &  C  & $-0.100$ & $-2.812$ & 1.5$\times10^{-6}$ & 10  \\
T Cep	  &  1 & 386.6 &  M  & $-1.824$ & $-3.580$ & 9.1$\times10^{-8}$ & 17  \\
$o$ Cet	  &  1 & 332.7 &  M  & $-2.420$ & $-5.899$ & 2.5$\times10^{-7}$ & 14  \\
R Cet	  &  0 & 165.9 &  M  & $+2.553$ & $-1.130$ & 1.9$\times10^{-7}$ & 8   \\
S CMi	  &    & 334.0 &  M  & $+0.500$ & $-1.564$ & 4.9$\times10^{-8}$ & 17  \\
R Cnc	  &  0 & 362.0 &  M  & $-0.622$ & $-2.802$ & 2.3$\times10^{-8}$ & 6   \\
W Cnc	  &    & 393.6 &  M  & $+1.143$ & $-1.578$ & 3.1$\times10^{-8}$ & 6   \\
S CrB	  &    & 360.6 &  M  & $-0.168$ & $-2.929$ & 2.3$\times10^{-7}$ & 17  \\
V CrB	  &  1 & 358.0 &  C  & $+1.301$ & $-0.997$ & 3.3$\times10^{-7}$ & 17  \\
BG Cyg	  &    & 288.4 &  M  & $+1.055$ & $-0.975$ & 7.4$\times10^{-8}$ & 8   \\
$\chi$ Cyg & 1 & 408.2 &  S  & $-1.902$ & $-4.168$ & 3.8$\times10^{-7}$ & 15  \\
R Cyg	  &  1 & 426.6 &  S  & $+0.861$ & $-2.039$ & 9.5$\times10^{-7}$ & 17  \\
U Cyg	  &  1 & 464.8 &  C  & $+1.174$ & $-1.512$ & 9.0$\times10^{-7}$ & 9   \\
V Cyg	  &    & 420.9 &  C  & $+0.020$ & $-3.449$ & 1.6$\times10^{-6}$ & 16  \\
V1426 Cyg &    & 470.0 &  C  & $+0.831$ & $-2.667$ & 1.0$\times10^{-5}$ & 9   \\
V1549 Cyg &    & 510.0 &  C  & $+2.793$ & $-2.349$ & 5.1$\times10^{-6}$ & 8   \\
V1906 Cyg &    & 600.0 &  M  & $+2.996$ & $-3.349$ & 8.2$\times10^{-6}$ & 1   \\
V1965 Cyg &    & 577.0 &  C  & $+2.950$ & $-3.113$ & 1.0$\times10^{-5}$ & 12  \\
V1968 Cyg &    & 783.0 &  C  & $+5.640$ & $-3.719$ & 7.5$\times10^{-6}$ & 17  \\
Z Cyg	  &    & 263.7 &  M  & $+2.557$ & $-1.336$ & 4.0$\times10^{-8}$ & 6   \\
U Dor	  &    & 394.4 &  M  & $+1.170$ & $-2.319$ & 5.9$\times10^{-7}$ & 8   \\
T Dra	  &  1 & 422.7 &  C  & $+1.637$ & $-2.039$ & 1.3$\times10^{-6}$ & 1   \\
RT Eri	  &    & 373.0 &  M  & $+0.662$ & $-2.191$ & 3.3$\times10^{-8}$ & 8   \\
W Eri	  &  0 & 373.6 &  M  & $+1.780$ & $-0.979$ & 1.8$\times10^{-7}$ & 8   \\
R For	  &    & 388.1 &  C  & $+1.210$ & $-1.846$ & 1.6$\times10^{-6}$ & 15  \\
UU For	  &    & 471.4 &  M  & $+1.271$ & $-2.897$ & 1.1$\times10^{-6}$ & 5   \\
R Gem	  &  1 & 370.1 &  S  & $+1.664$ & $-0.415$ & 4.3$\times10^{-7}$ & 17  \\
ZZ Gem	  &  1 & 315.6 &  C  & $+3.236$ & $+1.516$ & 9.1$\times10^{-8}$ & 8   \\
CK Gru	  &    & 580.0 &  M  & $+1.730$ & $-2.768$ & 1.6$\times10^{-6}$ & 5   \\
RU Her	  &  1 & 485.6 &  M  & $+0.192$ & $-2.717$ & 7.5$\times10^{-7}$ & 6   \\
U Her	  &    & 406.8 &  M  & $-0.634$ & $-3.180$ & 5.9$\times10^{-7}$ & 6   \\
V821 Her  &    & 524.0 &  C  & $+1.850$ & $-3.612$ & 3.0$\times10^{-6}$ & 17  \\
V833 Her  &    & 540.0 &  C  & $+4.190$ & $-4.343$ & 2.3$\times10^{-5}$ & 1   \\
V1076 Her &    & 609.0 &  C  & $+6.770$ & $-4.078$ & 1.7$\times10^{-5}$ & 1   \\
WY Her	  &    & 376.7 &  M  & $+2.940$ & $-0.975$ & 1.1$\times10^{-6}$ & 8   \\
R Hor	  &  1 & 404.6 &  MS & $-0.880$ & $-3.745$ & 5.9$\times10^{-7}$ & 17  \\
CZ Hya	  &  1 & 432.0 &  C  & $+2.434$ & $-0.451$ & 9.0$\times10^{-7}$ & 9   \\
IY Hya	  &    & 409.0 &  C  & $+1.964$ & $-2.888$ & 4.1$\times10^{-6}$ & 1   \\
R Hya	  &  1 & 376.6 &  M  & $-2.663$ & $-4.433$ & 2.1$\times10^{-7}$ & 15  \\
RU Hya	  &  0 & 333.2 &  M  & $+1.600$ & $-1.406$ & 1.3$\times10^{-7}$ & 8   \\
W Hya	  &  0 & 388.6 &  M  & $-3.215$ & $-5.202$ & 1.1$\times10^{-7}$ & 15  \\
X Hya	  &    & 300.8 &  M  & $+1.130$ & $-1.447$ & 3.0$\times10^{-8}$ & 6   \\
R Leo	  &    & 305.0 &  M  & $-2.550$ & $-4.773$ & 1.1$\times10^{-7}$ & 17  \\
R Lep	  &    & 435.4 &  C  & $+0.070$ & $-2.930$ & 8.7$\times10^{-7}$ & 17  \\
RT Lep	  &    & 393.0 &  M  & $+2.208$ & $-1.086$ & 4.4$\times10^{-7}$ & 8   \\
T Lep	  &    & 369.1 &  M  & $-0.266$ & $-2.300$ & 7.3$\times10^{-9}$ & 8   \\
RS Lib	  &    & 218.1 &  M  & $-0.598$ & $-2.237$ & 3.7$\times10^{-8}$ & 6   \\
R LMi	  &    & 373.2 &  M  & $-0.340$ & $-3.279$ & 2.6$\times10^{-7}$ & 17  \\
GI Lup	  &    & 465.8 &  S  & $+1.796$ & $+0.266$ & 5.5$\times10^{-7}$ & 13  \\
II Lup	  &    & 576.0 &  C  & $+1.790$ & $-4.769$ & 1.7$\times10^{-5}$ & 17  \\
V358 Lup  &    & 632.0 &  C  & $+4.360$ & $-4.160$ & 1.0$\times10^{-5}$ & 12  \\
Y Lup	  &    & 401.6 &  M  & $+1.360$ & $-1.762$ & 1.5$\times10^{-7}$ & 8   \\
AP Lyn	  &    & 730.0 &  M  & $+0.984$ & $-3.899$ & 6.5$\times10^{-6}$ & 15  \\
R Lyn	  &    & 378.8 &  S  & $+2.140$ & $+0.553$ & 3.3$\times10^{-7}$ & 15  \\
S Lyr	  &    & 438.3 &  S  & $+3.930$ & $-1.223$ & 2.0$\times10^{-6}$ & 15  \\
BQ Mic	  &    & 560.4 &  M  & $+2.120$ & $-1.792$ & 2.4$\times10^{-6}$ & 5   \\
CL Mon	  &  1 & 483.0 &  C  & $+1.940$ & $-1.234$ & 2.5$\times10^{-6}$ & 4   \\
GX Mon	  &    & 527.0 &  M  & $+1.317$ & $-4.085$ & 8.4$\times10^{-6}$ & 17  \\
V688 Mon  &    & 653.0 &  C  & $+4.278$ & $-2.456$ & 6.1$\times10^{-6}$ & 17  \\
RZ Mus	  &    & 333.3 &  M  & $+2.422$ & $+0.009$ & 1.4$\times10^{-6}$ & 8   \\
V2548 Oph &    & 747.0 &  C  & $+5.610$ & $-3.875$ & 2.9$\times10^{-5}$ & 3   \\
V Oph	  &  1 & 297.3 &  C  & $+1.689$ & $+0.111$ & 5.0$\times10^{-8}$ & 8   \\
X Oph	  &    & 333.9 &  M  & $-0.824$ & $-3.038$ & 2.5$\times10^{-7}$ & 11  \\
R Ori	  &    & 379.8 &  C  & $+4.385$ & $+2.368$ & 3.1$\times10^{-7}$ & 8   \\
S Ori	  &  1 & 433.4 &  M  & $-0.500$ & $-2.473$ & 2.1$\times10^{-7}$ & 6   \\
U Ori	  &    & 371.1 &  M  & $-0.263$ & $-3.999$ & 2.8$\times10^{-7}$ & 6   \\
V1259 Ori &    & 696.0 &  C  & $+7.730$ & $-3.118$ & 8.8$\times10^{-6}$ & 17  \\
V351 Pav  &    & 462.0 &  M  & $+2.174$ & $-1.552$ & 2.5$\times10^{-6}$ & 5   \\
IZ Peg	  &    & 486.0 &  C  & $+7.090$ & $-3.207$ & 2.1$\times10^{-5}$ & 2   \\
LL Peg	  &    & 696.0 &  C  & $+10.500$& $-5.170$ & 1.1$\times10^{-5}$ & 15  \\
R Peg	  &    & 377.8 &  M  & $+0.520$ & $-2.090$ & 4.1$\times10^{-7}$ & 11  \\
RZ Peg	  &  1 & 436.9 &  C  & $+2.127$ & $+0.019$ & 4.6$\times10^{-7}$ & 15  \\
TU Peg	  &    & 323.2 &  M  & $+1.277$ & $-0.683$ & 8.4$\times10^{-8}$ & 8   \\
W Peg	  &  0 & 343.9 &  M  & $+0.007$ & $-2.357$ & 9.8$\times10^{-8}$ & 6   \\
V384 Per  &    & 535.0 &  C  & $+1.150$ & $-3.798$ & 2.3$\times10^{-6}$ & 17  \\
Y Per	  &    & 250.8 &  C  & $+3.679$ & $+2.060$ & 1.0$\times10^{-7}$ & 8   \\
S Pic	  &    & 424.7 &  M  & $+0.300$ & $-2.670$ & 5.9$\times10^{-7}$ & 6   \\
W Pic	  &    & 326.0 &  C  & $+1.130$ & $-0.570$ & 2.3$\times10^{-7}$ & 15  \\
AW Psc	  &    & 730.0 &  M  & $+2.293$ & $-2.825$ & 4.6$\times10^{-6}$ & 5   \\
WX Psc	  &    & 650.0 &  M  & $+2.217$ & $-5.022$ & 4.0$\times10^{-5}$ & 15  \\
R Pyx	  &    & 367.4 &  C  & $+2.961$ & $+0.437$ & 2.1$\times10^{-7}$ & 8   \\
RR Sco	  &  0 & 278.9 &  M  & $-0.305$ & $-2.131$ & 1.1$\times10^{-8}$ & 8   \\
RS Sco	  &    & 319.6 &  M  & $+0.362$ & $-2.184$ & 2.2$\times10^{-8}$ & 6   \\
RT Sco	  &    & 449.2 &  S  & $+0.286$ & $-2.265$ & 4.5$\times10^{-7}$ & 15  \\
RW Sco	  &    & 389.6 &  M  & $+1.704$ & $-1.576$ & 2.1$\times10^{-7}$ & 8   \\
R Ser	  &  1 & 353.6 &  M  & $+0.733$ & $-1.864$ & 4.5$\times10^{-8}$ & 6   \\
RR Sgr	  &    & 334.3 &  M  & $+0.656$ & $-1.503$ & 5.4$\times10^{-8}$ & 8   \\
RZ Sgr	  &    & 207.0 &  S  & $+1.362$ & $-0.399$ & 3.0$\times10^{-6}$ & 15  \\
ST Sgr	  &    & 394.2 &  S  & $+1.625$ & $-0.772$ & 2.0$\times10^{-7}$ & 15  \\
T Sgr	  &  1 & 392.8 &  S  & $+1.134$ & $-0.377$ & 1.4$\times10^{-7}$ & 13  \\
V342 Sgr  &    & 372.0 &  M  & $+0.471$ & $-3.194$ & 1.9$\times10^{-6}$ & 8   \\
V2234 Sgr &    & 450.4 &  M  & $+2.015$ & $-2.907$ & 1.8$\times10^{-6}$ & 5   \\
V5104 Sgr &    & 655.0 &  C  & $+3.520$ & $-3.520$ & 9.3$\times10^{-6}$ & 3   \\
IK Tau	  &    & 450.0 &  M  & $-0.935$ & $-5.880$ & 2.0$\times10^{-5}$ & 16  \\
R Tau	  &    & 323.1 &  M  & $+0.889$ & $-1.613$ & 3.1$\times10^{-8}$ & 8   \\
RU Vir	  &  1 & 437.3 &  C  & $+1.882$ & $-2.258$ & 9.5$\times10^{-6}$ & 1   \\
RS Vir	  &  0 & 352.0 &  M  & $+1.150$ & $-2.024$ & 1.9$\times10^{-7}$ & 8   \\
S Vir	  &  1 & 370.4 &  M  & $+0.300$ & $-1.777$ & 8.4$\times10^{-8}$ & 7   \\
AI Vol	  &    & 511.0 &  C  & $+2.122$ & $-4.024$ & 4.9$\times10^{-6}$ & 17  \\
R Vol	  &    & 452.0 &  C  & $+1.710$ & $-2.206$ & 1.7$\times10^{-6}$ & 15  \\
\end{longtable}
\tablefoot{References for gas mass-loss rates:
1: \citet{KM85}; 2: \citet{Kna85}; 3: \citet{Kna86};
4: \citet{Olo88}; 5: \citet{Ny92}; 6: \citet{You95};
7: \citet{Kna98}; 8: \citet[][geometric mean]{Groen99};
9: \citet{Schoi01}; 10: \citet{Schoi02}; 11: \citet{Win03};
12: \citet{Schoi06}; 13: \citet{Ram09}; 14: \citet{DeBe10};
15: \citet{Schoi13}; 16: \citet{RO14}; 17: \citet{Dani15};
18: \citet{Ram18}.}
}

\end{appendix}

\end{document}